\documentclass{aa}

\usepackage[varg]{txfonts}
\usepackage[para,online,flushleft]{threeparttable}

\newcommand{\kms}{km~s$^{-1}$}
\newcommand{\arcs}{$^{\prime\prime}$}

\newcommand{\mspcc}{\ensuremath{\rm M_{\odot} pc^{-3} }}
\newcommand{\HI}{H{\sc i}}

\begin{document}

\title{Theoretical modelling of two-component molecular discs in spiral galaxies}
%\subtitle{Two-component molecular discs in spiral galaxies}
\author{Narendra Nath Patra\inst{\ref{inst}}}
\institute{Raman Research Institute, C. V. Raman Avenue, Sadashivanagar, Bengaluru - 560 080, India.\\ \email{narendra@rri.res.in}}
\label{inst}

%\author [Patra N. N]{  Narendra Nath Patra$^{1}$ \thanks {E-mail: narendra@ncra
%.tifr.res.in} \\
%       $^{1}$ NCRA-TIFR, Post Bag 3, Ganeshkhind, Pune 411 007, India \\       
%}

%\correspondingauthor{Narendra Nath Patra}
%\email{narendra@rri.res.in, naren.ncra@gmail.com}

%\author[0000-0002-9981-296X]{Narendra Nath Patra}

\abstract{As recent observations of the molecular discs in spiral galaxies point to the existence of a diffuse, low-density thick molecular disc along with the prominent thin one, we investigate the observational signatures of this thick disc by theoretically modelling two-component molecular discs in a sample of eight nearby spiral galaxies. Assuming a prevailing hydrostatic equilibrium, we set up and solved the joint Poisson's-Boltzmann equation to estimate the three-dimensional distribution of the molecular gas and the molecular scale height in our sample galaxies. The molecular scale height in a two-component molecular disc is found to vary between $50-300$ pc, which is higher than what is found in a single-component disc. We find that this scale height can vary significantly depending on the assumed thick disc molecular gas fraction. We also find that the molecular gas flares as a function of the radius and follows a tight exponential law with a scale length of $\left(0.48 \pm 0.01 \right) r_{25}$. We used the density solutions to produce the column density maps and spectral cubes to examine the ideal observing conditions to identify a thick molecular disc in galaxies. We find that unless the molecular disc is an edge-on system and imaged with a high spatial resolution ($\lesssim 100$ pc), it is extremely hard to identify a thick molecular disc in a column density map. The spectral analysis further reveals that at moderate to high inclination ($i \gtrsim 40^o$), spectral broadening can fictitiously introduce the signatures of a two-component disc into the spectral cube of a single-component disc. Hence, we conclude that a low inclination molecular disc imaged with high spatial resolution would serve as the ideal site for identifying the thick molecular disc in galaxies.}


\maketitle

\section{Introduction}

The molecular gas in galaxies is the place of birth of stars. Gravitational collapse and fragmentation of molecular clouds leads to the formation of stars \citep{toomre64,mckee07,ostriker10,hopkins11c,kim11,kennicutt12,krumholz14,agertz15b,
hayward17,krumholz17a}. In galaxies, molecular gas serves as the immediate precursor to star formation. In contrast, atomic gas plays the role of long-term supply for the star formation. The molecular gas in spiral galaxies dominates the inner parts of the interstellar medium (ISM), whereas the atomic gas dominates the outer regions. As the formation of stars must go through the molecular phase, this phase of the ISM is of immense importance in relation to galaxy formation and evolution \citep{myers86,shu87,scoville89,saintonge13,agertz15a}. Not only that, but the abundance and the distribution of this component can also lead to a better understanding of the processes and conditions by which the gas is converted into stars. For example, it is well known that in spiral galaxies, the molecular gas surface density directly correlates to the star formation rate density \citep{kennicutt98b,bigiel08,leroy09a,onodera10,schruba10,feldmann11,saintonge11c,leroy13b,
kruijssen14,kreckel18}; however, the same phenomenon in dwarf galaxies is not apparent \citep{schruba12}. Hence, investigating the structure and the distribution of the molecular gas in galaxies could lead to a better understanding of the interplay between gas and star formation.  

Estimating the three-dimensional distribution of the molecular and the atomic gas in external galaxies is difficult because of line-of-sight projection effects \citep{patra18a,patra19a}. In the Milky Way, however, it is possible to estimate the three-dimensional distribution of the molecular gas by locating the positions of the molecular clouds and remapping them into a common three-dimensional grid \citep{grabelsky87,solomon87,wouterloot90,rathborne09b,roman-duval10,roman-duval16}. However, the estimated distance to the clouds is very often found to be ambiguous as the kinematic distances introduce significant fractional errors \citep[e.g.][]{sanders84,scoville87,bronfman88}.

To overcome these errors, there have been several remarkable studies in recent years to explore the molecular gas in external galaxies. These studies have mostly focused on the following two aspects of the molecular discs: firstly, the global distribution of the molecular gas and its connection to the star formation \citep[see, e.g. ][]{leroy09a,schruba12,utomo18,sun18,schinnerer19}; and secondly, the phases of the molecular gas \citep{stark84,scoville90,combes97b,calduprimo13,mogotsi16,calduprimo16,roman-duval16}. The studies of the spectral properties in connection to the phases in the molecular discs reveal that the molecular gas in galaxies might exist in the following two phases: one with a low kinetic temperature and hence a narrow spectral width ($\sim 7$ \kms); the other with a larger spectral width, which is comparable to that of the atomic gas ($\sim 12$ \kms)\citep{wilson90,wilson94,calduprimo13,mogotsi16}. This high spectral width component is much more diffused and not easily detectable in the individual spectrum of localised CO emission, though it has a significant share in the total molecular mass \citep[see, e.g. ][]{goldsmith08,liszt10,pety13,roman-duval16}. It should be mentioned here that these velocity widths refer to a much higher kinetic temperature than what is expected in the molecular clouds and hence, these widths indicate the turbulent motions of the clouds rather than their pure thermal widths.

Interestingly, the diffuse component of the molecular disc found to have a velocity dispersion similar to that of the atomic gas (\HI) indicates that they might occupy the same dynamical phase space with similar scale heights \citep{calduprimo13,mogotsi16,calduprimo16}. This demands a theoretical understanding of the origin and the sustenance of this diffuse molecular component in environments where it is much harder to protect the molecular gas via shelf-shielding. An explanation for the existence of the diffuse molecular gas arises in the context of the presence of the molecular gas in the arm and the inter-arm regions of spiral galaxies. Observationally, it has been found that the giant molecular clouds (GMCs) are detected in higher numbers in the arm regions than the inter-arm regions in spiral galaxies \citep{dame86,hunter97,sawada04,nakanishi06}. To explain this, two different propositions are suggested. Firstly, it is proposed that, during the passage of the spiral density wave, there is an accelerated phase transition that converts the atomic gas in the inter-arm region to molecular gas in the arm very rapidly. The molecular gas again gets converted into atomic and ionised gas when the density wave gets passed through the inter-arm region \citep{blitz80,cohen80, scoville79,heyer15}. In the second scenario, it is proposed that there is a little phase transition. Instead, in the inter-arm regions (or in the low-density environment), there already exist tiny molecular clouds which can self-shield themselves. The existence of the GMCs in the arm region is then attributed to the dynamically driven coagulation and fragmentation of these tiny molecular clouds in the arms and the inter-arms, respectively \citep{scoville79,vogel88,hartmann01,kawamura09,miura12}. This, in turn, generates large molecular clouds in the arm region by the merger of smaller clouds, and the more massive clouds get fragmented dynamically once the spiral density wave passes through \citep{koda16}. For the first scenario to be true, no CO emission is expected to be observed in the inter-arm regions, which the earliest observations with insufficient sensitivity confirmed \citep{engargiola03,heyer04,fukui09,vogel88,rand90}. However, recent sensitive observations have revealed a significant amount of CO in the inter-arm regions as well \citep{liszt10,koda11,pety13}.

The existence of the molecular gas at large heights from the midplane can be easily explained under the assumption of the second scenario. The molecular gas does not get destroyed at these low-density environments; instead, it is protected within small clouds. In that case, the observed molecular spectra in these regions is decided by the turbulent motions of these clouds, which is much higher than the turbulent motions of the GMCs. This, in turn, results in a two-component molecular disc with two different molecular velocity dispersions. Not only that, but as this component is made up of tiny molecular clouds, it is diffused in nature and would be easily resolved out in high-resolution interferometric observations.

Several recent observations, in fact, strongly suggest the existence of this diffuse low-density molecular gas, which has a vertical thickness comparable to that of the atomic gas. For example, \citet{garcia-burillo92} found a thick molecular disc in the edge-on galaxy NGC 891, which extends up to $\sim 1.4$ kpc from the midplane. The existence of a thick molecular disc in this galaxy was further strengthened by the sensitive observations of the cold dust by \citet{bocchio16} using Spitzer telescope. They detected cold dust emission in NGC 891 with a vertical scale height of $\sim 1.44 $ kpc \citep[see also,][]{kreysa93,rejkuba09}. On the other hand, for the Galaxy, \citet{dame94} used the 1.2 m telescope of the Harvard-Smithsonian Center for Astrophysics to conduct a very sensitive survey of diffuse molecular gas in the inner Galaxy beyond 100 pc from the midplane. They found a faint, thick molecular disc approximately three times as wide as that of the central dense CO layer and which is comparable to the width of the \HI~layer. Likewise, many other spectral studies have found a similar velocity dispersion in the molecular discs, as observed in the atomic discs. For example, \citet{combes97b} measured the velocity dispersions in two nearly face-on galaxies, NGC 628 and NGC 3938, to find a very similar velocity dispersion of the molecular and the atomic gas. Using the data from The \HI~Nearby Galaxy Survey (THINGS) \citep{walter08} and the HERA CO Line Extra-galactic Survey (HERACLES) \citep{leroy09a}, \citet{calduprimo13} stacked the line-of-sight of \HI~and CO spectra to produce high signal to noise ratio (S/N) stacked spectra. Using these high S/N spectra, they also observed similar velocity dispersions in both the molecular and the atomic discs \citep[see,][also for a similar analysis in Andromeda galaxy]{calduprimo16}. Later, \citet{mogotsi16} used the same data to study the \HI~and the CO spectra of bright individual regions to find a somewhat lower velocity dispersion in the molecular disc than what is observed in the atomic disc. This is consistent with the scenario that diffuse high dispersion molecular gas cannot be detected in a low S/N individual spectrum. \citet{pety13} studied the molecular gas in M 51 using both single-dish and interferometric observations and found that the single-dish observation recovers much more flux than the interferometric observation. They conclude that at least 50\% of gas in the molecular disc of M 51 is in the diffused state, which is resolved out in the interferometric observation. A similar fraction of the diffuse molecular gas in other galaxies was also found in several other studies \citep[see, e.g. ][]{goldsmith08,liszt10,calduprimo15,roman-duval16}.

With these observational inferences and theoretical propositions, it is worth investigating in detail the observational aspects due to the presence of two-component molecular discs in galaxies and, in particular, the impact of the diffused molecular component on the observed thickness and the spectral properties of the molecular discs. To do that, we modelled the galactic disc as a four-component system with a stellar disc, an atomic disc, and a two-component molecular disc in hydrostatic equilibrium under their mutual gravity in the external force field of the dark matter halo. Under this assumption, the density distributions of different baryonic components can be described by the joint Poisson's-Boltzmann equation, which upon solving, would provide a detailed three-dimensional distribution of the molecular and atomic gas. Furthermore, these density distributions can be used to build dynamical models of the molecular and atomic discs, which again can be used to produce the column density maps and the spectral cubes. These simulated maps and cubes can then be studied to understand the observational signatures of a two-component molecular disc as compared to a single-component one. In this paper, we formulate the joint Poisson's-Boltzmann equation for a four-component galactic system and solve it numerically to understand the distribution of the molecular gas and its observational consequences in a sample of eight nearby spiral galaxies from the HERACLES and THINGS survey. 

\section{Sample}

We chose our sample galaxies from the parent sample of the HERACLES survey. As part of the HERACLES survey, 18 galaxies were observed in CO using the 30-m IRAM telescope. These galaxies were chosen to be a part of the THINGS (for available \HI~data) and {\it Spitzer} Infrared Nearby Galaxies Survey (SINGS) \citep{kennicutt03} (for available stellar surface density data). In hydrostatic equilibrium, the gravity in the vertical direction is balanced by the pressure. Hence, the surface densities of different baryonic discs are essential inputs to the hydrostatic equation. Not only that, but the dark matter halo also provides a significant amount of gravity, which decides the vertical distribution of gas under hydrostatic equilibrium. Given these circumstances, the surface density of different baryonic discs and the mass model, which can be used to obtain the dark matter density distribution, for a galaxy are primary requirements in solving the hydrostatic equation. However, all of the 18 galaxies observed in the HERACLES survey did not have the mass models due to complications in extracting their rotation curves (e.g. low and high inclination, the presence of strong non-circular motion in \HI~data, etc.) \citep[see][for more details]{deblok08}. These constraints restrict ten galaxies for which hydrostatic modelling could not be performed. We hence solved the hydrostatic equation for the eight remaining galaxies for which all the necessary data are available. In Table~\ref{tab:samp}, we list the basic properties of our sample galaxies. In column (1), we list the names of our sample galaxies, column (2) shows the distance to the galaxy in megaparsec, columns (3) and (4) represent the inclination and the position angle of the optical discs, respectively, and column (5) shows the optical radius, $r_{25}$, the 25$^{th}$ magnitude B-band isophote. Column (6) represents the radial extent of the molecular discs in our galaxies as obtained by \citet[][]{schruba11} (See their Fig. A.1).

\begin{table}[h]
\caption{Basic properties of our sample galaxies}
\begin{center}
\begin{threeparttable}
\begin{tabular}{lccccc}

\hline
Name & Dist$^a$ & Incl$^a$ & PA$^a$ & $r_{25}^a$ & $R_{CO}^b$\\
     &(Mpc) & ($^o$) & ($^o$) & ($^\prime$) & (kpc)\\
\hline
NGC 925   &  9.2   &  66  &  287  &  5.3  &  6.4\\
NGC 2841  &  14.1  &  74  &  154  &  5.3  &  11.9\\
NGC 2976  &  3.6   &  65  &  335  &  3.6  &  1.9\\
NGC 3198  &  13.8  &  72  &  215  &  3.2  &  8.6\\
NGC 3521  &  10.7  &  73  &  340  &  4.2  &  12.1\\
NGC 5055  &  10.1  &  59  &  102  &  5.9  &  15.7\\
NGC 6946  &  5.9   &  33  &  243  &  5.7  &  9.7\\
NGC 7331  &  14.7  &  76  &  168  &  4.6  &  19.8\\
\hline

\end{tabular}
\begin{tablenotes}
\item[a]\citet{walter08}, \item[b]\citet{schruba11}
\end{tablenotes}
\end{threeparttable}
\end{center}
\label{tab:samp}
\end{table}

\section{Modelling the galactic discs}
\subsection{Setting up the hydrostatic equation}

We assume the galactic disc to be a four-component system that consists of a stellar disc, an atomic disc, and a two-component molecular disc. The molecular disc considered here has the following two components: one with high velocity dispersion (thick disc) and diffuse gas that has large scale heights; the other with low velocity dispersion (thin disc), which is settled close to the midplane. All of these baryonic discs are considered to be in hydrostatic equilibrium under their mutual gravity in the external force field of the dark matter halo. For simplicity, we assume that these baryonic discs are concentric, coplanar, and azimuthally symmetric. Under these assumptions, the joint Poisson's equation for an elemental mass can be written in cylindrical polar coordinate as  

\begin{equation}
\frac{1}{R} \frac{\partial }{\partial R} \left( R \frac{\partial \Phi_{total}}{\partial R} \right) + \frac{\partial^2 \Phi_{total}}{\partial z^2} = 4 \pi G \left( \sum_{i=1}^{4} \rho_{i} + \rho_{h} \right)
,\end{equation}

\noindent where $\Phi_{total}$ is the total gravitational potential due to all of the disc components and the dark matter halo. We note that $\rho_i$ denotes the volume density, where $\it i$ runs for stars, atomic gas, thin disc molecular gas, and thick disc molecular gas. Additionally, $\rho_h$ denotes the mass density of the dark matter halo, which can be described by either an NFW profile \citep{navarrofrenkwhite97}  

\begin{equation}
\label{nfw}
\rho_{h} (R) = \frac {\rho_0}{\frac{R}{R_s} \left( 1 + \frac{R}{R_s}\right)^2}
\end{equation}

\noindent or by an isothermal profile

\begin{equation}
\label{eq_iso}
\rho_h(R) = \frac {\rho_0}{1 + \left(\frac{R}{R_s}\right)^2}
,\end{equation}

\noindent where $\rho_0$ denotes the characteristic density, and $R_s$ represents the scale radius. These two halo parameters completely describe a spherically symmetric dark matter halo.

For an individual component in hydrostatic equilibrium, the gradient in pressure (in the $z$ direction) is balanced by the gradient in potential. Thus, 

\begin{equation}
\frac{\partial }{\partial z} \left(\rho_i {\langle {\sigma}_z^2 \rangle}_i \right) + \rho_i \frac{\partial \Phi_{total}}{\partial z} = 0
,\end{equation}

\noindent where ${\langle {\sigma}_z \rangle}_i$ is the vertical velocity dispersion of the $i^{th}$ component, which is an input parameter.

\noindent By eliminating ${\Phi}_{total}$ from Equation (1) and (4), we get 

\begin{equation}
\begin{split}
{\langle {\sigma}_z^2 \rangle}_i \frac{\partial}{\partial z} \left( \frac{1}{\rho_i} \frac{\partial \rho_i}{\partial z} \right) &= \\ 
&-4 \pi G \left( \rho_s + \rho_{HI} + \rho_{H2} + \rho_{H2}^{\prime} + \rho_h \right)\\ 
&+ \frac{1}{R} \frac{\partial}{\partial R} \left( R \frac{\partial \Phi_{total}}{\partial R} \right)
\end{split}
,\end{equation}

\noindent where $\rho_s$, $\rho_{HI}$, $\rho_{H2}$, and $\rho_{H2}^{\prime}$ are the mass densities of stars, \HI, molecular gas in the thin disc, and molecular gas in the thick disc, respectively. 

The above equation can be further simplified using 

\begin{equation}
{\left( R \frac{\partial \Phi_{total}}{\partial R} \right)}_{R,z} = {(v_{rot}^2)}_{R,z}
.\end{equation}

\noindent Assuming that the vertical gradient in ${(v_{rot})}_{R,z}$ is small, one can approximate ${(v_{rot})}_{R,z}$ by the observed rotation curve $v_{rot}$, which is a function of $R$ alone. Thus Equation (5) reduces to 
 
\begin{equation}
\begin{split}
{\langle {\sigma}_z^2 \rangle}_i \frac{\partial}{\partial z} \left( \frac{1}{\rho_i} \frac{\partial \rho_i}{\partial z} \right) &= \\
&-4 \pi G \left( \rho_s + \rho_{HI} + \rho_{H2} + \rho_{H2}^{\prime} + \rho_h \right)\\ 
&+ \frac{1}{R} \frac{\partial}{\partial R} \left( v_{rot}^2 \right)
\end{split}
\label{eq_hydro}
.\end{equation}

\noindent Eq.~\ref{eq_hydro} represents four coupled, second-order ordinary partial differential equations in the variables ${\rho}_s$, $\rho_{HI}$, $\rho_{H2}$, and $\rho_{H2}^{\prime}$. The solutions of these equations at any radius $R$, provide the density distributions as a function of $z$. Thus solving these equations at all radii renders a detailed three-dimensional density distribution of different baryonic discs.

\begin{figure*}
\begin{center}
\begin{tabular}{c}
\resizebox{1.\textwidth}{!}{\includegraphics{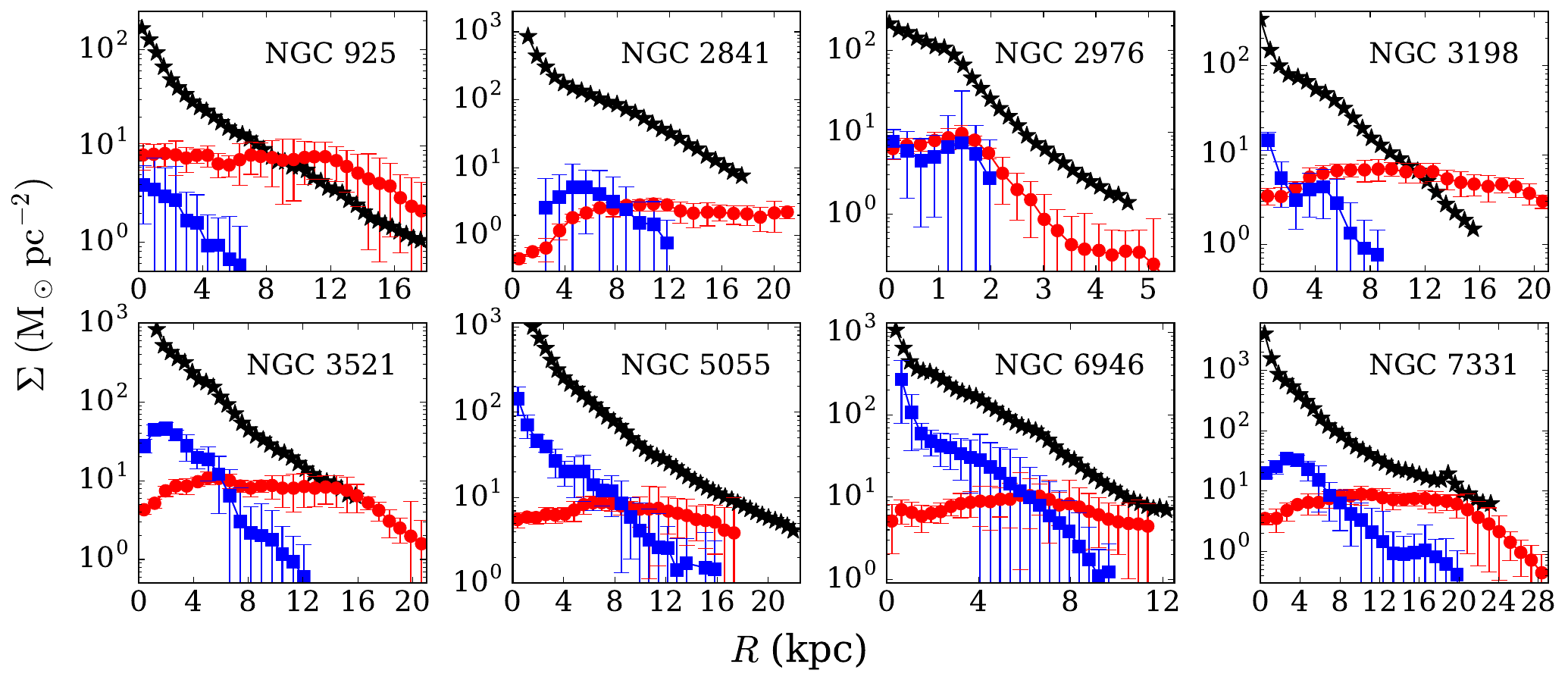}}
\end{tabular}
\end{center}
\caption{Surface density profiles for our sample galaxies. Different panels indicate the surface densities for different galaxies, as mentioned in the top left corner of each panel. The black asterisks, the red circles, and the blue squares represent the stellar, the atomic, and the molecular surface density profiles, respectively. The stellar surface density profiles are taken from \citet{leroy09a}, whereas the atomic and the molecular surface density profiles are obtained from \citet{schruba11}.}
\label{sden}
\end{figure*}

\subsection{Input parameters}

In vertical hydrostatic equilibrium, the gravity produced by the baryonic discs and the dark matter halo is balanced by the pressure. As the baryons in the galactic discs provide a significant amount of gravity, the surface density profiles of different baryonic components are one of the primary inputs to Eq.~\ref{eq_hydro}. We obtained the stellar surface density profiles for our sample galaxies as derived by \citet{leroy09a} using the 3.6 $\mu m$ data from the SINGS survey. We used the atomic and the molecular surface density profiles from \citet{schruba11}. \citet{schruba11} derived the \HI~and the molecular surface density profiles using the THINGS and HERACLES survey data, respectively. The atomic and the molecular surface density profiles were determined by averaging the \HI~and the CO spectra, respectively, in the tilted rings of width 15\arcs as extracted by the process of the rotation curve fitting \citep[see][for more details]{deblok08}. To enhance the sensitivity of the molecular surface density profiles, the CO spectra within a tilted ring were co-added after shifting their central velocities to zero. These co-added and stacked spectra have a much higher S/N than the individual spectrum, which is crucial for the inclusion of the low-density thick disc molecular gas. In Fig.~\ref{sden}, we plotted the surface density profiles for different baryonic discs in our sample galaxies. As can be seen from the figure, for all our sample galaxies, the stellar surface density dominates the atomic or the molecular surface densities, though, the gas discs tend to extend to a larger radius than the stellar disc.

The vertical velocity dispersions of different baryonic components are the other important inputs to Eq.~\ref{eq_hydro}. They provide the necessary pressure support to balance gravity. As the vertical velocity dispersion of the stellar disc does not significantly influence the vertical scale height (defined as the half width at half maxima) of the gas discs \citep{banerjee11}, we calculated the stellar velocity dispersion analytically assuming the stellar disc to be an isothermal single component system in hydrostatic equilibrium \citep[see][for more details]{leroy08}. The velocity dispersions in the gaseous discs, on the other hand, can be estimated using spectroscopic observations with radio telescopes. For example, early low-resolution \HI~observations resulted in an \HI~velocity dispersion ($\sigma_{HI}$) between 6-13 \kms~in nearby spiral galaxies \citep{shostak84,vanderkruit84,kamphuis93}. In an extensive analysis of the second moments of the \HI~spectral cubes of THINGS galaxies, \citet{tamburro09} found a mean velocity dispersion of $\sim$ 10 \kms~at the optical radius, $r_{25}$. Later, \citet{ianjamasimanana12} used the same sample to stack line-of-sight \HI~spectra and found a $\sigma_{HI} = 12.5 \pm 3.5$ \kms~($\sigma_{HI} = 10.9 \pm 2.1$ \kms~for galaxies with inclination less than 60$^o$). \citet{calduprimo13} studied the $\sigma_{HI}$ and $\sigma_{CO}$ in a sample of 12 nearby galaxies from the THINGS survey and found a $\sigma_{HI}/\sigma_{CO} = 1.0 \pm 0.2$ with a median  of $\sigma_{HI} = 11.9 \pm 3.1$ \kms. However, \citet{mogotsi16} used the same sample to study the high S/N individual spectra and found a $\sigma_{HI}/\sigma_{CO} = 1.4 \pm 0.2$ with $\sigma_{HI} = 11.7 \pm 2.3$ \kms. As can be seen from these studies, a $\sigma_{HI} \sim$ 12 \kms~is reasonable to assume in \HI~discs of our sample galaxies.

\cite{stark84} observed molecular clouds in the Galaxy and found that the velocity dispersions of the low mass clouds are higher than that of the high mass clouds. The low mass clouds have a $\sigma_{CO} \sim $ 9.0 \kms~, whereas the high mass clouds have a $\sigma_{CO} \sim$ 6.6 \kms, which is also reconfirmed by their subsequent observations \citep{stark05}. In investigating the spectral cubes of the THINGS and the HERACLES survey data, \citet{calduprimo13} and \citet{mogotsi16} conclude that due to its diffuse and faint nature, the thick disc molecular gas could only be detected in high S/N stacked spectra. They found a $\sigma_{CO} \sim 7.3 \pm 1.7$ \kms~for the thin disc molecular gas. The $\sigma_{CO}$ for the thick disc, on the other hand, is expected to be the same as the $\sigma_{HI}$. 

Given these results, for our sample galaxies, we assume velocity dispersions of $\sim 7$ \kms~and $\sim 12$ \kms~in the thin and the thick molecular discs, respectively. The fraction of the total molecular gas in the thick disc is not well constrained by observations. \citet{pety13} estimated that at least $\sim 50\%$ of the molecular gas in M51 is in the thick disc. Here, in this analysis, we assume that the molecular gas is equally divided into the thin and the thick disc. However, we also explore the effect of a different assumed thick disc molecular gas fraction on our results.

\begin{figure*}
\begin{center}
\begin{tabular}{c}
\resizebox{1.\textwidth}{!}{\includegraphics{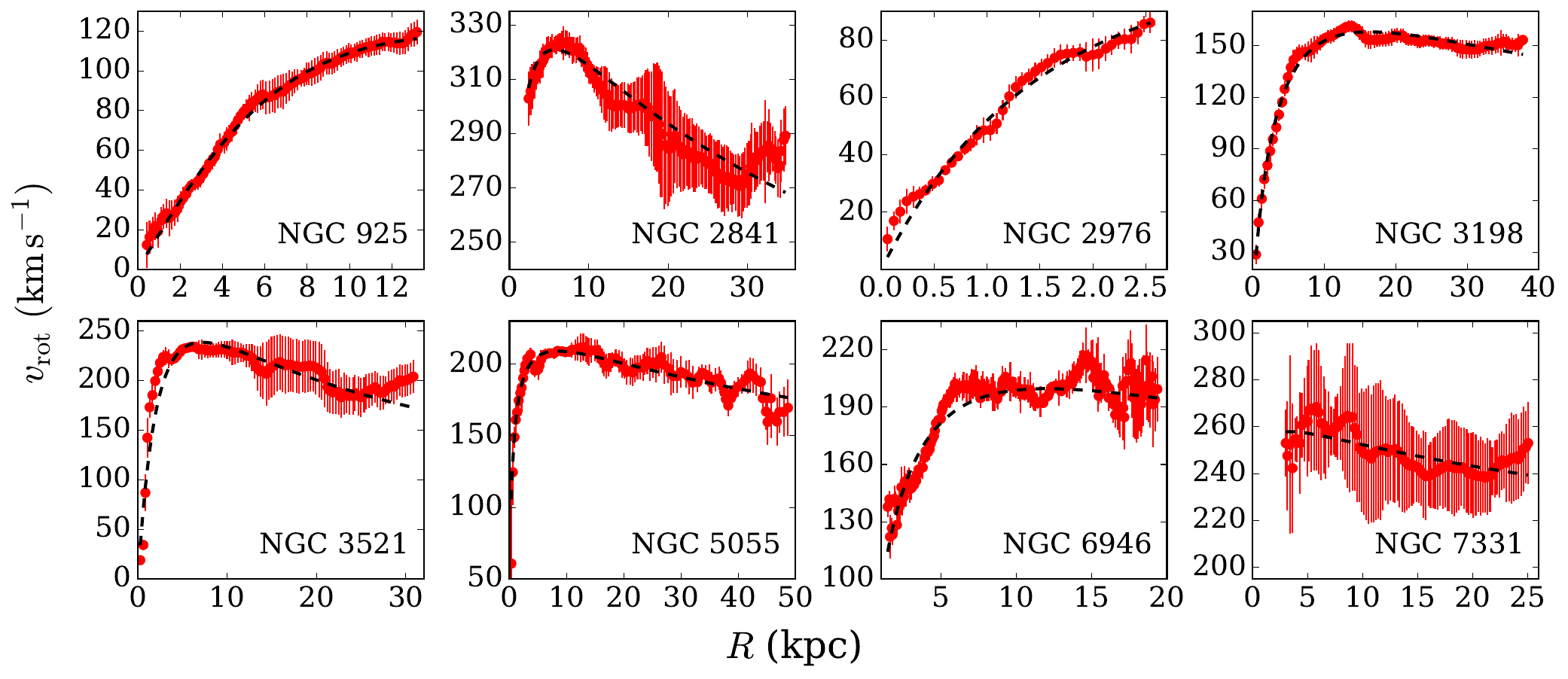}}
\end{tabular}
\end{center}
\caption{Rotation curves of our sample galaxies as derived using the \HI~data from the THINGS survey \citep{deblok08}. The rotation curves were fitted using a Brandt profile (Eq. 8) for parametrisation. In each panel, the red circles with error bars represent the rotation curve, whereas the black dashed line represents the Brandt profile fit. The best-fit parameters can be found in Table~\ref{tab:rotcur}.}
\label{rotcur}
\end{figure*}

The next input parameter is the rotation curve. The last term on the RHS of Eq.~\ref{eq_hydro} represents the radial term, which might influence the solution of Eq.~\ref{eq_hydro} considerably in the inner parts of a galaxy where the rotation curve is not flat. This term can be calculated using the knowledge of the observed rotation curve. In Fig.~\ref{rotcur}, we plotted the rotation curves of our sample galaxies as extracted by \citet{deblok08} using the THINGS survey data. As we use a second-order derivative of the rotation curve in Eq.~\ref{eq_hydro}, we parameterised it to prevent any possible divergence while solving the hydrostatic equation. We employed a commonly used Brandt's profile to represent the rotation curves. The Brandt's profile can be given as \citep{brandt60}

\begin{equation}
v_{rot} (R) = \frac{V_{max}\left(R/R_{max} \right)}{\left(1/3 + 2/3 \left(\frac{R}{R_{max}}\right)^n\right)^{3/2n}}
,\end{equation}

\noindent where $V_{max}$ is the maximum rotational velocity and $R_{max}$ is the radius at which this maximum velocity is attained. Additionally, $n$ represents the power-law index, which decides how fast the rotation curve rises as a function of the radius. In Table~\ref{tab:rotcur}, we show the best fit parameter values for the rotation curves of our sample galaxies. 

\begin{table}
\caption{Best fit parameters of the rotation curves}
\begin{center}
\begin{tabular}{lccc}

\hline
Name & $V_{max}$ & $R_{max}$ & $n$ \\
     &(\kms) & (kpc) & \\
\hline
NGC 925   & 117.96$\pm$1.39 & 16.42$\pm$0.87 & 1.80$\pm$0.10\\
NGC 2841  & 320.89$\pm$0.51 & 5.69$\pm$0.20  & 0.36$\pm$0.02\\
NGC 2976  & 103.45$\pm$13.61& 6.86$\pm$3.03  & 1.05$\pm$0.28\\
NGC 3198  & 158.09$\pm$0.68 & 16.36$\pm$0.31 & 0.82$\pm$0.03\\
NGC 3521  & 238.61$\pm$1.73 & 7.22$\pm$0.23  & 1.19$\pm$0.07\\
NGC 5055  & 209.07$\pm$0.63 & 8.45$\pm$0.18  & 0.36$\pm$0.01\\
NGC 6946  & 199.68$\pm$0.50 & 12.03$\pm$0.58 & 0.69$\pm$0.05\\
NGC 7331  & 257.86$\pm$1.65 & 3.45$\pm$1.27  & 0.12$\pm$0.04\\
\hline

\end{tabular}
\end{center}
\label{tab:rotcur}
\end{table}

The last input required to solve the hydrostatic equation is the density distribution of the dark matter halo. The dark matter halo provides a considerable amount of gravity, which influences the hydrostatic structure in the vertical direction, especially at the outer regions where the dark matter dominates the gravity. We used the dark matter distributions from the mass models of our sample galaxies as derived by \citet{deblok08}. For our sample galaxies, neither an NFW profile nor an ISO profile describes all the galaxies systematically better than the other. Hence, for a particular galaxy, we selected a dark matter profile that describes the observed rotation curve better \citep[see][for more details]{deblok08}. In Table~\ref{tab:dmpar}, we present the dark matter halo types and their profile parameters, which we used to solve Eq.~\ref{eq_hydro} for our sample galaxies.

\begin{table}
\caption{Dark matter halo parameters}
\begin{center}
\begin{tabular}{lccc}
\hline
Name & DM halo & $R_c$ & $\rho_0$ \\
     &         & (kpc) & ($\times 10^{-3} $ \mspcc)\\
\hline
NGC 925   & ISO & 9.67  & 5.90\\
NGC 2841  & NFW & 20.55 & 12.40\\
NGC 2976  & ISO & 5.09  & 35.50\\
NGC 3198  & ISO & 2.71  & 47.50\\
NGC 3521  & ISO & 1.32  & 370.20\\
NGC 5055  & ISO & 11.73 & 4.80\\
NGC 6946  & ISO & 3.62  & 45.70\\
NGC 7331  & NFW & 60.20 & 1.05\\
\hline

\end{tabular}
\end{center}
\label{tab:dmpar}
\end{table}

\subsection{Solving the hydrostatic equilibrium equation}

With these inputs, Eq.~\ref{eq_hydro} can be solved to obtain detailed three-dimensional density distributions of different baryonic discs in our sample galaxies. A comprehensive description on how to solve Eq.~\ref{eq_hydro} assuming the galactic disc to be a three-component system (stars + atomic gas + single-component molecular gas) can be found in \citet{patra18a,patra19a}. Here we used a similar approach, but with one extra component in the molecular disc. In this subsection, we briefly illustrate the scheme we used to solve a four-component galactic disc. Eq.~\ref{eq_hydro} refers to four second-order partial differential equations in stars, atomic gas, and two molecular gas components coupled through gravity-term in the RHS. 

As these equations are second-order differential equations, one needs at least two initial conditions to solve them. We chose the following conditions to solve our equations:

\begin{equation}
\left( \rho_i \right)_{z = 0} = \rho_{i,0} \ \ \ \ {\rm and} \ \ \ \left(\frac{d \rho_i}{dz}\right)_{z=0} = 0
\label{init_cond}
.\end{equation}

\noindent These two conditions signify that the volume densities of individual baryonic discs and their derivatives (along $z$) at the midplane are known. The second condition is a result of the force balance in the midplane \citep{spitzer42}. Due to symmetry in the vertical direction, the gravitational potential at the midplane is expected to be maximum. This, in turn, would result in a density maxima. Hence, the first derivative of the densities, for all the baryonic discs, would be zero at $z=0$. On the other hand, the first condition is not generic and assumes that the midplane density is known. However, it is not a directly measurable quantity, though, it can be estimated using the knowledge of the observed surface density, $\Sigma_i$. Since it is an unknown quantity, for an individual component, we first assumed a trial value, $\rho_{0,i,t}$, and solved Eq.~\ref{eq_hydro} to obtain a trial solution, $\rho_{i,t}(z)$. This trial solution was then integrated to calculate the trial surface density, $\Sigma_i = \int \rho_i (z) dz$, and compared with the observed value, $\Sigma_{i,o}$, to update the trial midplane density in the next iteration. We repeated this exercise until we found a trial midplane density, which produces a surface density matched to the observed value with better than 0.1\% accuracy.

Using the scheme mentioned above, Eq.~\ref{eq_hydro} can be solved for a disc without any coupling, that is, in the absence of any other baryonic discs. However, as Eq.~\ref{eq_hydro} refers to a coupled equation, all of the components should be solved simultaneously. We implemented this using an iterative approach similar to what has been used in many earlier studies \citep{banerjee08,patra14,patra18a,patra19a}. In the first iteration, we solved for individual components assuming no coupling (i.e. solved separately). In this case, the densities of the other components are assumed to be zero. For example, if we are solving for stars ($\rho_s$), then $\rho_{HI}$, $\rho_{H2}$, and $\rho_{H2}^\prime$ are considered to be zero. Thus at the end of the first iteration, we obtained the individual solutions, that is, $\rho_{s,1}(z)$, $\rho_{HI,1}(z)$, $\rho_{H2,1}(z)$, and $\rho_{H2,1}^\prime(z)$ (the subscript `1' indicates the iteration number). In the next iteration, we solved individual components in a similar way, but this time in the presence of the other components, that is, using the density solutions from the previous iteration. For example, when we solved for stars in the second iteration, instead of considering $\rho_{HI} = \rho_{H2} = \rho_{H2}^\prime = 0$, we used $\rho_{HI} = \rho_{HI,1}$, $\rho_{H2} = \rho_{H2,1}$, and $\rho_{H2}^\prime = \rho_{H2,1}^\prime$ and produce $\rho_{s,2}$. This $\rho_{s,2}$ is a slightly better coupled-solution than $\rho_{s,1}$. $\rho_{HI,2}$, $\rho_{H2,2}$, and $\rho_{H2,2}^\prime$, which were also obtained in a similar way.  We continued this iterative process until the density solutions for all the components converge with better than 0.1\% accuracy. Thus, in this scheme, we iteratively approach towards a coupled solution starting from a single-disc one. We note that for our sample galaxies, the solutions quickly converge within a couple of iterations.

It can be seen from Fig.~\ref{sden} that the molecular discs in our sample galaxies are shorter than the stellar or the atomic discs. Hence, we solved Eq.~\ref{eq_hydro} up to a radius where the CO is detected with confidence (See Table~\ref{tab:samp}). The central regions of spiral galaxies are very often found to host highly energetic activities (e.g. enhanced star formation, outflow, etc.), which might disturb the hydrostatic condition. The dark matter density of an NFW halo also rises very sharply in the central region. Due to these reasons, we did not solve Eq.~\ref{eq_hydro} in a region, $R < 1$ kpc, to avoid any possible divergences. The atomic discs in our sample galaxies extend much further than the molecular discs. At these radii (beyond the molecular disc), the galactic discs reduce to two-component systems (stars + atomic gas), and subsequently Eq.~\ref{eq_hydro} should be modified before solving. As Eq.~\ref{eq_hydro} cannot be solved analytically, we solved it numerically using an eighth-order Runge-Kutta method as implemented in the python package {\tt scipy}. We solved Eq.~\ref{eq_hydro} in our sample galaxies at every 100 pc radial distance. As the mean resolution of the HERACLES survey is $\sim$ 500 pc, a resolution of 100 pc is expected to be good enough to capture the radial variation of the molecular surface density. However, in the vertical direction, one needs a much higher spatial resolution to capture the rapid variation of the molecular gas density, which varies in 10s of the parsec-scale. To achieve this, we employed an adaptive resolution in the vertical direction and always find it to  be better than a few parsecs for our sample galaxies. It should be noted that as we investigate the observational signatures of a two-component molecular disc against a single-component one, we solved the hydrostatic equation for our sample galaxies assuming their molecular discs to be both a single-component and a two-component system.

\begin{figure}
\begin{center}
\begin{tabular}{c}
\resizebox{0.4\textwidth}{!}{\includegraphics{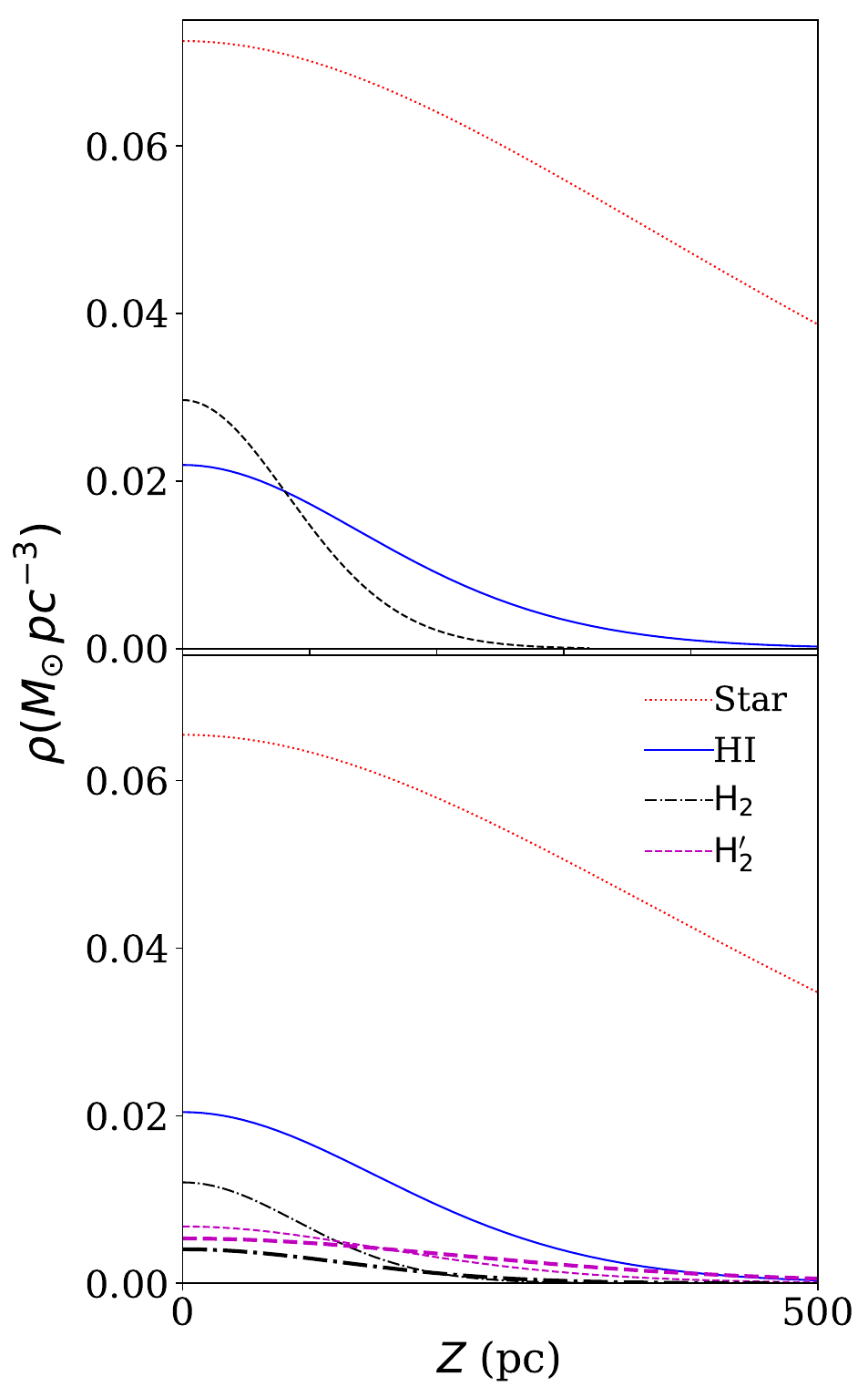}}
\end{tabular}
\end{center}
\caption{Sample solutions of Eq.~\ref{eq_hydro} for NGC 7331 at $R=8$ kpc. 
The top panel shows the density solutions for a three-component galactic disc having a single-component molecular disc. The bottom panel shows the same for a four-component disc. In this case, the molecular disc is assumed to be a two-component system with a thin and a thick disc. The red dotted and the solid blue lines in both of the panels represent the volume densities of the stellar and atomic discs, respectively. The black dashed line in the top panel represents the volume density of the thin disc molecular gas in a three-component disc. In the bottom panel, the thin dashed-dotted black and dashed magenta lines represent the volume density of molecular gas in the thin and thick discs, respectively, for an assumed thick disc gas fraction of 0.5. Whereas, the thick dashed-dotted black and the dashed magenta lines represent the same for an assumed thick disc molecular gas fraction of 0.7. See the text for more details.}
\label{samp_sol}
\end{figure}

\begin{figure}
\begin{center}
\begin{tabular}{c}
\resizebox{0.4\textwidth}{!}{\includegraphics{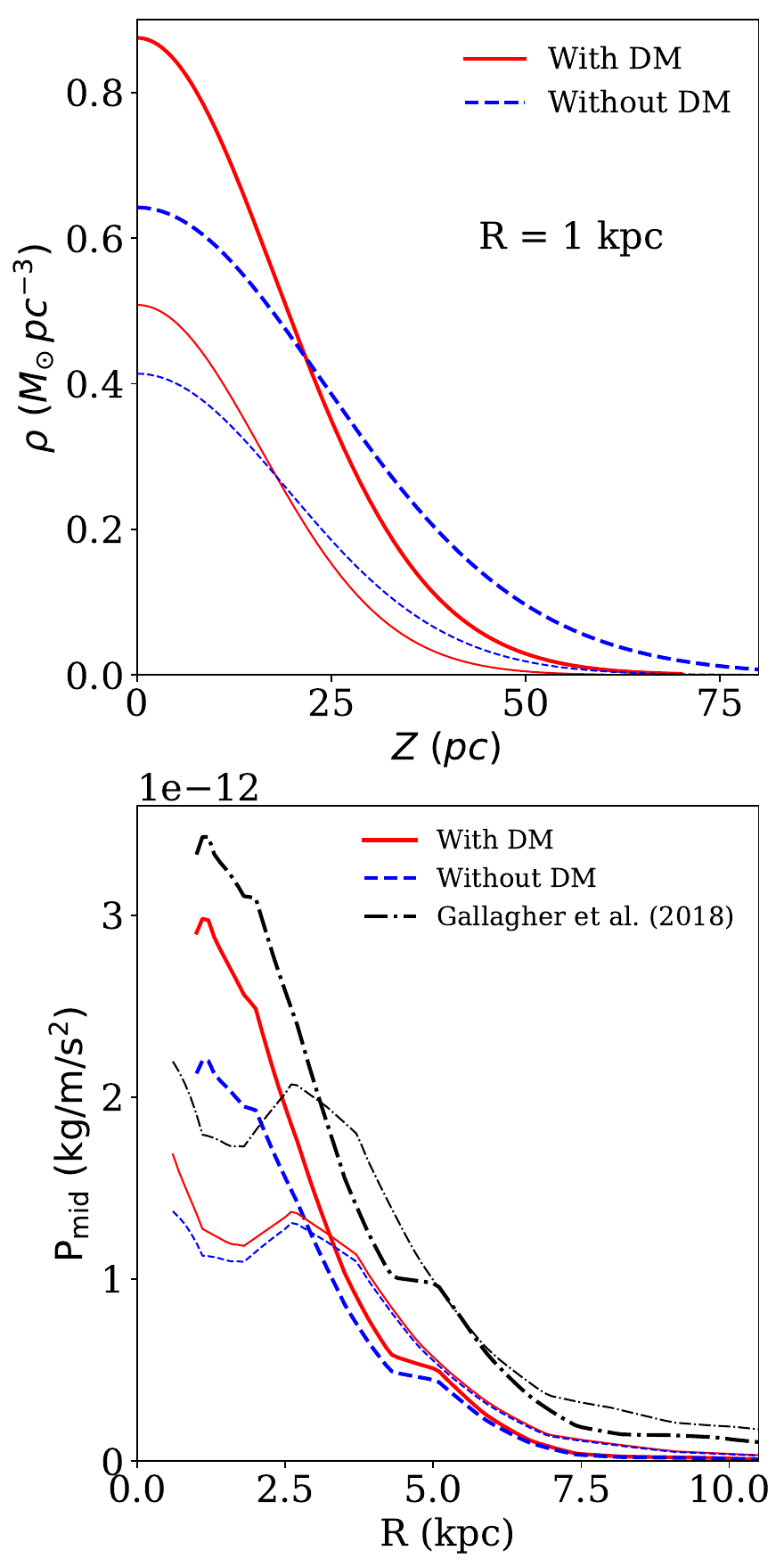}} 
\end{tabular}
\end{center}
\caption{Top panel: Comparison of the vertical molecular density solutions with and without dark matter halo. The thick lines represent the density solutions for NGC 3521, and the thin lines represent the same for NGC 7331. Bottom panel: Effect of the dark matter halo on the midplane pressure. The thick and thin lines represent the same as the above panel. The solid red line represents the calculated midplane pressure, including the gravity of the dark matter halo, whereas, the blue dashed line represents the same by excluding the dark matter halo. The black dashed-dotted lines represent the calculated midplane pressure using the prescription of \citet{gallagher18}. See the text for more details.}
\label{mdpsr}
\end{figure}

\section{Results and discussion}

\begin{figure*}
\begin{center}
\begin{tabular}{c}
\resizebox{1.\textwidth}{!}{\includegraphics{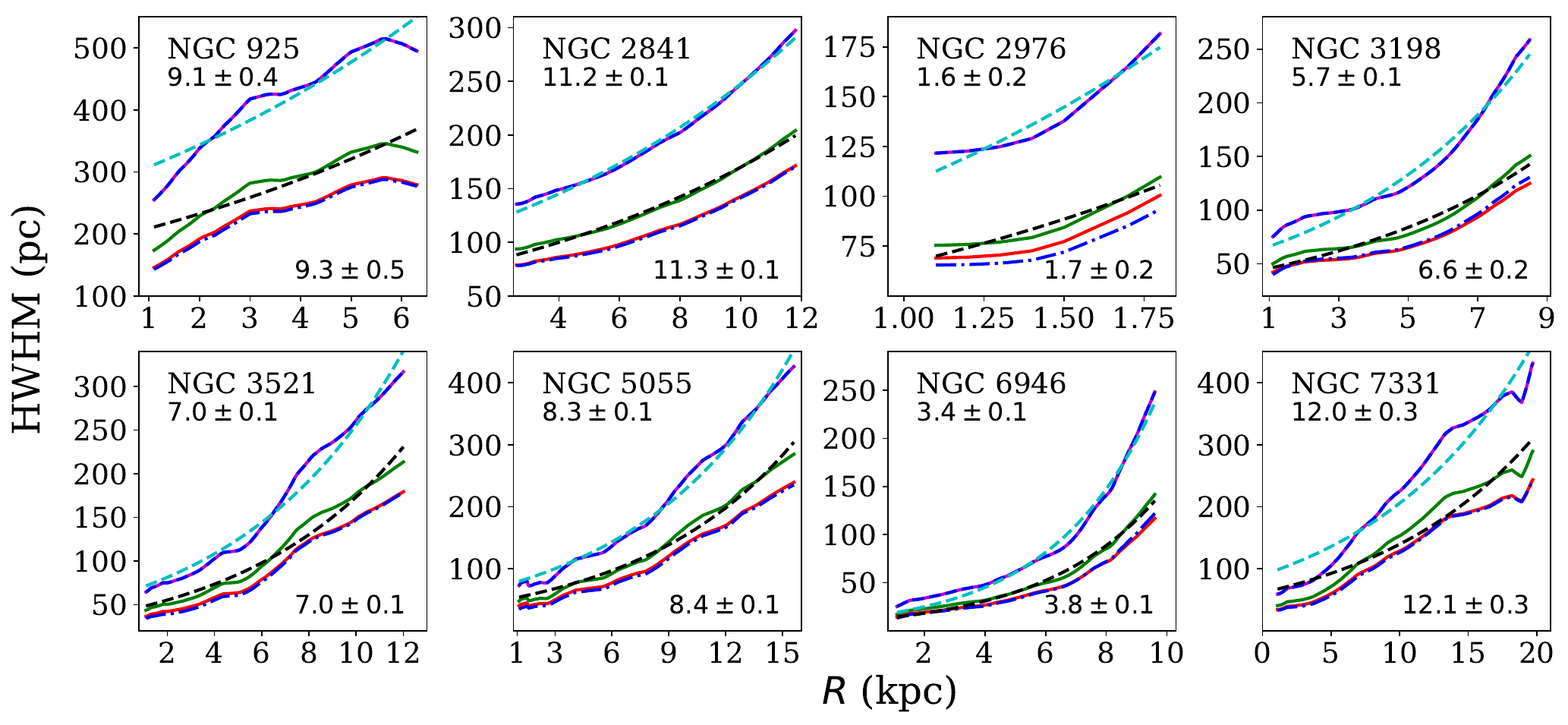}}
\end{tabular}
\end{center}
\caption{Atomic and molecular scale heights in our sample galaxies, assuming their molecular discs to be a two-component system. Each panel in the figure shows the scale heights for different galaxies, as mentioned in the top left corners of the respective panels. The solid red and magenta lines represent the scale heights for the thin and thick molecular discs, respectively, whereas the dashed blue lines show the atomic scale heights. The solid green lines represent the scale heights of the total molecular disc, that is, thin+thick. The cyan and black dashed lines show the exponential fits to the atomic and molecular scale heights. The respective scale lengths of the exponential fits are quoted in the top left (for atomic scale height) and the bottom right (for molecular scale height) corners of each panel in the units of kiloparsecs. For comparison purposes, the molecular scale heights assuming a single-component molecular disc in our sample galaxies are shown by the blue dashed-dotted lines.}
\label{hwhm}
\end{figure*}

\begin{figure}
\begin{center}
\begin{tabular}{c}
\resizebox{0.45\textwidth}{!}{\includegraphics{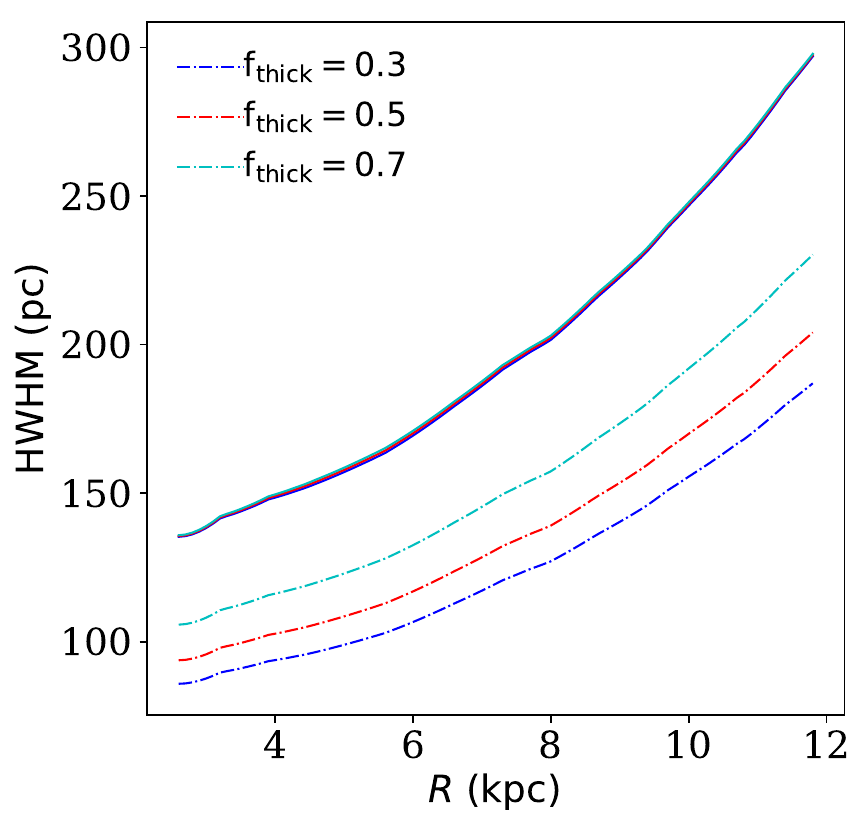}} 
\end{tabular}
\end{center}
\caption{Atomic and molecular scale height profiles for NGC 2841 for different assumed thick disc molecular gas fraction. We assumed three different thick disc molecular gas fractions of 0.3, 0.5, and 0.7. The solid lines represent the scale height of the atomic disc, whereas the dashed-dotted lines denote the scale height for the molecular disc. The scale height of the total molecular disc can change by as much as two times due to the variation of the thick disc molecular gas fraction from 0.3 to 0.7.}
\label{sclh_n2841}
\end{figure}

In Fig.~\ref{samp_sol}, we plotted sample solutions of the hydrostatic equilibrium equation for NGC 7331 at a radius of 8 kpc. The top panel shows the solutions for a three-component galactic disc, whereas the bottom panel shows the same for a four-component disc. Ideally, in the absence of coupling altogether, the solution of the hydrostatic equation for an isothermal disc should follow a $sech^2$ law \citep[see, e.g.][]{bahcall84a,bahcall84b}. However, we find that due to coupling between the baryonic discs and the dark matter halo, the solution of Eq.~\ref{eq_hydro} deviates from a $sech^2$ law and behaves more like a Gaussian. It can be seen from the figure that the stellar disc extends to a much larger height than the gaseous discs in both cases. As compared to a three-component disc, in a four-component disc, there is a lower amount of low-velocity dispersion molecular gas, settling close to the mid-plane. This effectively reduces the total gravity near the midplane, resulting in thicker atomic and molecular discs in a four-component system (bottom panel) than what is seen in a three-component system (top panel). The molecular gas in the thick disc can be seen to extend to much larger heights than the molecular gas in a thin disc. This height is often comparable to that of the atomic disc (magenta dotted and blue-dashed lines in the bottom panel), leading to a hypothesis that the thick disc molecular gas and the atomic gas are part of the same dynamical component. Here, we solved the hydrostatic equation assuming an equal amount of gas in the thin and thick disc of the two-component molecular disc. For comparison purposes, we also solved Eq.~\ref{eq_hydro} for a thick disc molecular gas fraction of 0.7, and the respective solutions for the molecular gas are plotted in the bottom panel of Fig.~\ref{samp_sol} (thick dashed-dotted black and dashed magenta lines). As can be seen from the figure, an increased fraction of the thick disc molecular gas fraction leads to a thicker molecular disc, thus increasing its effective scale height.

Density solutions can often provide meaningful insights into several physical processes. For example, by considering a hydrostatic equilibrium in 12 nearby star-forming disc galaxies, \citet{bacchini19a,bacchini19b} show that the star formation law exhibits much less scatter when it is determined using volumetric gas density instead of gas surface density. Not only that, but using the density solutions and the observed velocity dispersion, one can estimate the pressure at the midplane, which is an important parameter when deciding if the gas participates in star formation or not \citep[see, e.g.][]{usero15,bigiel16,helfer97}. However, many of these studies do not solve the hydrostatic equilibrium equation self-consistently to estimate the relevant physical quantities. For example, \citet{bacchini19a,bacchini19b} calculated the volume density of the gas discs by considering a preassumed vertical profile for the stellar (exponential) and gas (Gaussian) distribution. 

Similarly, \citet{gallagher18} estimated the midplane pressure in a sample of nearby, large galaxies by ignoring the dark matter halo and assuming a constant stellar scale height. They calculated the midplane pressure by summing the gas self-gravity and the weight of the gas in the potential of the stellar disc by assuming an exponential vertical profile. The midplane pressure can be given as 

\begin{equation}
P_{mid} \approx \frac{\pi G \Sigma_{gas}^2}{2} + \Sigma_{gas} \left(2G\rho_*\right)^{1/2} \sigma_{gas,z} 
\label{pmid}
,\end{equation}

\noindent where $\Sigma_{gas}$ is the total gas surface density (atomic+molecular), $\rho_*$ is the volumetric mass density of stars and dark matter together, and the $\sigma_{gas,z}$ is the vertical velocity dispersion of the gas. \citet{gallagher18} ignored the dark matter contribution to $\rho_*$, arguing that the total mass of the dark matter within the stellar layer at any radius would be negligible. Thus, they estimated the stellar density at the midplane assuming a known scale height, $h_*$, given by \citep{vanderkruit88}

\begin{equation}
\rho_* \approx \frac{\Sigma_*}{2h_*}
,\end{equation}

\noindent where $\Sigma_*$ is the stellar surface density. They further assumed a constant scale height for the stellar disc, as computed by \citet{leroy08}, using the knowledge of the flattening ratio of the stellar discs in galaxies \citep{kregel02}. The scale height can be given as $l_*/h_* = 7.3 \pm 2.2$. Where $l_*$ represents the exponential scale length of the stellar disc. It should be noted here that the observed flattening ratio might not capture the true flaring of the stellar disc as the observations are only sensitive to the high luminosity stellar component. In that case, this prescription can considerably underestimate the stellar scale height and hence overestimate the $\rho_*$. 

To examine the impact of this assumed constant stellar scale height and the gravity of the dark matter halo, we solved Eq.~\ref{eq_hydro} without dark matter and calculated the gas volume densities and the midplane pressure. In the top panel of Fig.~\ref{mdpsr}, we show the effect of the dark matter halo on the estimated vertical density distribution of the molecular gas. For simplicity, here, we solved Eq.~\ref{eq_hydro} assuming a single-component molecular disc; however, the conclusions also remain the same for a two-component molecular disc. We solved the hydrostatic equilibrium equation for two galaxies from our sample, that is, NGC 3521 (thick lines) and NGC 7331 (thin lines). These two galaxies represent two extreme ends of the dark matter dominance in our sample (see, e.g. Table~\ref{tab:dmpar}). As can be seen from the top panel of the figure, the presence of the dark matter halo significantly alters the vertical density profiles. Ignoring the dark matter would result in an underestimation of the volume density close to the midplane and an overestimation of it at more considerable heights. The effect of the dark matter halo is expected to be more pronounced at outer radii, where the dark matter dominates the gravity. At these radii, it can play a significant role in regulating star formation by modulating the volume density and/or by providing the mass required to produce gravitational instabilities \citep[see, e.g.][]{das20}.

In the bottom panel of Fig.~\ref{mdpsr}, we show the estimated midplane pressures for NGC 3521 (thick lines) and NGC 7331 (thin lines) using the following three prescriptions: firstly, by solving the hydrostatic equation self-consistently and including the dark matter halo (the solid red lines); secondly, by solving it self-consistently without the dark matter halo (the blue dashed lines); and thirdly, by using the same method used by \citet{gallagher18} (as given by Eq.~\ref{pmid}) (black dashed-dotted line). As can be seen from the figure, the formalism of \citet{gallagher18} significantly overestimates the midplane pressure. For NGC 3521 and NGC 7331, we find that this overestimation could be by a factor of a few (at the midplane) to as large as by $\sim$ 20 (at the outskirts). A substantial contribution to this discrepancy could be attributed to the fact that they overestimate the $\rho_*$ significantly by underestimating the flaring of the stellar disc. They also did not solve the hydrostatic equation in a self-consistent manner by considering the gravitational coupling of the baryonic discs. Further, it can also be seen from the figure that the absence of the dark matter halo (hence reduced potential) underestimates the midplane pressure. For our two sample galaxies, the dark matter gravity can reduce the midplane pressure by $\sim 20\%$ at an inner radius to $\sim 5-10\%$ at outer radii.

We solved Eq.~\ref{eq_hydro} self-consistently for all of our sample galaxies to obtain the vertical density distribution of different baryonic components. We further used these density solutions to asses the thickness of the gas discs by estimating the vertical scale heights, defined as the half width at half maxima of the vertical density distribution. In Fig.~\ref{hwhm}, we plotted the scale height profiles as a function of the radius. Observationally, for a two-component molecular disc, both the disc components are combined together to produce a single measurement. Hence, for these molecular discs, we estimated the total molecular density by combining the densities of the thin and thick disc. It can be seen from the figure that the atomic and molecular discs flare as a function of the radius. For NGC 7331, the atomic scale height is found to be a factor of $\sim$ 2 higher than the molecular scale height at all radii. Even so, the scale height of the thick molecular disc and the atomic disc are found to be the same due to their same assumed velocity dispersion. Nonetheless, it should be emphasised that the scale height alone is not a quantitative measure of the observable thickness of a baryonic disc. The measurable thickness also depends on the midplane density of a baryonic component, and hence, in spite of having the same scale heights, the atomic and thick molecular discs could have a different observable thickness. 

It can also be seen from Fig.~\ref{hwhm} that the scale height of a thin disc (solid red lines) and the single-component molecular disc (dashed-dotted blue lines) are very similar, though the scale height of the thick disc is much higher. This illustrates that the velocity dispersion is much more influential than the surface density in deciding the scale height. We find that the scale height of the thin and thick molecular discs in our sample galaxies vary between $\sim 20-200$ pc and $100-500$ pc, respectively, depending on the radius. In contrast, the total molecular scale height varies between $\sim 50-300$ pc. This is higher than the scale height of the single-component molecular disc. To quantify the flaring of the atomic and the molecular discs, we fitted their scale height profiles using a rising exponential function. As can be seen, an exponential function describes the flaring of the scale height reasonably well in our sample galaxies. The respective scale lengths of the exponential scale heights are quoted in the top left (for atomic discs) and bottom right (for molecular discs) corners of each panel in Fig.~\ref{hwhm} in the kiloparsecs.

The scale height profiles presented in Fig.~\ref{hwhm} were produced by assuming an equal amount of molecular gas in the thin and the thick disc \citep{pety13}. However, the observational estimation of scale hight profiles is not well constrained, and a different fractional abundance is very well possible in galaxies \citep[see, e.g.][]{goldsmith08,liszt10}. Hence, we explored the effect of this thick disc molecular gas fraction on the scale height of the gas discs by solving Eq.~\ref{eq_hydro} in one of our sample galaxies, NGC 2841, by assuming a thick disc gas fraction of 0.3, 0.5, and 0.7. In Fig.~\ref{sclh_n2841}, we plotted the atomic and molecular scale height profiles for this galaxy for a different thick disc molecular gas fraction. As can be seen, the atomic scale heights (solid lines) do not change considerably. However, the scale height of the molecular discs varies significantly if one assumes a different amount of molecular gas in the thin and thick disc. For NGC 2841, the molecular scale height has been found to vary by as much as a factor of $\sim$ 2 at the centre to $\sim 20\%$ at the outer radii. 

\begin{figure}
\begin{center}
\begin{tabular}{c}
\resizebox{0.45\textwidth}{!}{\includegraphics{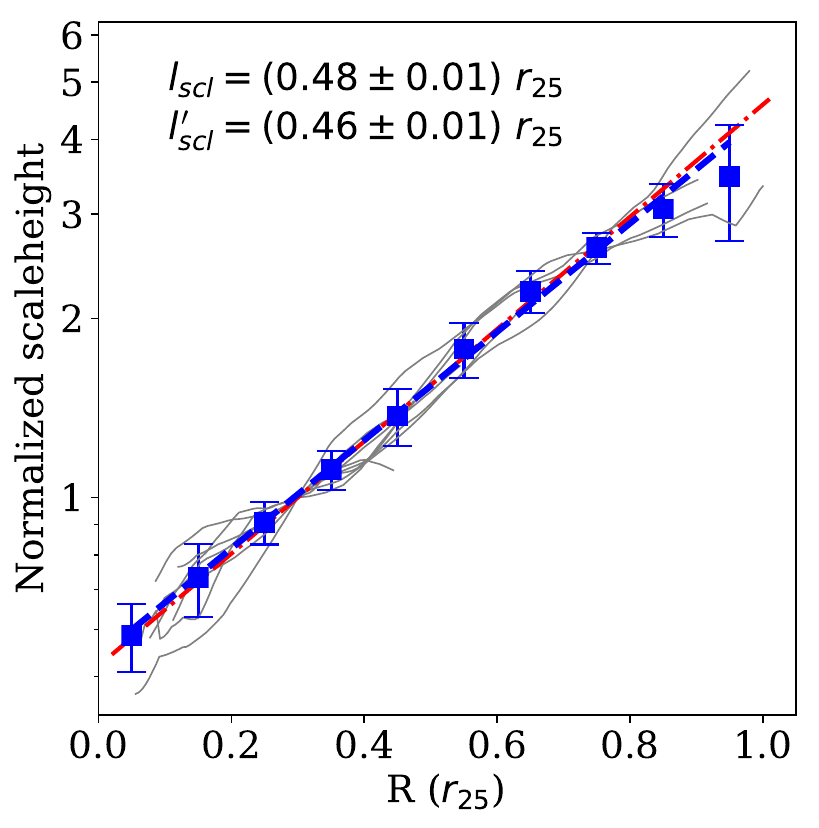}} 
\end{tabular}
\end{center}
\caption{Normalised scale height profiles of our sample galaxies for two-component molecular discs. The solid thin lines represent the scale height profiles normalised to 1 at 0.3$r_{25}$. The blue squares with error bars and the thick blue dashed line indicate the mean normalised scale heights and an exponential fit to it, respectively. The thick red dashed-dotted line shows an exponential fit to the normalised scale height profiles assuming single-component molecular discs in our sample galaxies. The respective scale lengths of the exponential fits are quoted in the top left of the figure. See the text for more details.}
\label{sclh_norm}
\end{figure}

Next, we used the molecular scale height profiles (Fig.~\ref{hwhm}) to investigate the universality of the flaring of the molecular discs. To do that, we normalised the molecular scale heights of our sample galaxies to unity at a radius of 0.3$r_{25}$ by adopting a similar approach as used by \citet{schruba11}. In Fig.~\ref{sclh_norm}, we plotted these normalised molecular scale height profiles. As can be seen from the figure, despite having very different surface density profiles and different dark matter halos, the flaring is considerably general and follows a fairly tight relation with the normalised radius, $r_{25}$. To quantify the flaring further, we averaged the normalised scale height values within a radial bin of 0.1$r_{25}$ and fitted this mean with an exponential function of the form $h_{scl} = h_0 \exp(R/l_{scl})$, where $l_{scl}$ represents the scale length of the exponential flaring. We find a $l_{scl} = (0.48 \pm 0.01$)$r_{25}$. For comparison purposes, in Fig.~\ref{sclh_norm}, we also plotted the exponential fit to the normalised scale height profile as obtained for single-component molecular discs in our sample galaxies (red dashed-dotted line). For single-component molecular discs, we find a flaring scale length of $l^\prime_{scl} = (0.46 \pm 0.01$)$r_{25}$. This value differs marginally from the value obtained for two-component discs, which indicates that the nature of the flaring is very similar in both the single-component and the two-component molecular discs. We also examined the effect of a different assumed thick disc molecular gas fraction on this flaring scale length. To do that, we carried out the above exercise for thick disc molecular gas fractions of 0.3 and 0.7 as well. We find $l^\prime_{scl} = (0.47 \pm 0.01$)$r_{25}$ and $l^\prime_{scl} = (0.48 \pm 0.01$)$r_{25}$ for a thick disc fraction of 0.3 and 0.7, respectively, which are consistent with the results for the thick disc fraction of 0.5 within error bars. Hence, the nature of the flaring seems to be intrinsic and not profoundly dependent on the assumed thick disc gas fraction.

\begin{figure}
\begin{center}
\begin{tabular}{c}
\resizebox{0.45\textwidth}{!}{\includegraphics{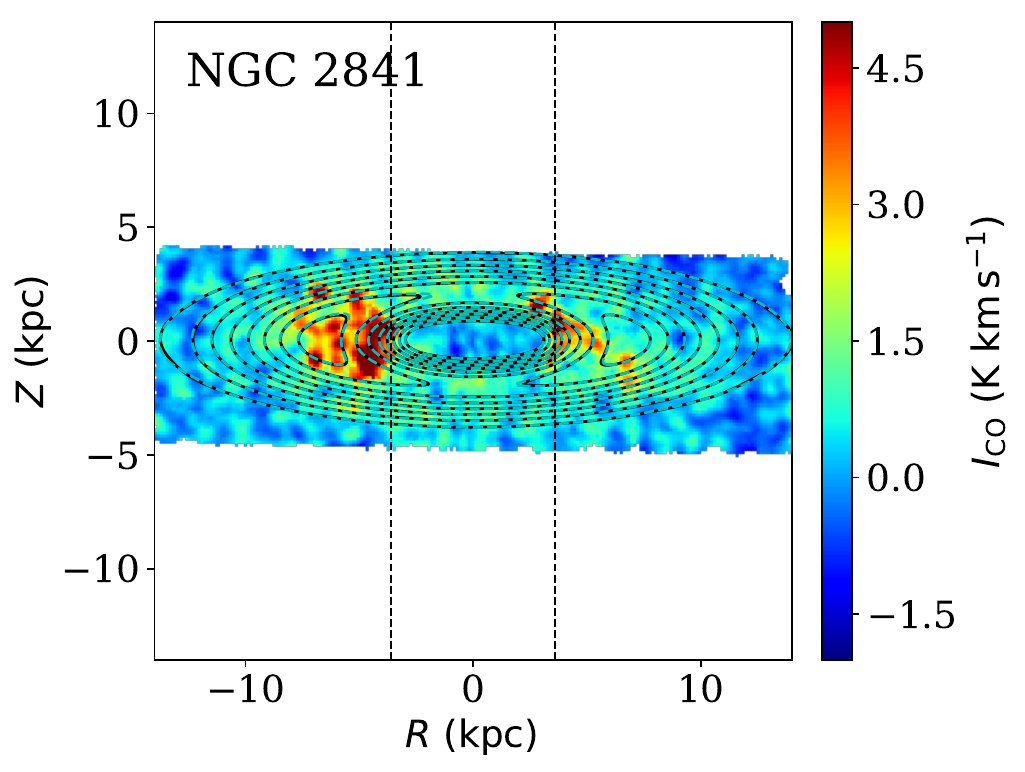}} \\
\resizebox{0.45\textwidth}{!}{\includegraphics{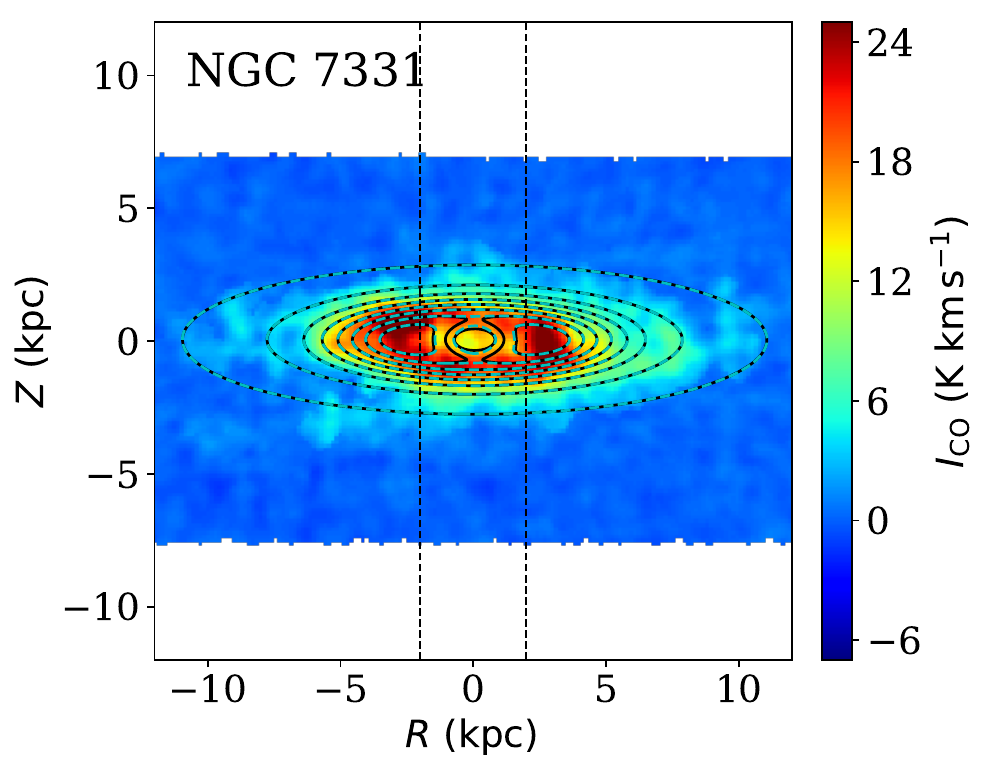}}
\end{tabular}
\end{center}
\caption{Modelled and observed molecular column density maps for NGC 2841 (top panel) and NGC 7331 (bottom panel). The colour scales in both of the panels represent the CO intensity distribution as observed in the HERACLES survey, whereas the contours represent the modelled intensity distribution. The contour levels are (1, 1.3, 1.6, 1.9, ...) $\rm K \thinspace km \thinspace s^{-1}$ for NGC 2841 and (1.5, 4.4, 7.4, ...) $\rm K \thinspace km \thinspace s^{-1}$ for NGC 7331. The solid contours represent the modelled intensity distribution for a two-component molecular disc, whereas the dashed contours indicate the same for a single-component molecular disc. The vertical dashed lines in each panel enclose a region where the solutions cannot be trusted due to a possible violation of the hydrostatic condition.}
\label{mom_4comp}
\end{figure}

However, the molecular scale height is not a directly measurable quantity. Instead, the molecular column density and the spectral cubes are what is observed with radio telescopes. Hence, building total column density maps and spectral cubes for our sample galaxies will be illuminating to investigate the difference between a single-component and a two-component molecular disc. To do that, using the density solutions of the molecular discs and the observed rotation curves, we built three-dimensional dynamical models for our sample galaxies. These models were then inclined to the observed inclinations and projected into the sky plane to generate the spectral cubes and the moment maps. These spectral cubes and the moment maps were further convolved with the telescope beam (13\arcs $\times$ 13\arcs) to mimic the real observations. Thus, we produced model molecular column density maps and spectral cubes for our sample galaxies. To see how these models compare with observations, in Fig.~\ref{mom_4comp}, we overplotted the model molecular column density maps (in solid black contours) on top of the observed maps (colour scale) for two representative galaxies, NGC 2841 (top panel) and NGC 7331 (bottom panel). From the top panel of the figure, it can be seen that the observed molecular column density distribution in NGC 2841 is somewhat patchy and lacks a clear azimuthal symmetry. However, as our modelling explicitly assumes an azimuthal symmetry, the modelled molecular column density for NGC 2841 does not match the observation fully. On the other hand, for NGC 7331, the model matches the observation reasonably well (bottom panel). For comparison purposes, we also plotted the model intensity distribution for a three-component disc (single-component molecular disc) using broken contours in both of the panels. As can be seen from the figure, there is no practical difference in the intensity distributions of the molecular discs for a three-component and four-component galactic disc. 

\begin{figure}
\begin{center}
\begin{tabular}{c}
\resizebox{0.45\textwidth}{!}{\includegraphics{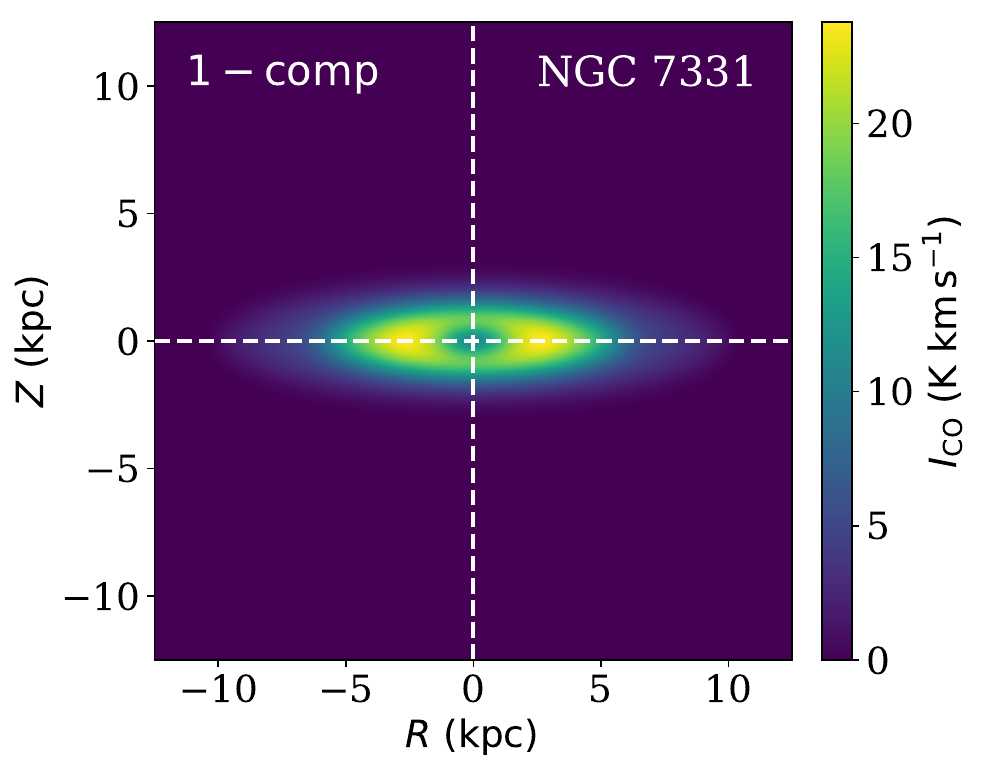}} \\
\resizebox{0.45\textwidth}{!}{\includegraphics{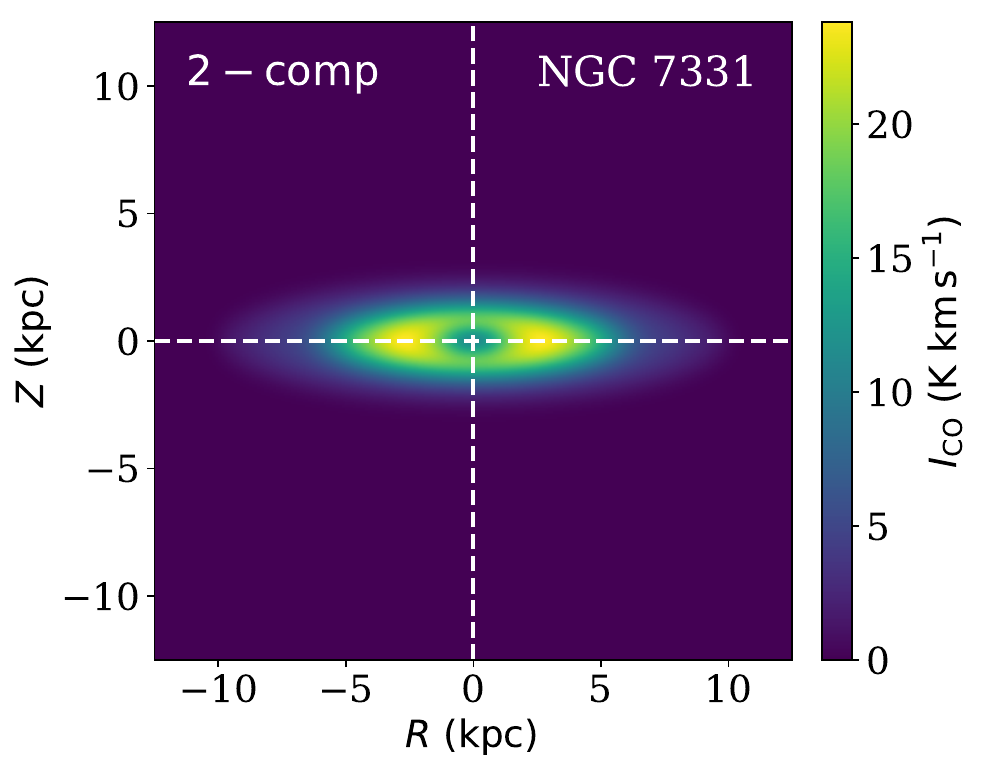}}
\end{tabular}
\end{center}
\caption{Modelled intensity maps for a single-component (top panel) and a two-component (bottom panel) molecular disc in NGC 7331. The colour scales in both of the panels indicate the observable CO intensity in the units of $\rm K \thinspace km \thinspace s^{-1}$. As can be seen, there is no apparent difference in the column density maps of the two discs. See the text for more details.}
\label{mom_ncomp}
\end{figure}

\begin{figure}
\begin{center}
\begin{tabular}{c}
\resizebox{0.45\textwidth}{!}{\includegraphics{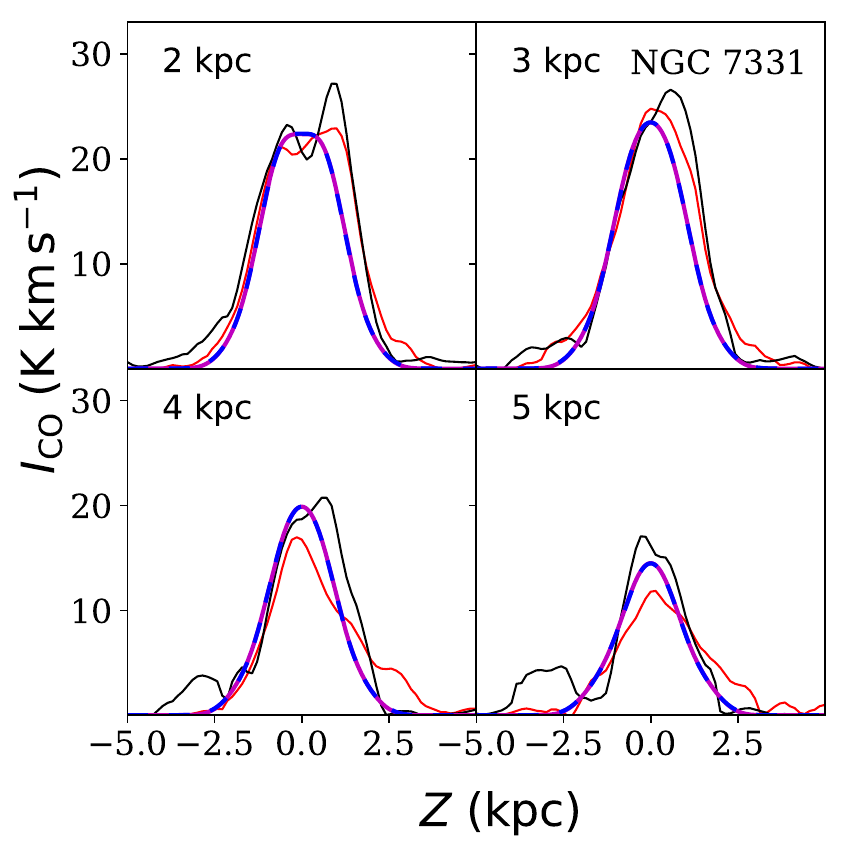}} 
\end{tabular}
\end{center}
\caption{Vertical column density profiles of the molecular discs in NGC 7331. The red and black thin solid lines are the observed vertical profiles for two different halves of the molecular disc. The thick blue dashed lines and the thick red solid lines represent the simulated vertical column density profiles for the single component and the two-component molecular discs, respectively. Different panels show the vertical profiles at different radii, as quoted in the top left corners of the respective panels.  As can be seen, there is no detectable difference in the vertical profiles of the single-component and two-component molecular discs. See the text for more details.}
\label{vprof}
\end{figure}

\begin{figure}
\begin{center}
\begin{tabular}{c}
\resizebox{0.4\textwidth}{!}{\includegraphics{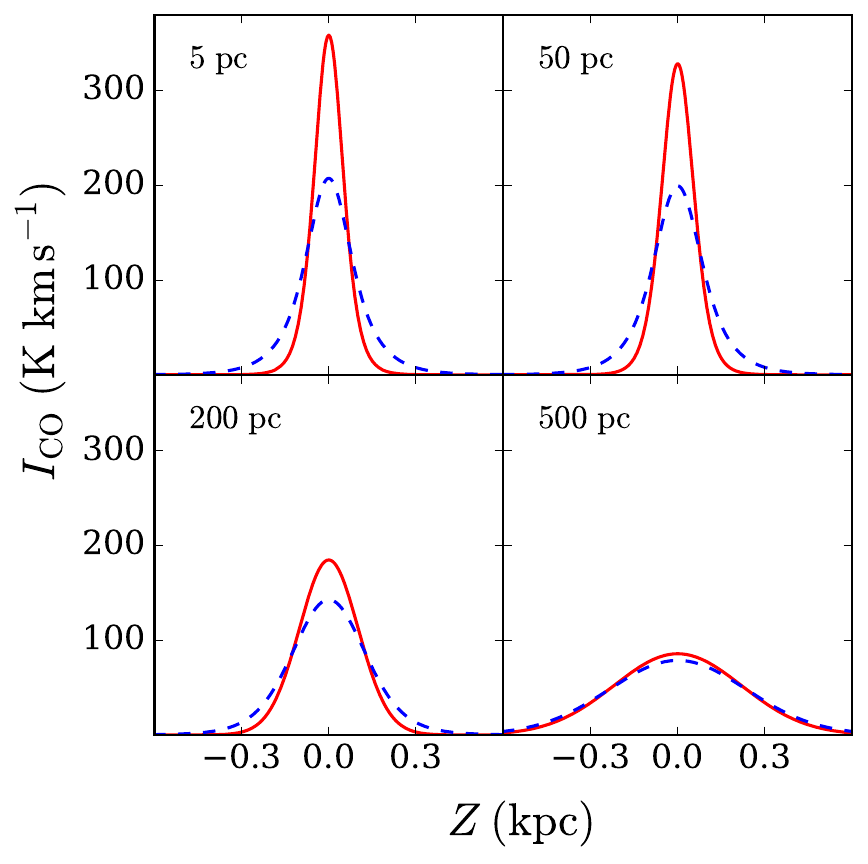}} \\
\resizebox{0.4\textwidth}{!}{\includegraphics{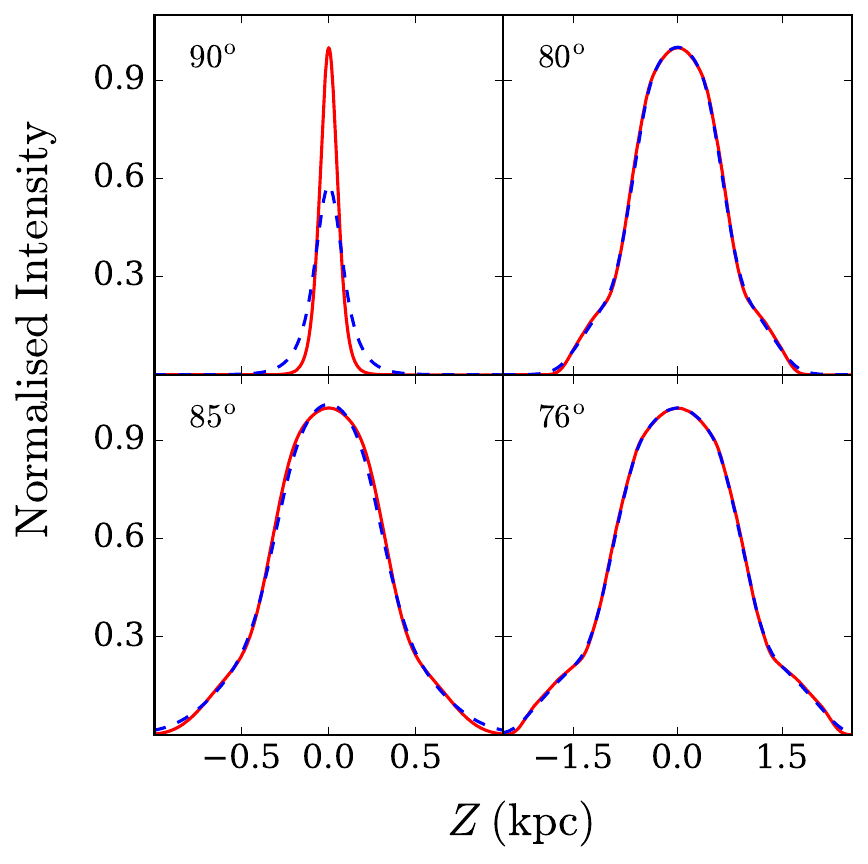}} 
\end{tabular}
\end{center}
\caption{Top panel: Vertical column density profiles of the edge-on molecular discs in NGC 7331 (single-component and two-component) at a radius of 4 kpc for different imaging resolutions. The spatial resolutions are quoted in the top left corners of each sub-panel. Bottom panel: Dependence of the normalised vertical column density profile on inclination. In both of the panels, the solid red and blue dashed lines represent the molecular column density profiles for a single-component and two-component molecular disc, respectively. See the text for more details.}
\label{vprof_incl90}
\end{figure}

For further comparison purposes, in Fig.~\ref{mom_ncomp}, we plotted the column density maps for a single-component (top panel) and two-component molecular disc (bottom panel) for NGC 7331. As can be seen from the figure, qualitatively, the intensity distribution of the molecular gas looks very similar in both of the panels. This similarity in the intensity maps is observed for other galaxies in our sample as well. We plotted similar figures for all of our sample galaxies in Fig.~\ref{mom_ncomp_all} in the appendix. For a better comparison of the column density maps, we took vertical cuts along the minor axis and extracted column density profiles. We estimated these column density profiles at different radii for both the simulated and observed molecular discs. In Fig.~\ref{vprof}, we show these column density profiles at four different radii for NGC 7331. As can be seen, the simulated profiles (the thick, solid red  and blue dashed lines) can be compared to the observed profiles nicely (the thin red and black solid lines), indicating fairly reasonable modelling of the molecular disc in NGC 7331. Other galaxies in our sample also exhibit a similar trend, which we show in Fig.~\ref{vprof_all} in the appendix. From these figures, we note that the two simulated molecular discs (single-component and two-component) produce almost the same vertical profiles, even though they have considerably different scale heights (see Fig.~\ref{hwhm}).

This could happen due to a combined effect of low spatial resolution and a non-edge-on inclination. To check how spatial resolution can influence the observed column density profiles, we imaged the molecular discs of NGC 7331 (single-component and two-component) at different spatial resolutions. For this, we first inclined the molecular discs to 90$^o$, since the maximum difference is expected at an edge-on orientation, and we convolved the resulting column density maps with different observing beams. These maps were then used to extract the column density profiles for comparison purposes. It should be emphasised here that the raw resolution we used is approximately a few parsecs, which is much higher than the adopted observing resolutions. In the top panel of Fig.~\ref{vprof_incl90}, we show the corresponding vertical column density profiles at $R=4$ kpc. The imaging resolutions are quoted in the top left corner of each sub-panel. As can be seen from the figure, for NGC 7331, at a high spatial resolution, the two molecular discs indeed produce different vertical column density profiles; however, the difference slowly washes away as one degrades the resolution. At the mean spatial resolution of the HERACLES survey, that is, $\sim$ 500 pc, it fails to produce any detectable difference between the column density profiles of the two discs. We observed the same trend for our other sample galaxies as well. We present their results in Fig.~\ref{vprof_incl90_all} (first and third rows) in the appendix.

We further investigated the effect of inclination on the column density profiles by inclining the molecular discs of NGC 7331 to several different inclinations. To decouple any influence from the imaging resolution, we used a spatial resolution of $\sim$ 5 pc for this purpose. In the bottom panel of Fig.~\ref{vprof_incl90}, we plotted the normalised vertical column density profiles for discs with different inclinations (quoted in the top left corner of each sub-panel). As can be seen from the figure, at an inclination of $90^o$, the discs could be discriminated, in principle; however, a slight change in the inclination eliminates any difference promptly. The effect of inclination on vertical profiles for our other sample galaxies are presented in the appendix,~\ref{vprof_incl90_all} (second and fourth rows). These results indicate that the vertical profiles are highly sensitive to inclination, and it is tough to identify the presence of a two-component molecular disc if the galaxy is not an edge-on system.

\begin{figure}
\begin{center}
\begin{tabular}{c}
\resizebox{0.43\textwidth}{!}{\includegraphics{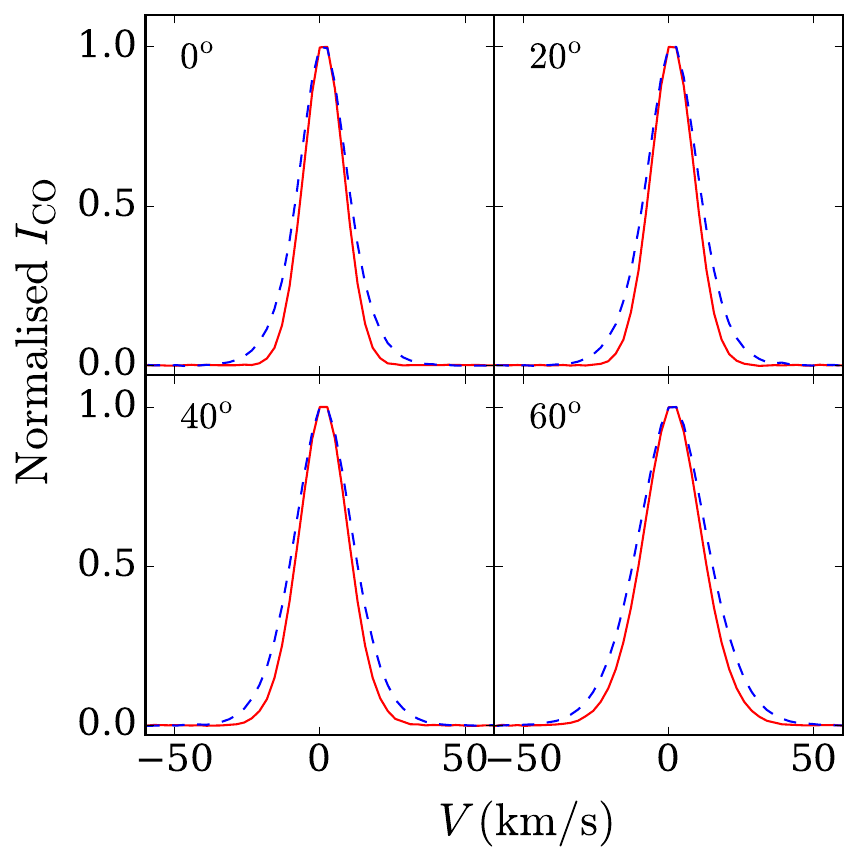}}
\end{tabular}
\end{center}
\caption{Comparison of the widths of normalised stacked spectra for a single-component and two-component molecular disc at different observing inclinations. The solid red lines indicate the stacked spectra for a single-component molecular disc, whereas the blue dashed lines show the same for a two-component disc. Different panels represent the effect of different assumed inclinations. The inclinations are quoted in the top left corners of the respective panels. The stacked spectra of a two-component molecular disc were found to produce a systematically higher width than that of a single-component disc. See the text for more details.}
\label{stack_incl}
\end{figure}

\begin{figure*}
\begin{center}
\begin{tabular}{c}
\resizebox{\textwidth}{!}{\includegraphics{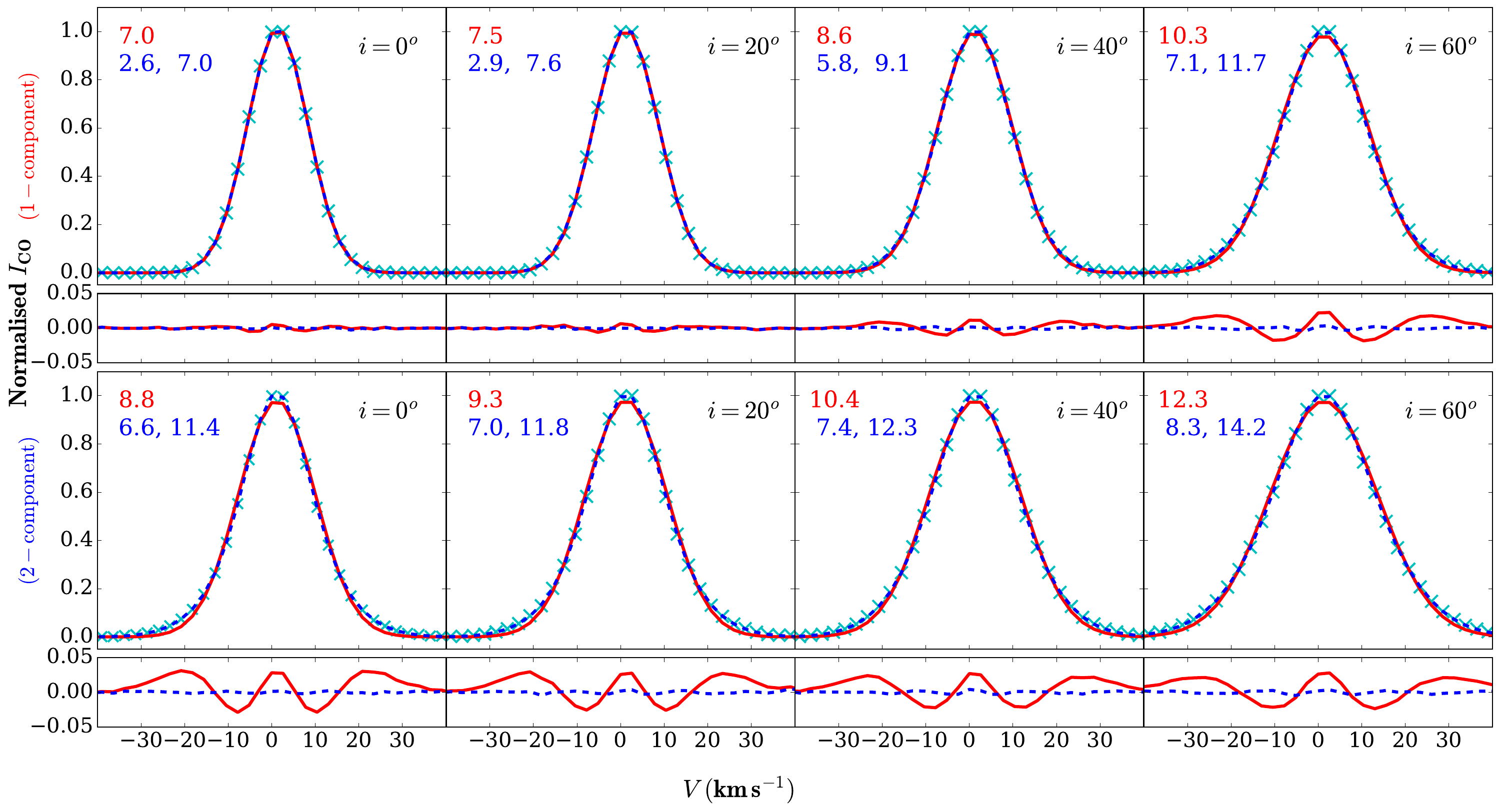}}
\end{tabular}
\end{center}
\caption{Result of Gaussian fittings to the stacked spectra of the single component and two-component molecular disc. The large top panels show the stacked spectra and their fits with single and double Gaussians for the single-component disc at different inclinations, whereas the large bottom panels show the same for a two-component molecular disc. The smaller panels in the top and bottom represent the residuals of the respective fits. The cyan crosses represent the stacked spectra, whereas the solid red lines and the blue dashed lines represent a single Gaussian and a double Gaussian fit to the spectra, respectively. The widths of the fitted Gaussian components (sigma) are quoted in the top left corners of each panel. The first number on the left (in red) is the sigma of a single Gaussian fit, whereas the blue numbers just below them are the widths for a double Gaussian fit. The assumed inclinations of the molecular discs are quoted at the top right corners of the respective panels.}
\label{stack_fitg}
\end{figure*}

Next, we explored the spectral properties of the discs to examine if they can be used to identify a two-component molecular disc in galaxies. Using the dynamical models of NGC 7331 (and other sample galaxies), we produced the spectral cubes for a single-component and two-component molecular disc for typical observing setups. We adopted the mean spatial resolution of the HERACLES survey, that is, $\sim$ 500 pc, and used a spectral resolution of 2.6 \kms. To simulate the real observing scenario, we added appropriate noise levels to the molecular spectral cubes. We note that the molecular surface densities of our sample galaxies were calculated by stacking the spectra within radial bins \citep{schruba11}. The stacking results in a much higher S/N on the final radial surface density profiles as compared to a moment map produced by adding the fluxes across the channels. In that sense, the noise in a spectral channel is much higher than the lowest surface density detected (in a radial bin) by stacking. Due to this reason, we did not simply add the noise as measured from the observed spectral cube. Instead, we added such an amount of noise to our simulated spectral cubes so that it produces an S/N of five at the farthest radial bin. A minimum S/N of five is necessary to determine the centroids of the spectra while stacking \citep{deblok08}. We emphasise that the choice of our noise levels does not affect our results in any significant way. Adopting the same technique, which led to the strong proposition for two-component molecular discs in galaxies, as used by \citet{calduprimo13}, we stacked the molecular spectra in both of the discs. As the original signature of a two-component molecular disc would have a larger spectral width than a single-component disc, an artificial broadening due to blending should be minimised for a more explicit inspection. Hence, we excluded a region of $0.2 R_{25}$ around $R=0$ from stacking to avoid significant blending due to the rotation of the galactic disc \citep{calduprimo13}. Even in doing so, there would be blending outside of this central region as well, which significantly depends on the disc inclination. To examine the effect of the inclination on the spectral widths, we produced the spectral cubes and stacked spectra for different molecular discs with different inclinations. 

Thus, generated stacked spectra for a single-component (solid red lines) and a two-component molecular disc (blue dashed lines) in NGC 7331 are compared in Fig.~\ref{stack_incl}. Different panels show the results for different inclinations, as quoted in the top left corner of each panel. As can be seen from the figure, a two-component disc systematically produces wider stacked spectra at all inclinations as compared to a single-component disc, though the difference narrows down as a function of inclination. However, we note that the difference in the widths ($\sim 1-2$ \kms) is not so extensive as to identify a two-component molecular disc just based on the absolute width of the stacked spectra. This result is consistent across all of our sample galaxies. We present the respective stacked spectra in Fig.~\ref{stack_incl_all} in the appendix. 

Along with the spectral width, we investigated the spectral shapes as well to examine the spectral behaviour of a two-component molecular disc. For a two-component disc, it is expected that the stacked spectra have the response of both the thin and thick disc velocity dispersion and hence are best fitted by a double-Gaussian. Whereas, a single-Gaussian is presumed to be the best to describe the stacked spectra of a single-component disc. However, spectral blending (inclination dependent) complicates the stacked spectra, and hence, it is not trivial to predict the expected spectral shape a priori. To get a more in-depth insight, we fitted the stacked spectra of both of the discs with a single and double-Gaussian and study them as a function of inclination. In the top panel of Fig.~\ref{stack_fitg}, we show the fits for a single-component disc in NGC 7331. Ideally, in the absence of any blending, the spectra should be best fitted by a single-Gaussian with the same width as the velocity dispersion of the molecular disc (7 \kms). However, as can be seen from the figure, at high inclinations, there is a significant blending, and as a result, the stacked spectra are best fitted by a double-Gaussian. Surprisingly, the width of the two Gaussian components for a single-component disc at an inclination of $\simeq 60^o$ (rightmost panel on the top row) are found to be $\sim 7$ \kms~and $\sim 12$ \kms~, which accurately mimics a two-component molecular disc. Hence, the spectral blending at high inclinations can induce the signatures of a two-component disc in the spectral cube of a single-component disc. However, at inclinations $\lesssim 40^o$, the blending is less, and the residual of a single-Gaussian fit to the stacked spectra does not show significant features. The spectral width is also recovered reasonably well by a single-Gaussian fit. On the other hand, the stacked spectra of a two-component molecular disc are always best fitted by double-Gaussians, irrespective of the disc inclination (bottom row in Fig.~\ref{stack_fitg}); however, the recovered spectral widths substantially deviate from the assumed velocity dispersions at high inclinations ($\gtrsim 40^o$). Nevertheless, at inclinations $\lesssim 40^o$, the recovery of the velocity dispersions is satisfactory. These results indicate that for a cleaner inspection of the presence of a two-component molecular disc in galaxies, the spectral cubes should be stacked for discs with an inclination $\lesssim 40^o$. We note that while building the spectral cubes, we assumed a spectral resolution of 2.6 \kms~and a spatial resolution of $\sim$ 500 pc. These kinds of resolutions are typical and not very coarse so as to consider these results non-typical. The widening of the stacked spectra indeed would depend on the observed spatial resolution and the rotation curve, however, and the parameters for NGC 7331 are typical to those of normal spiral galaxies. In fact, in many physical properties, NGC 7331 is so similar to the Galaxy that very often, it is called the `Milky Way's twin.' For consistency purposes, we also carried out the same exercise for all of our sample galaxies (see Fig.~\ref{stack_fitg_all} in the appendix). Despite having a wide range of rotation curves and physical properties, they show the same signatures in their spectral shapes, as we find for NGC 7331. This confirms that the spectral shape of molecular discs with low inclination could be used to identify the presence of a two-component molecular disc in galaxies. We note that we assume an equal amount of molecular gas in the thin and thick disc to produce the spectral cubes. Nevertheless, due to the high S/N of the stacked spectra, our conclusion is not sensitive to the thick disc molecular gas fraction in a significant way. However, a detailed analysis is required to understand the dependence of the absolute molecular spectral widths on the thick disc molecular gas fraction in galaxies, which we plan to do next.

\section{Summary and conclusion}

We consider the molecular discs in spiral galaxies to be a single component or a two-component (thin+thick) system in hydrostatic equilibrium under the mutual gravity of the baryonic discs in the external potential of the dark matter halo. We set up the respective joint Poisson's-Boltzmann equation of hydrostatic equilibrium for the baryonic discs and solved them numerically in a sample of eight spiral galaxies from the HERACLES and THINGS survey to determine the three-dimensional density distribution of the molecular gas in these galaxies. We assumed a $\sigma_{CO} = 7$ and 12 \kms~for the thin and thick molecular discs, respectively. From the solutions of the hydrostatic equation, we find that the molecular gas in the thick disc extends to more significant heights than the molecular gas in a single-component disc or thin disc. We also find the thick disc molecular gas to have a very similar scale height to that of the atomic gas, driving the possibility of forming a common dynamical component together.

Exploiting the joint Poisson's-Boltzmann equation, we evaluate the impact of the dark matter halo on deciding the vertical density distribution and the midplane pressure. In our two sample galaxies, NGC 3521 and NGC 7331, we find that ignoring the dark matter halo potential can significantly underestimate the vertical density close to the midplane and overestimate the same at more significant heights. We also find that the dark matter can considerably influence the midplane pressure as well. Ignoring dark matter gravity can reduce the midplane pressure by as much as $\sim 20\%$ at the inner radii and $\sim 5-10\%$ at the outer radii. A comparison of the midplane pressure with previous studies \citep[e.g.][]{gallagher18} reveals that solving the Poisson's-Boltzmann equation self-consistently is crucial in estimating the midplane pressure correctly.

We further used the density solutions of the hydrostatic equation to estimate the vertical atomic and molecular scale heights in our sample galaxies. We find that the molecular scale height for a single-component molecular disc varies between $20-200$ pc, whereas the molecular scale height is found to vary between $50-300$ pc for a two-component disc. While solving the hydrostatic equilibrium equation, we primarily assumed that the molecular gas is equally distributed between the thin and thick discs. Even so, we also find that the vertical density profile and the molecular scale height is sensitive to the assumed thick disc molecular gas fraction and can change by a factor of approximately two for a change in the thick disc molecular gas fraction from 0.3 to 0.7. We further find that both the molecular discs flare as a function of the radius, and the normalised scale heights follow a tight exponential law. The scale length of this exponential flaring is found to be $\left(0.48 \pm 0.01 \right) r_{25}$ and $\left(0.46 \pm 0.01\right) r_{25}$ for two-component and single-component molecular discs, respectively. However, as the molecular scale height is not a directly measurable quantity, we used the density solutions to build dynamical models, column density maps, and spectral cubes for our sample galaxies. We examined the column density maps of our sample galaxies in detail to find that a non-edge-on inclination and a low spatial resolution leads to no apparent difference in the column density distribution between the single component and the two-component molecular discs.  

We study the column density maps of our sample galaxies at different assumed inclinations and spatial resolutions. Using the column density maps at $90^o$ inclination, we demonstrate that at this edge-on orientation, with a high spatial resolution ($\lesssim 100$ pc), a single component molecular disc produces a considerably thinner observed disc as compared to a two-component molecular disc. This difference quickly disappears as the spatial resolution reaches $\sim 500$ pc, which is typical of earlier large surveys. However, recent observational capabilities allow one to image the full disc of nearby galaxies with a much higher spatial resolution. For example, the PdBI Arcsecond Whirlpool Survey (PAWS) images the M51 galaxy with a spatial resolution of $\sim$ 40 pc \citep{pety13,schinnerer13}. The Physics at High Angular resolution in Nearby GalaxieS with ALMA survey (PHANGS-ALMA) aims to image the molecular gas in 74 nearby galaxies with a spatial resolution of $\sim 45-120$ pc \citep{sun18,utomo18,schinnerer19,chevance19}. The high-quality data from these surveys will be particularly useful to distinguish between a single-component and two-component molecular disc in galaxies. Besides the spatial resolution, we also find that the inclination is also very critical in imprinting any observable difference in the column density maps. A small departure from the edge-on orientation ($\sim 5^o - 10^o$) washes away any difference between the observed vertical column density profiles of the two discs.

Furthermore, in using the dynamical models of our sample galaxies, we simulated the spectral cubes for a single-component and two-component molecular disc for different inclinations. Using a traditional approach, we stacked the molecular spectra in these spectral cubes to produce global stacked spectra. Single and double Gaussian fittings to these spectra reveal that at inclinations $i \gtrsim 40^o$, spectral blending is significant and increases the spectral widths considerably. Not only that, but also at high inclinations, this broadening fictitiously produces a broad component into the stacked spectra, which exactly mimics the existence of a thick molecular disc in the spectral cube of a single-component molecular disc. However, at low inclinations ($i \lesssim 40^o$), the stacked spectra of a single and a two-component molecular disc are best fitted by a single and a double component Gaussian, respectively. Hence, we conclude that the spectral cubes of low inclination ($i \lesssim 40^o$) galaxies are ideal for examining the existence of a thick molecular disc in spiral galaxies. 

In summary, we conclude that given the typical spatial resolution and inclination, it is tough to distinguish between a single component and two-component molecular disc just by inspecting the column density maps or the spectral cubes. However, at an edge-on orientation with a high spatial resolution, it is possible to identify if the observed molecular disc is produced by a single or two-component molecular disc. Whereas for a face-on galaxy (or low inclination, $i \lesssim 40^o$), the spectral cube can be used more effectively to identify the presence of a two-component molecular disc. The stacked spectrum of a two-component disc in these cases is expected to be better fitted by a double Gaussian than a single Gaussian.

\section{Appendix}
\renewcommand\thefigure{A.\arabic{figure}}
\setcounter{figure}{0} 

\begin{figure*}[h]
\begin{center}
\begin{tabular}{cccc}
\resizebox{0.23\textwidth}{!}{\includegraphics{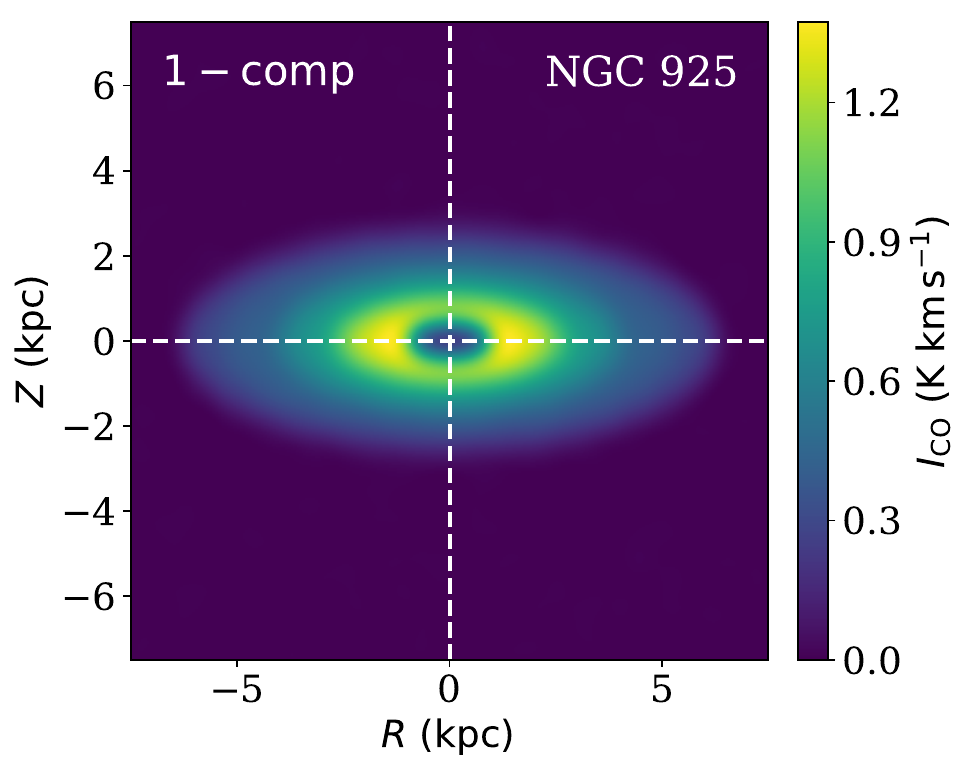}} &
\resizebox{0.23\textwidth}{!}{\includegraphics{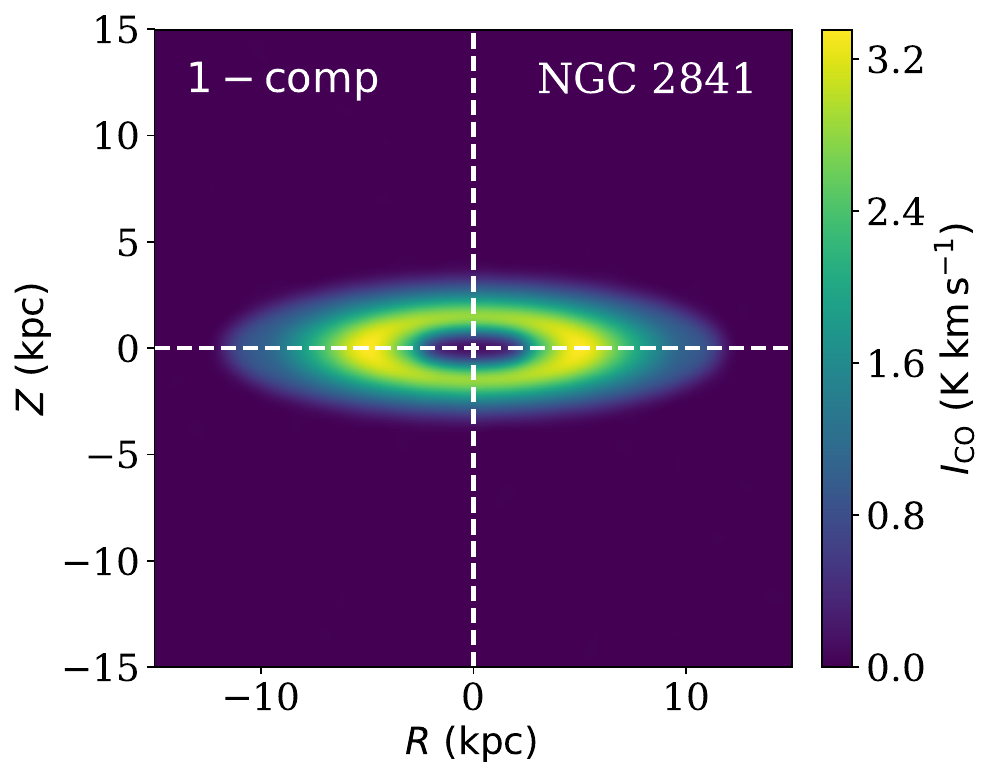}} &
\resizebox{0.23\textwidth}{!}{\includegraphics{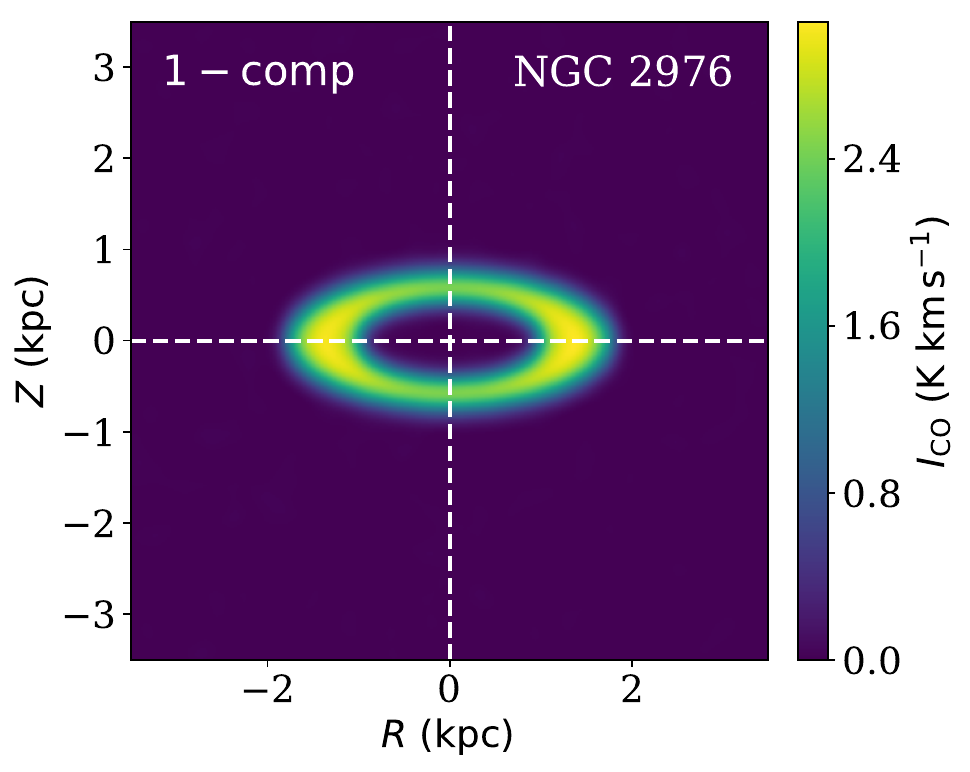}} &
\resizebox{0.25\textwidth}{!}{\includegraphics{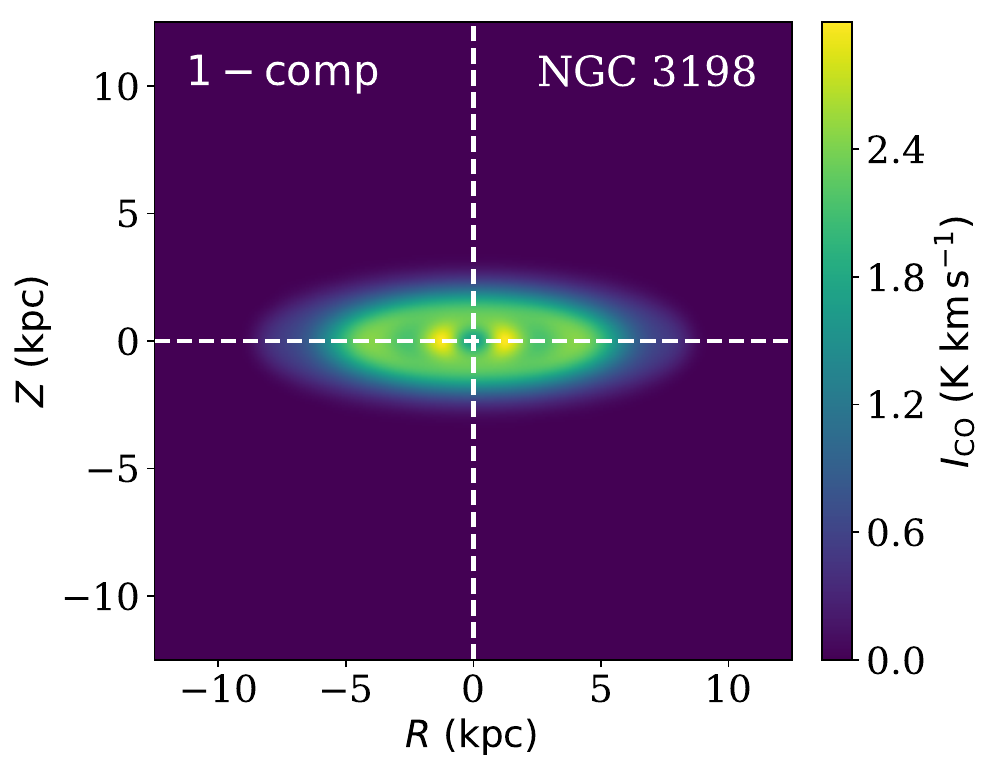}} \\

\resizebox{0.23\textwidth}{!}{\includegraphics{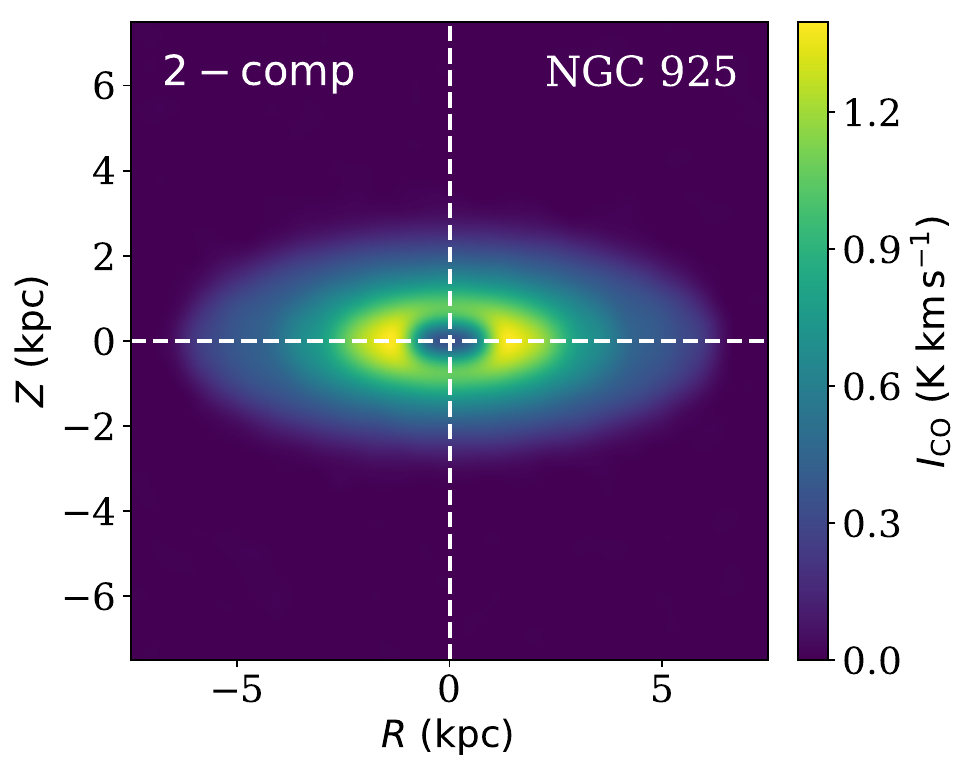}} &
\resizebox{0.23\textwidth}{!}{\includegraphics{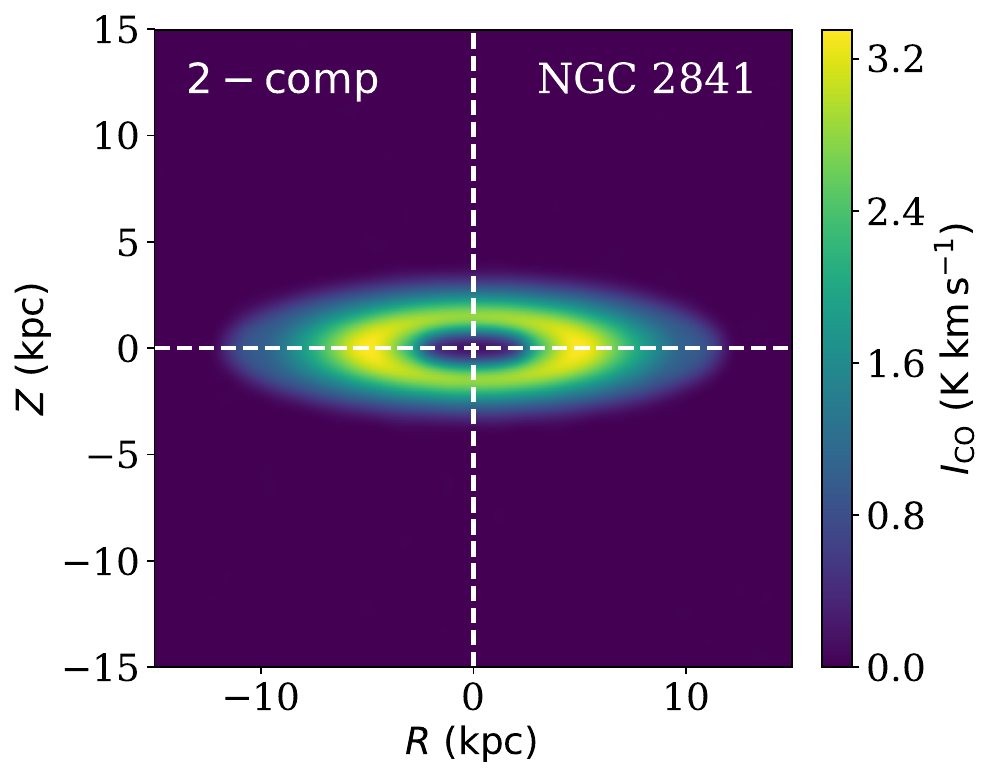}} &
\resizebox{0.23\textwidth}{!}{\includegraphics{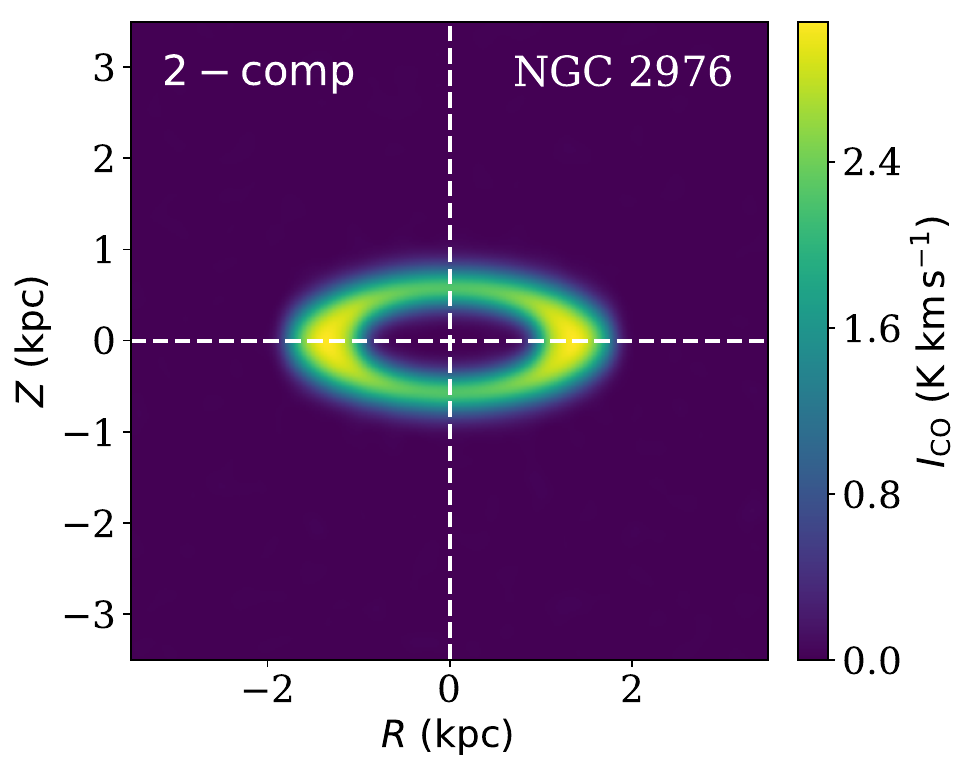}} &
\resizebox{0.23\textwidth}{!}{\includegraphics{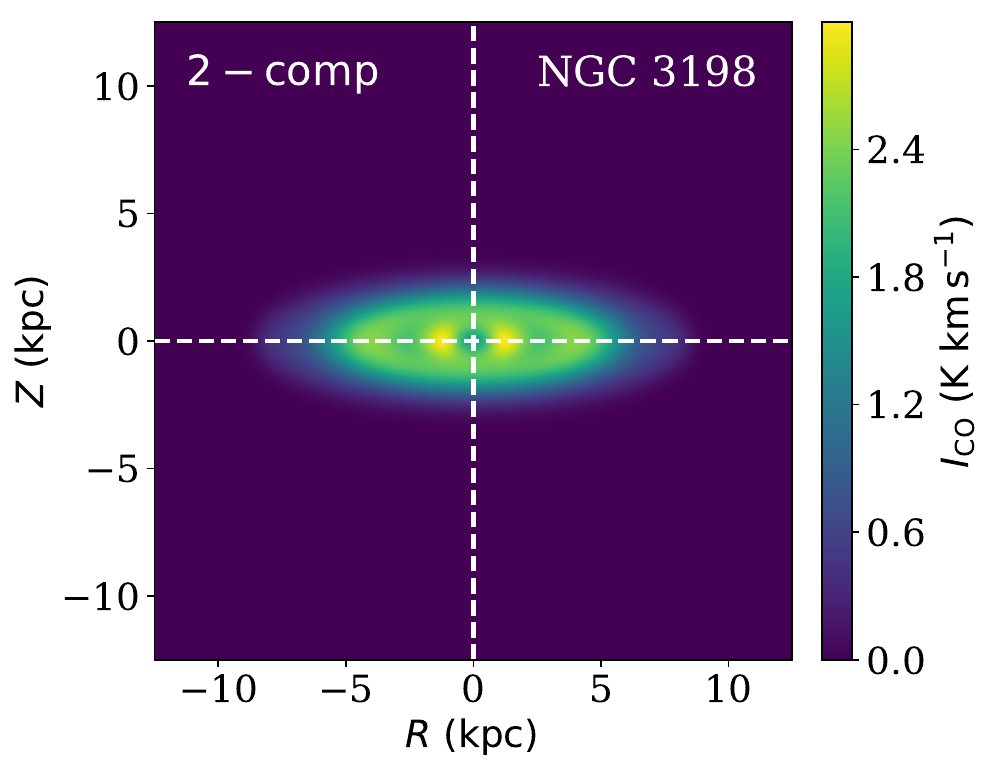}} \\

\resizebox{0.23\textwidth}{!}{\includegraphics{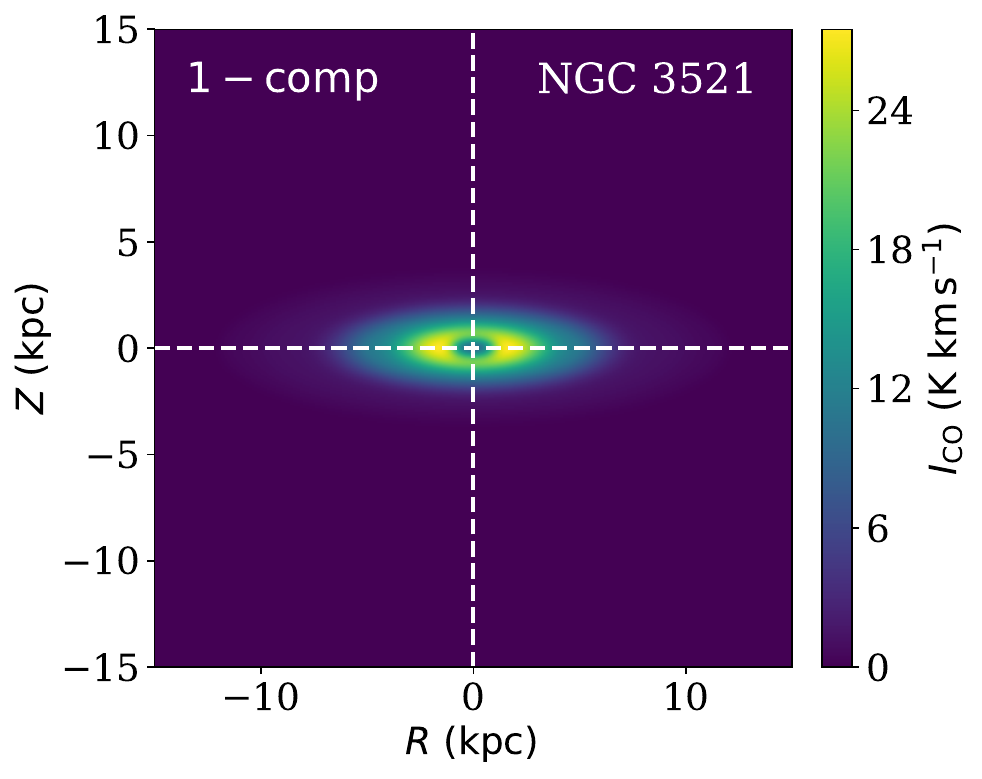}} &
\resizebox{0.23\textwidth}{!}{\includegraphics{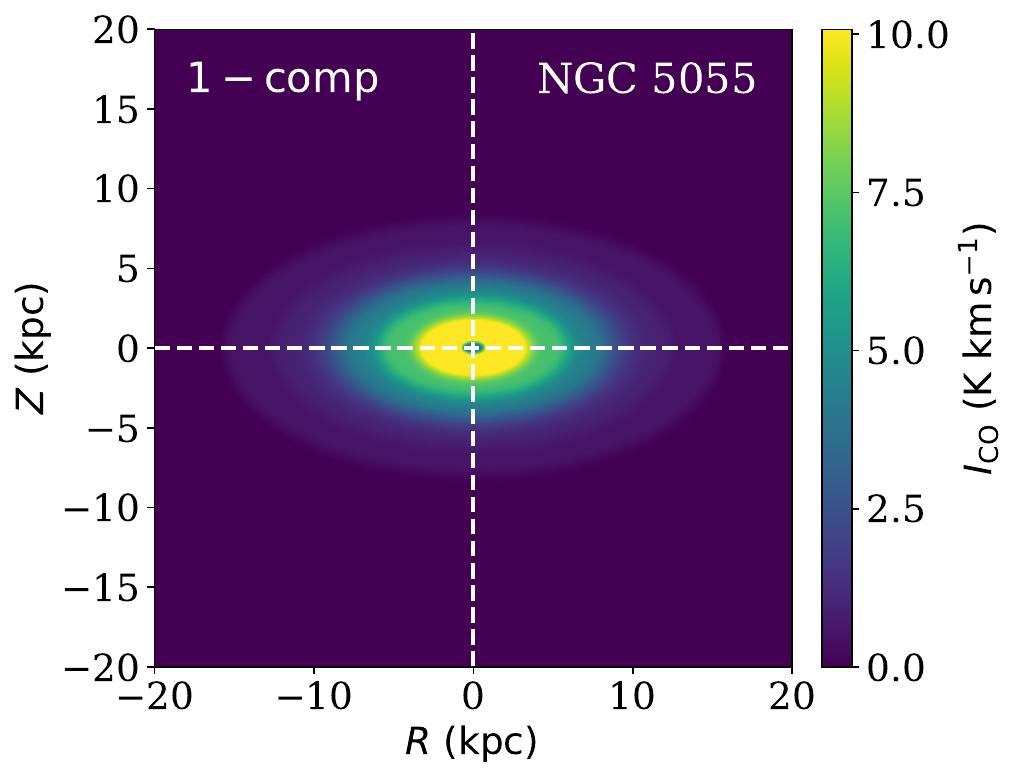}} &
\resizebox{0.23\textwidth}{!}{\includegraphics{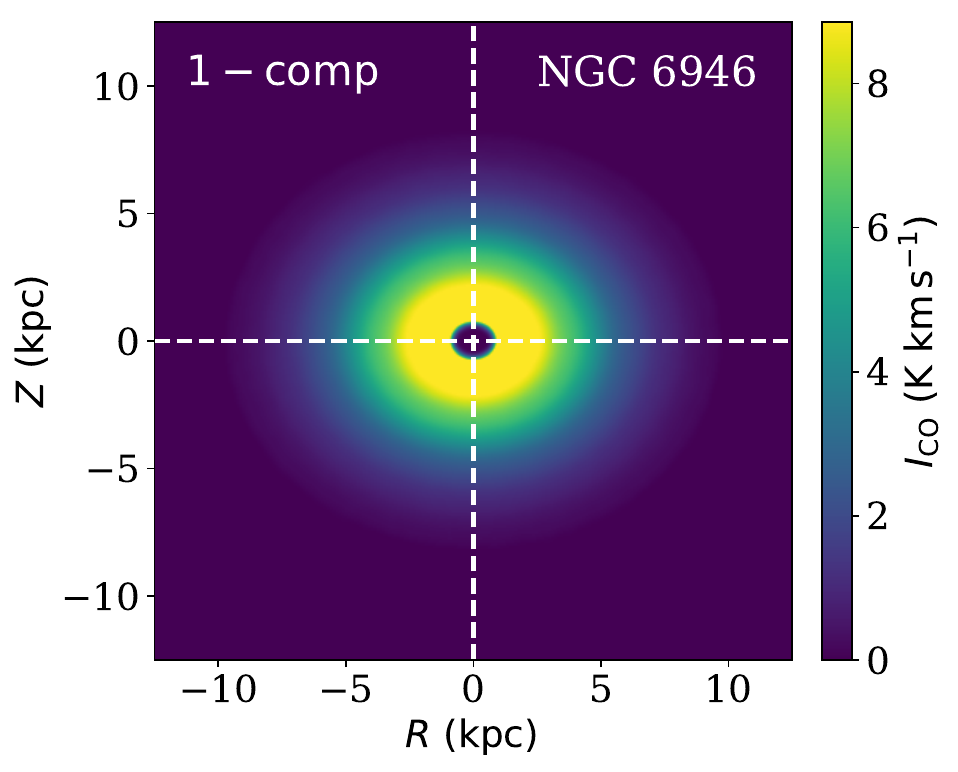}} &
\resizebox{0.23\textwidth}{!}{\includegraphics{n7331_1comp_incl76_mom0.pdf}} \\

\resizebox{0.23\textwidth}{!}{\includegraphics{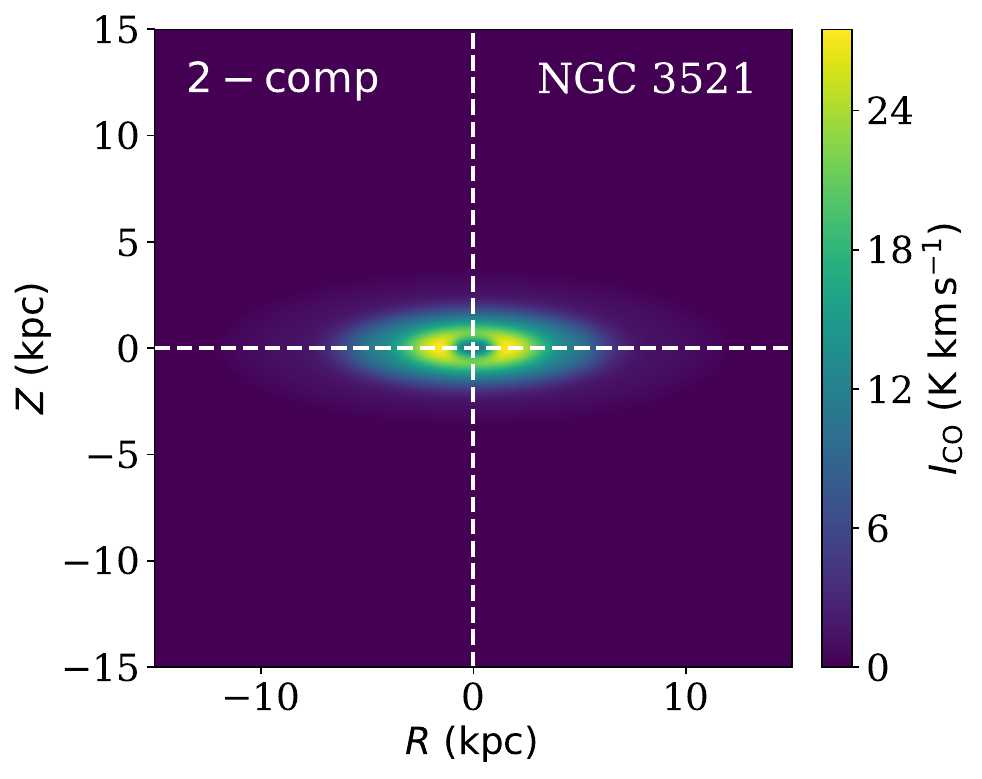}} &
\resizebox{0.23\textwidth}{!}{\includegraphics{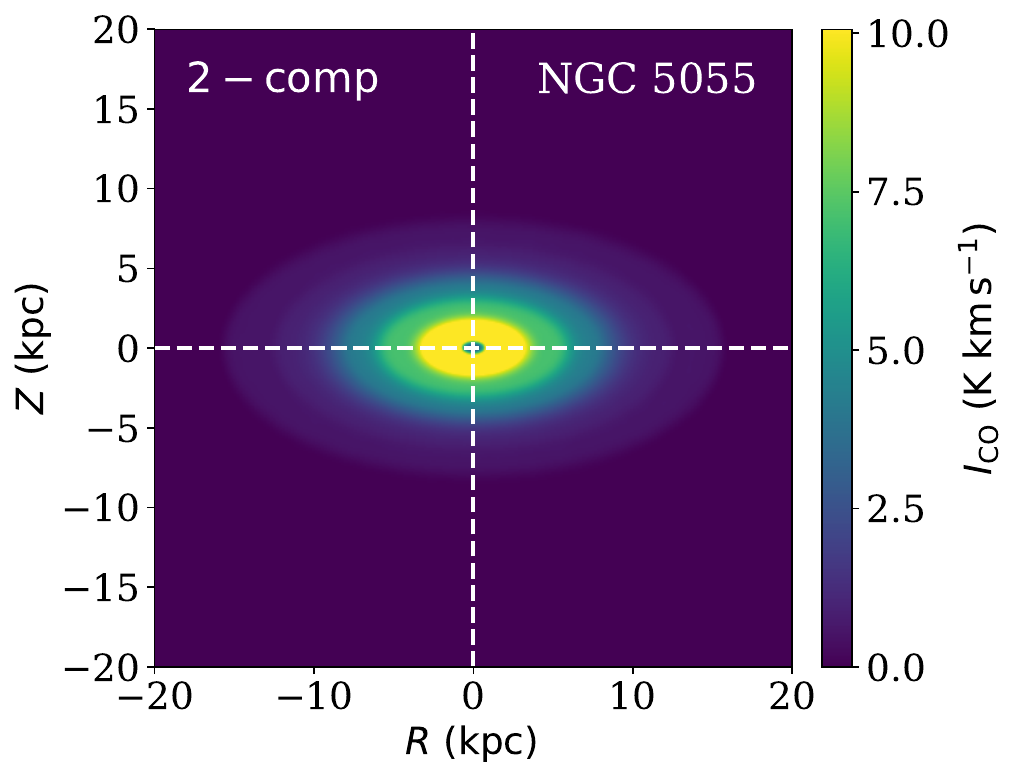}} &
\resizebox{0.23\textwidth}{!}{\includegraphics{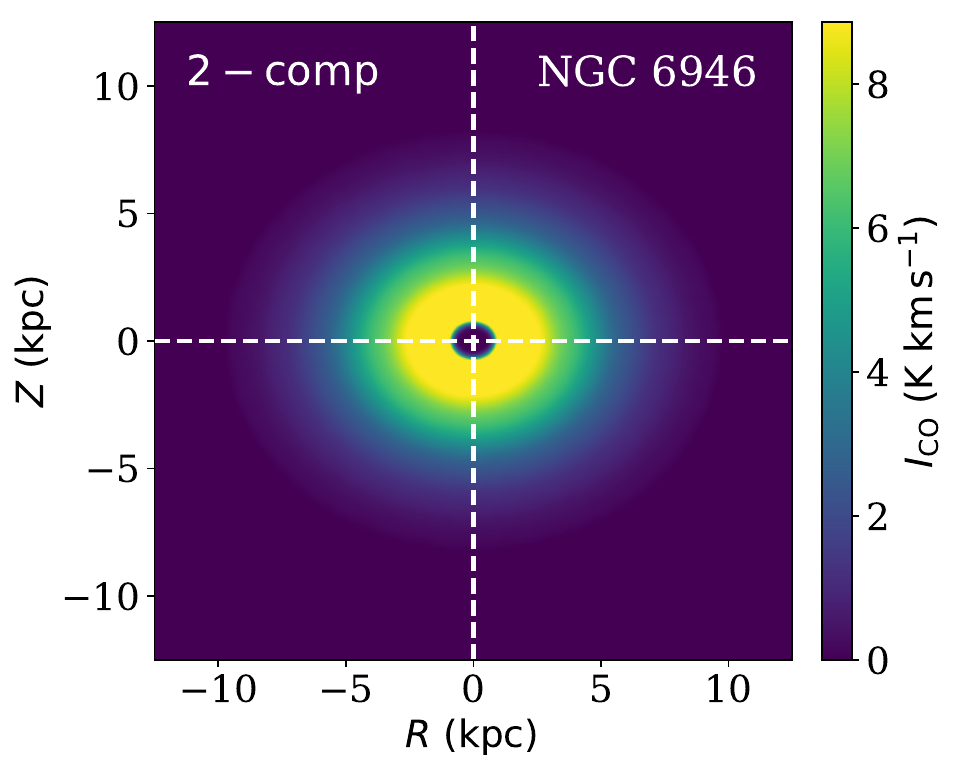}} &
\resizebox{0.23\textwidth}{!}{\includegraphics{n7331_2comp_incl76_mom0.pdf}} \\

\end{tabular}
\end{center}
\caption{Modelled molecular discs of all of our sample galaxies (Same as Fig.~\ref{mom_ncomp}). In the first and third rows from the top, we show the CO intensity maps for single-component molecular discs. In the second and the fourth rows from the top, we show the same for two-component molecular discs. The names of the galaxies are quoted in the top right corner of the respective panels. The colour bars represent the simulated CO intensities in the units of K \kms.}
\label{mom_ncomp_all}
\end{figure*}

\begin{figure*}[h]
\begin{center}
\begin{tabular}{cccc}
\resizebox{0.23\textwidth}{!}{\includegraphics{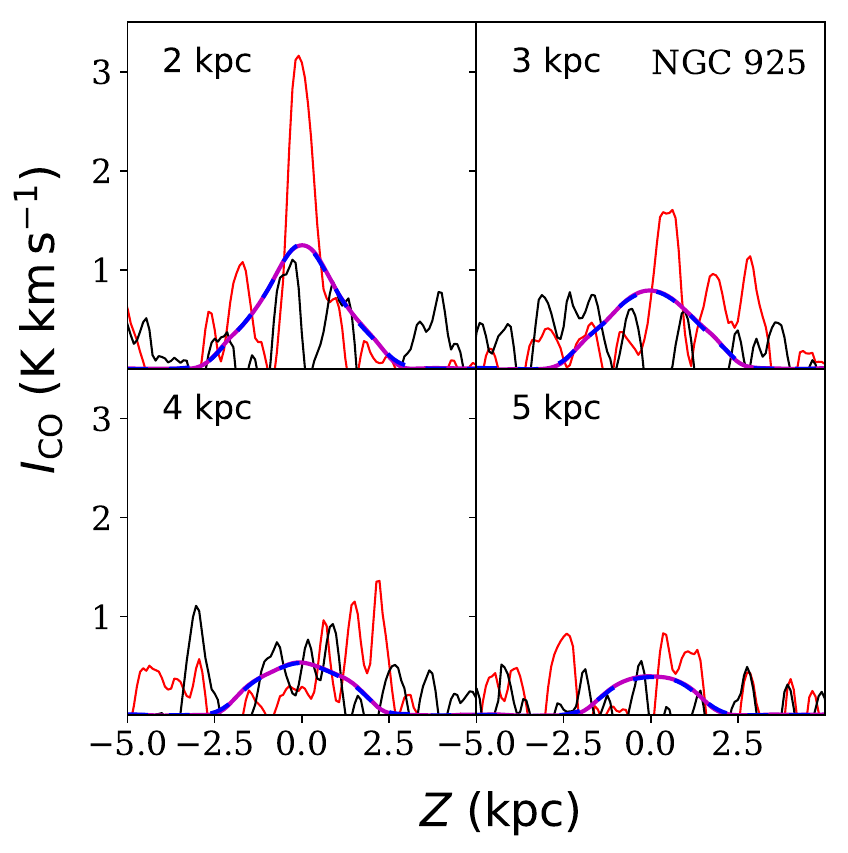}} &
\resizebox{0.23\textwidth}{!}{\includegraphics{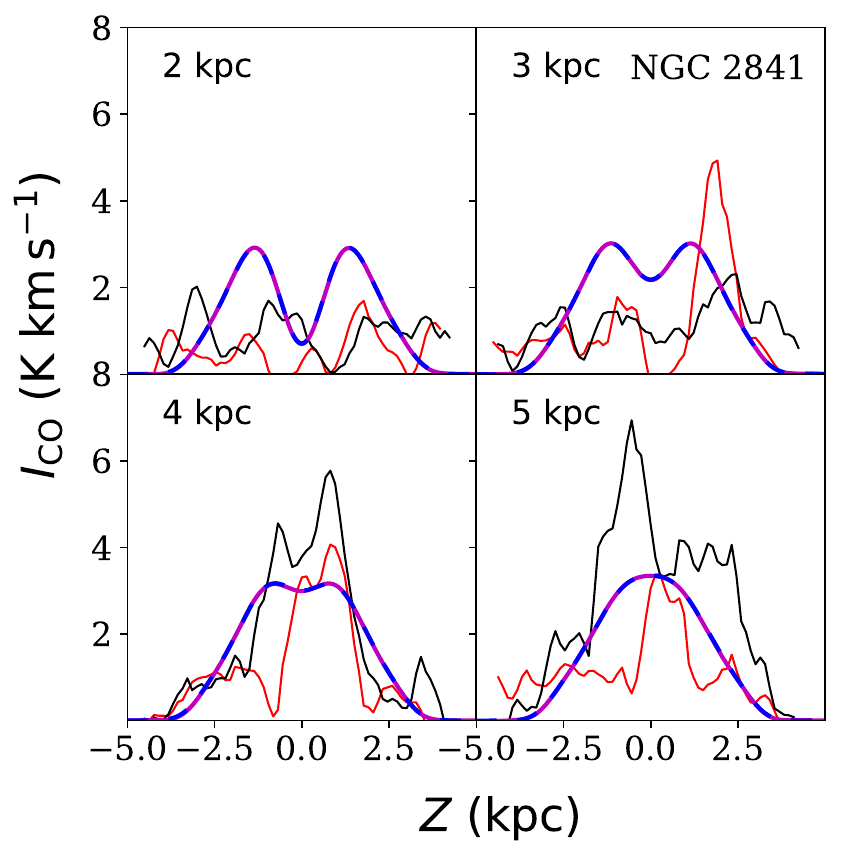}} &
\resizebox{0.23\textwidth}{!}{\includegraphics{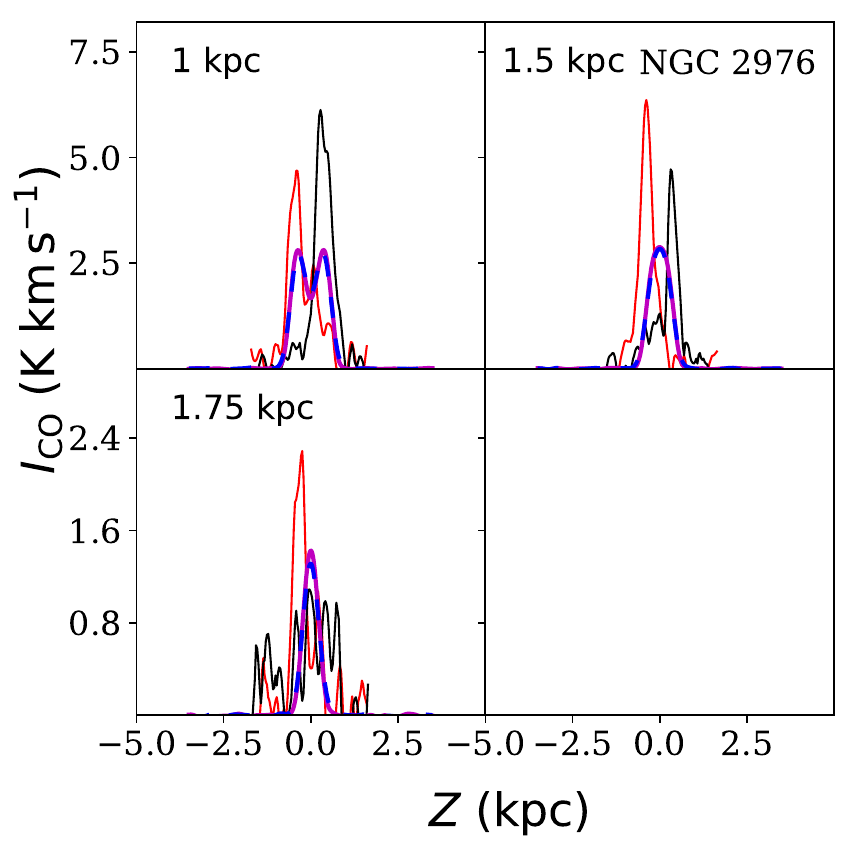}} &
\resizebox{0.23\textwidth}{!}{\includegraphics{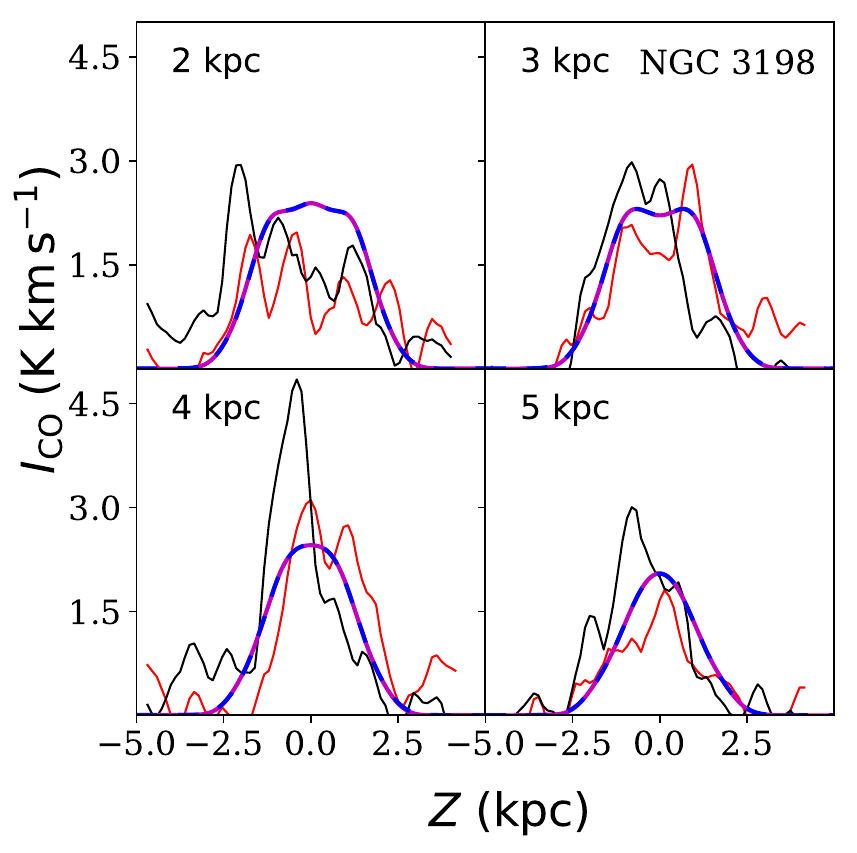}} \\

\resizebox{0.23\textwidth}{!}{\includegraphics{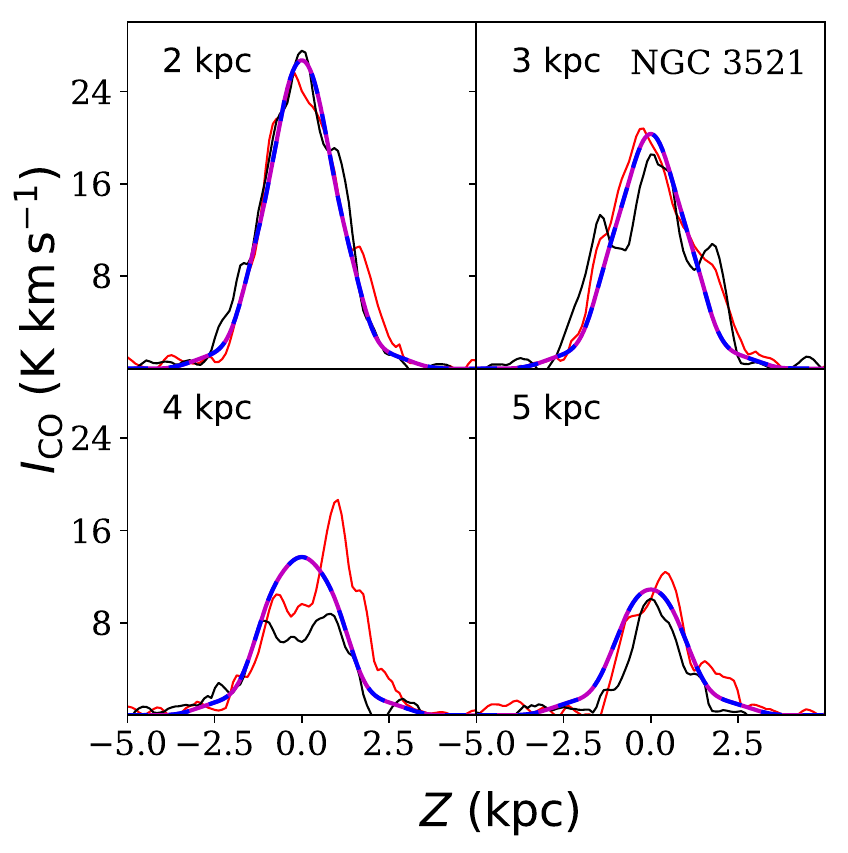}} &
\resizebox{0.23\textwidth}{!}{\includegraphics{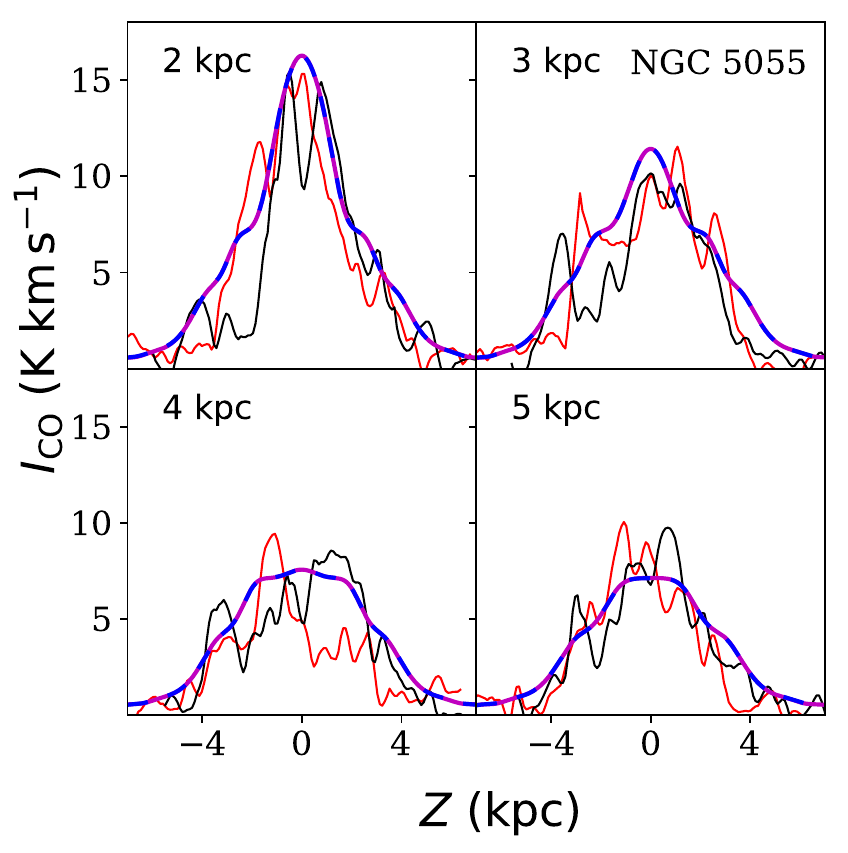}} &
\resizebox{0.23\textwidth}{!}{\includegraphics{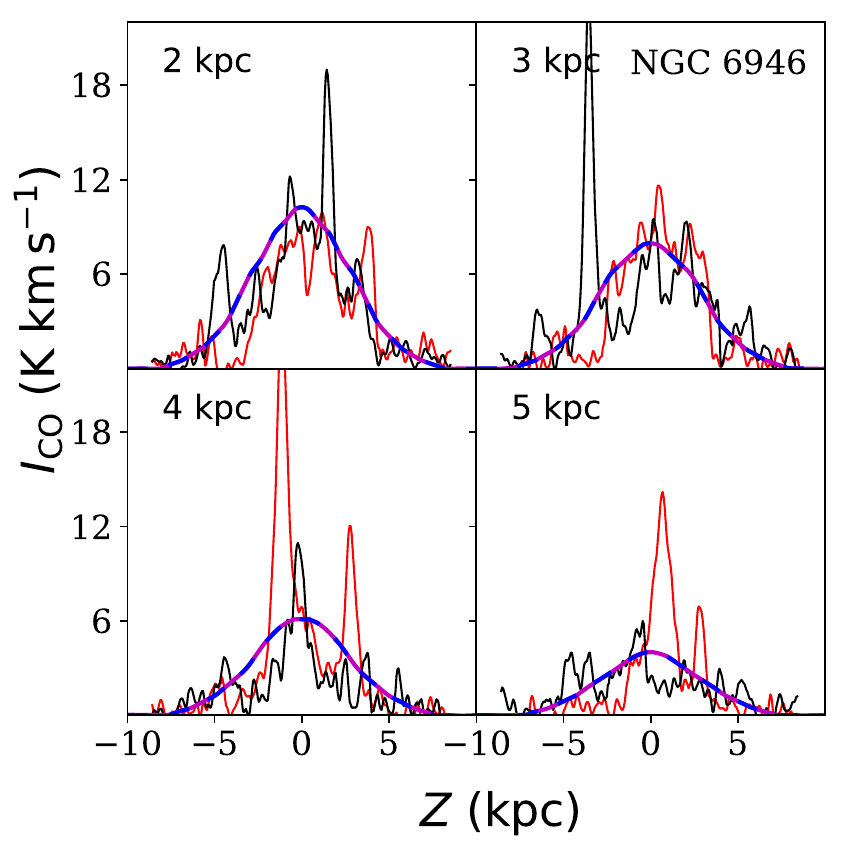}} &
\resizebox{0.23\textwidth}{!}{\includegraphics{n7331_comp_vprof.pdf}} \\
\end{tabular}
\end{center}
\caption{Vertical column density profiles of the molecular discs of all of our sample galaxies. The figure legends are the same as in Fig.~\ref{vprof}.}
\label{vprof_all}
\end{figure*}

\begin{figure*}[h]
\begin{center}
\begin{tabular}{cccc}
\resizebox{0.23\textwidth}{!}{\includegraphics{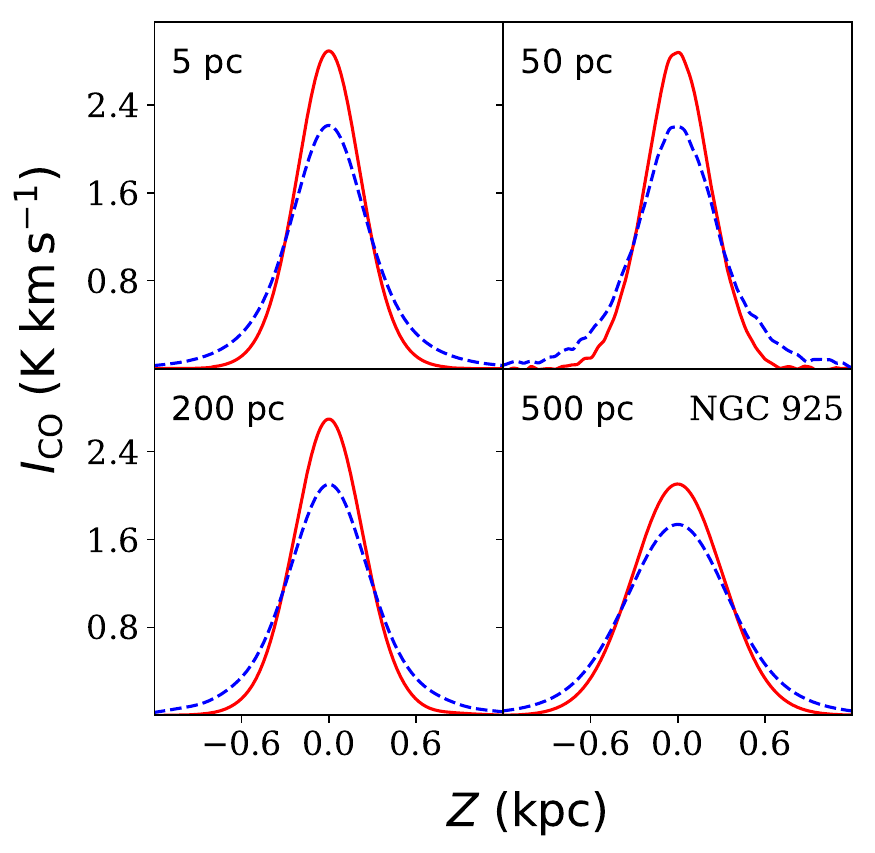}} &
\resizebox{0.23\textwidth}{!}{\includegraphics{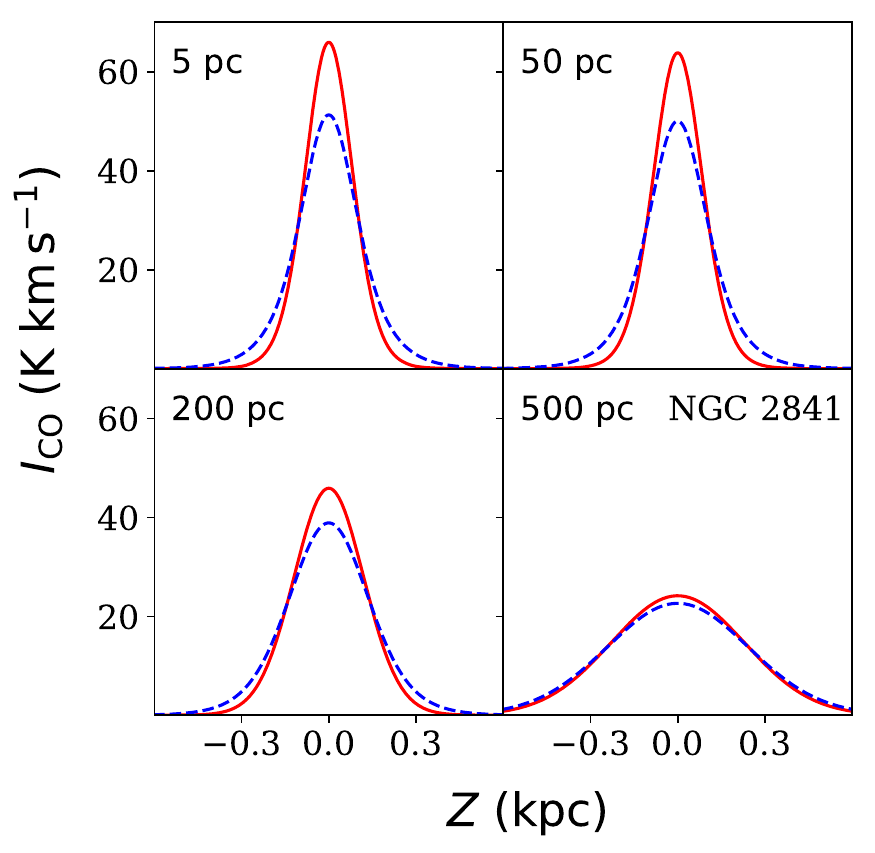}} &
\resizebox{0.23\textwidth}{!}{\includegraphics{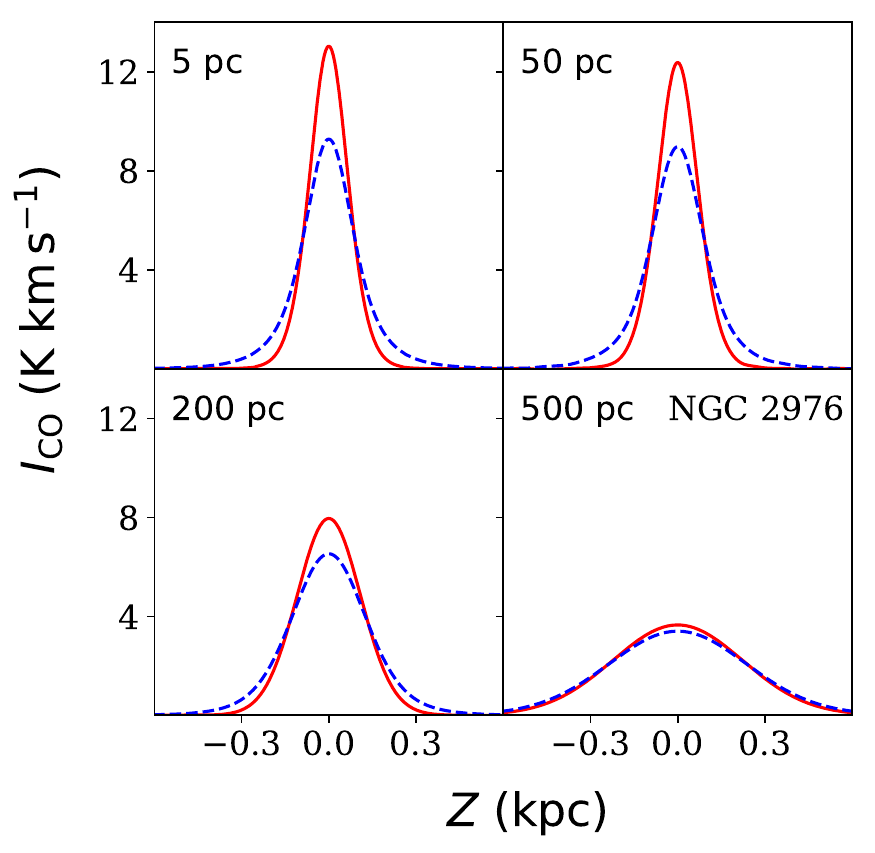}} &
\resizebox{0.23\textwidth}{!}{\includegraphics{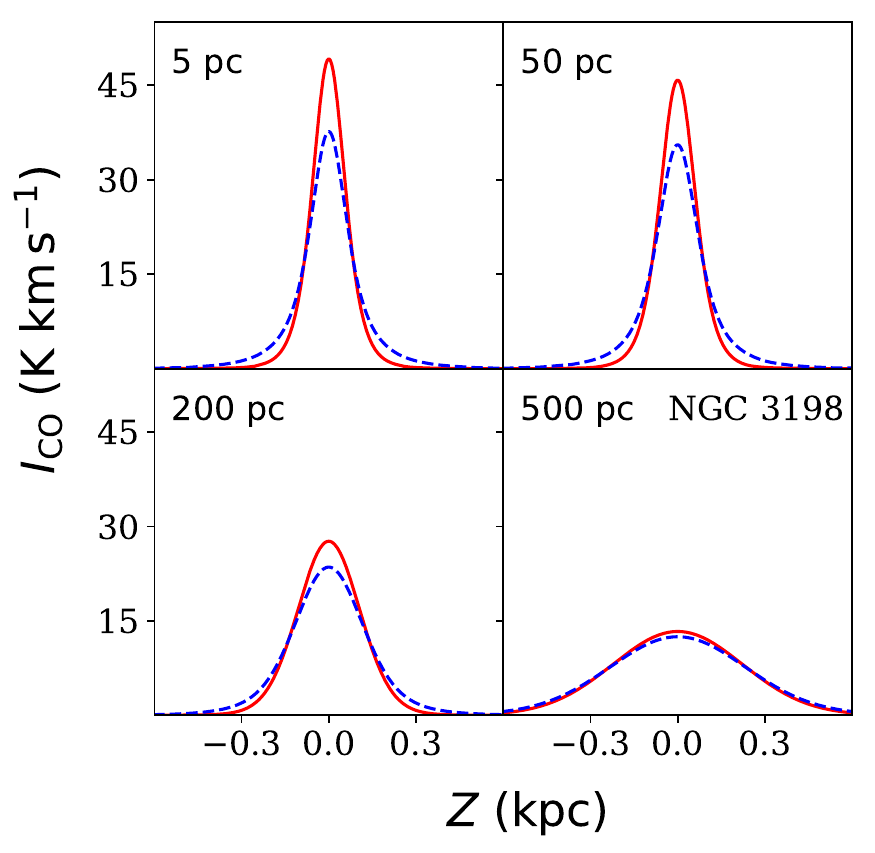}} \\

\resizebox{0.23\textwidth}{!}{\includegraphics{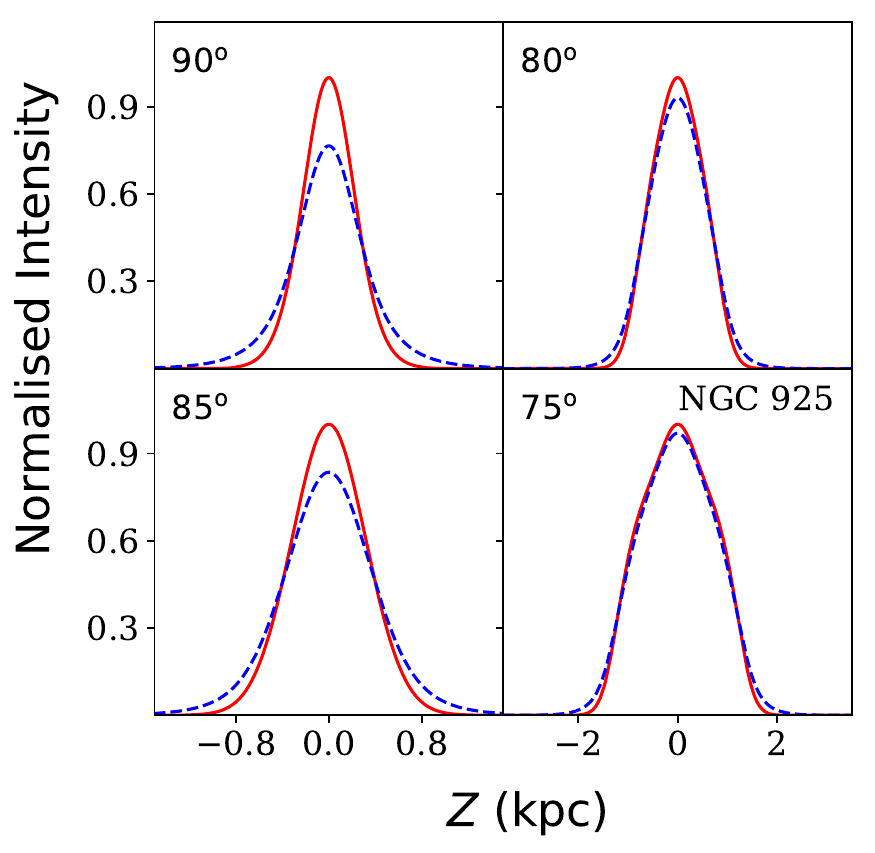}} &
\resizebox{0.23\textwidth}{!}{\includegraphics{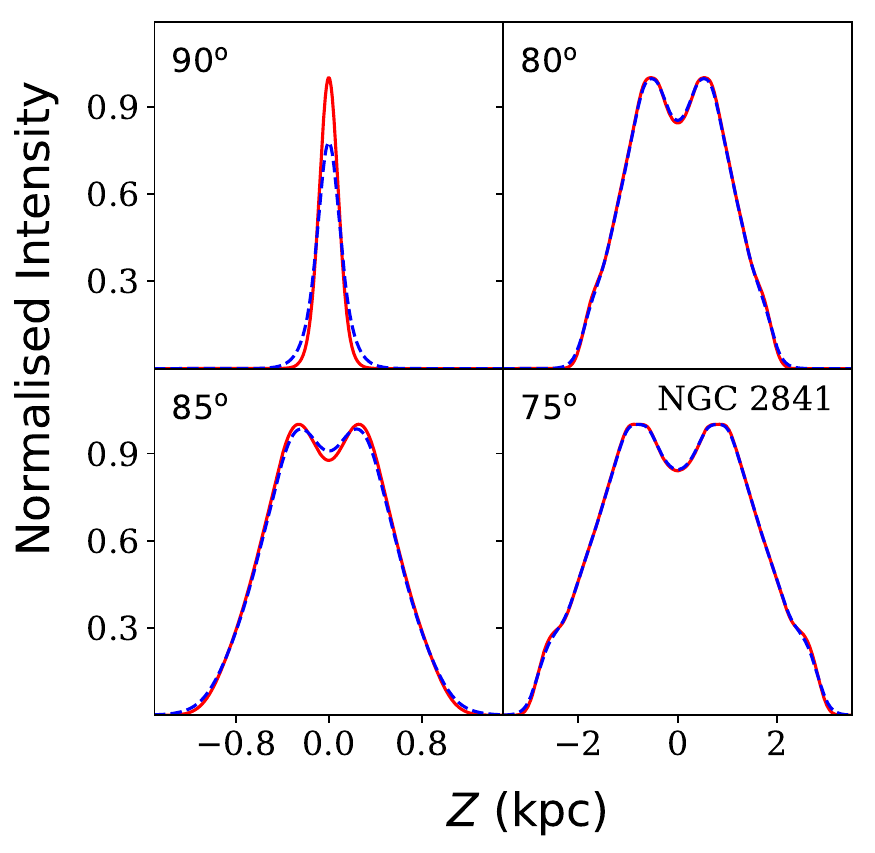}} &
\resizebox{0.23\textwidth}{!}{\includegraphics{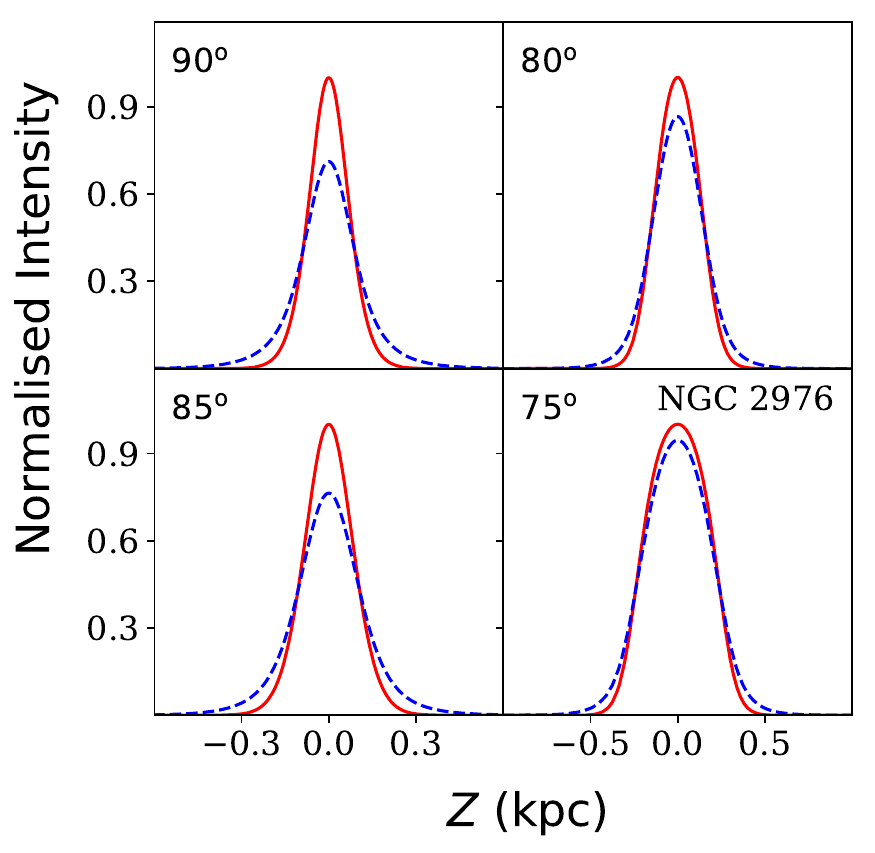}} &
\resizebox{0.23\textwidth}{!}{\includegraphics{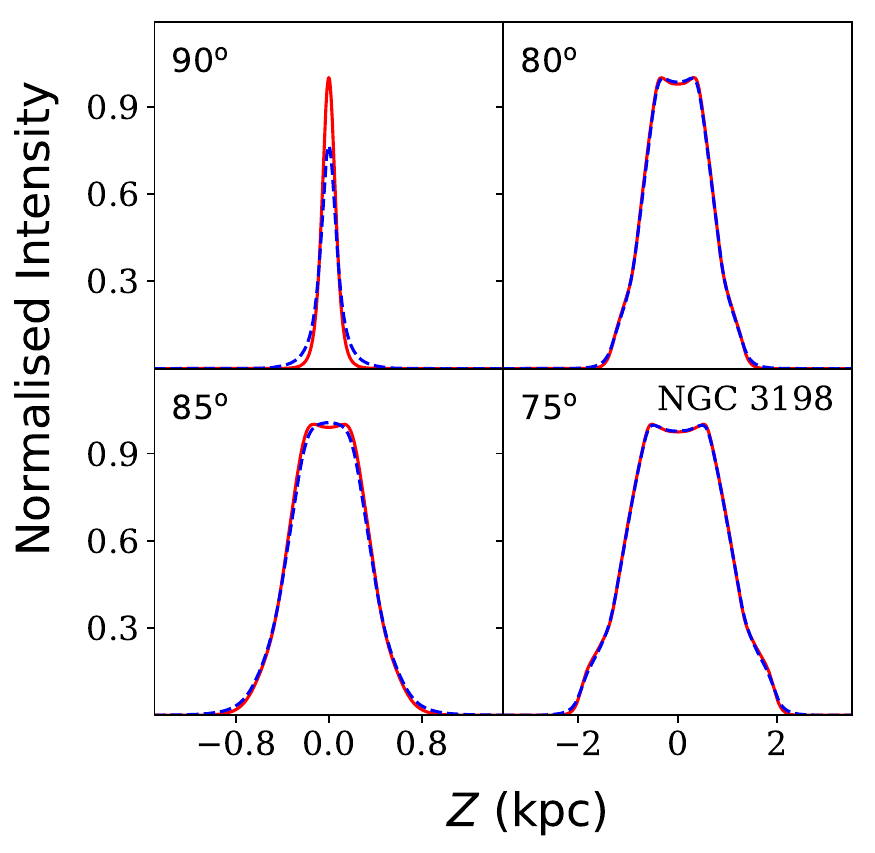}} \\

\resizebox{0.23\textwidth}{!}{\includegraphics{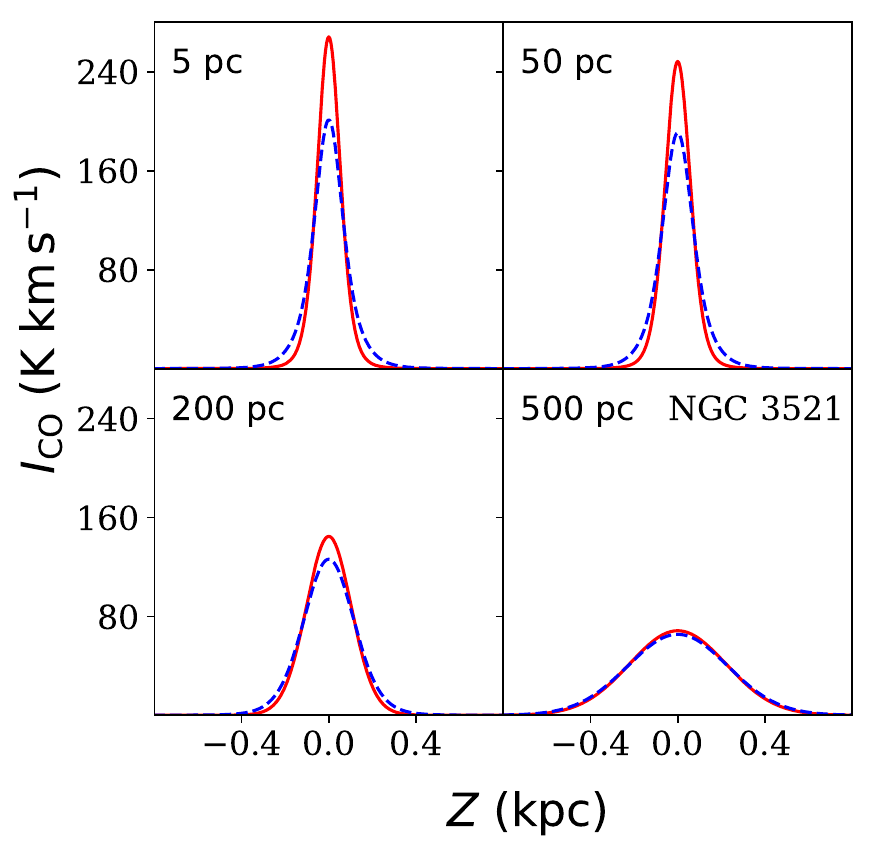}} &
\resizebox{0.23\textwidth}{!}{\includegraphics{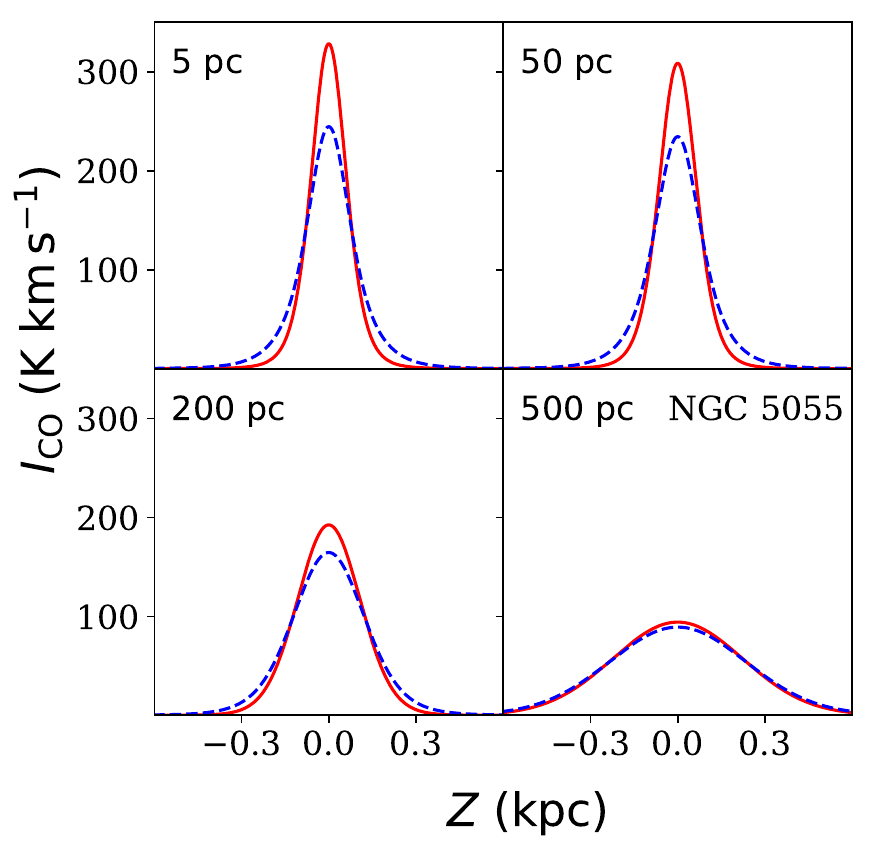}} &
\resizebox{0.23\textwidth}{!}{\includegraphics{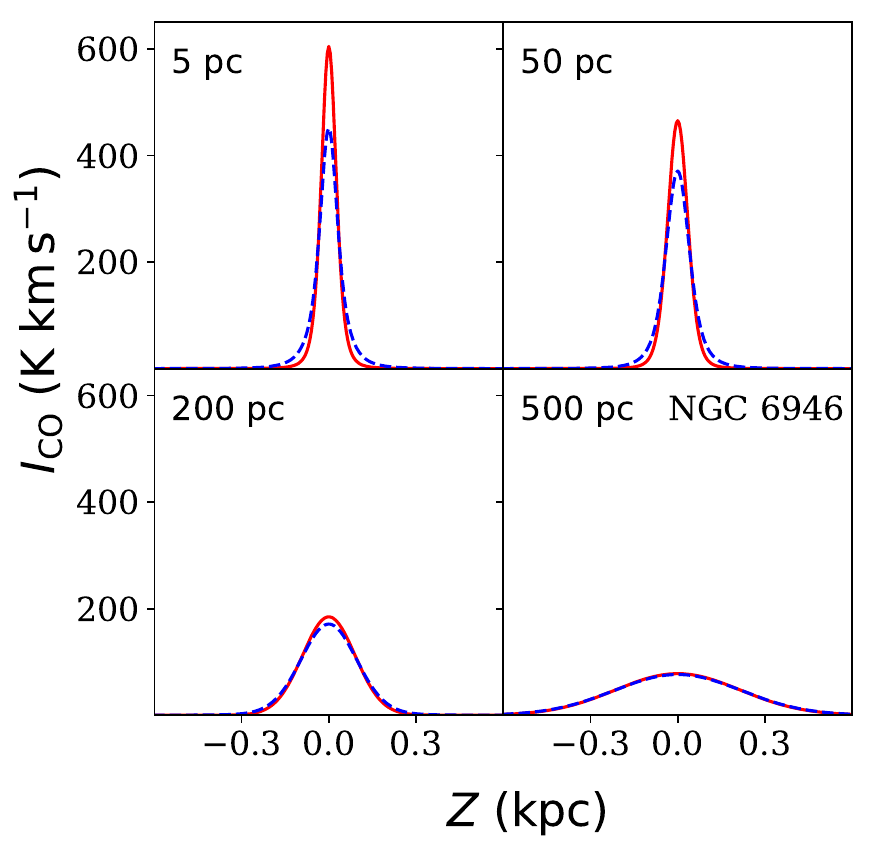}} &
\resizebox{0.23\textwidth}{!}{\includegraphics{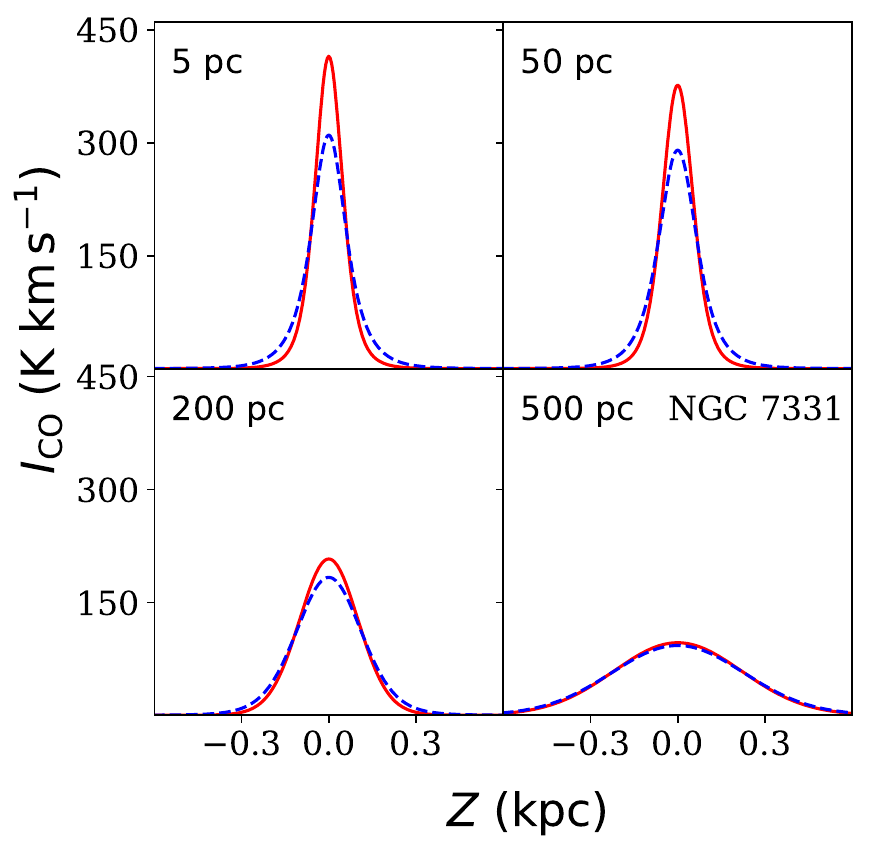}} \\

\resizebox{0.23\textwidth}{!}{\includegraphics{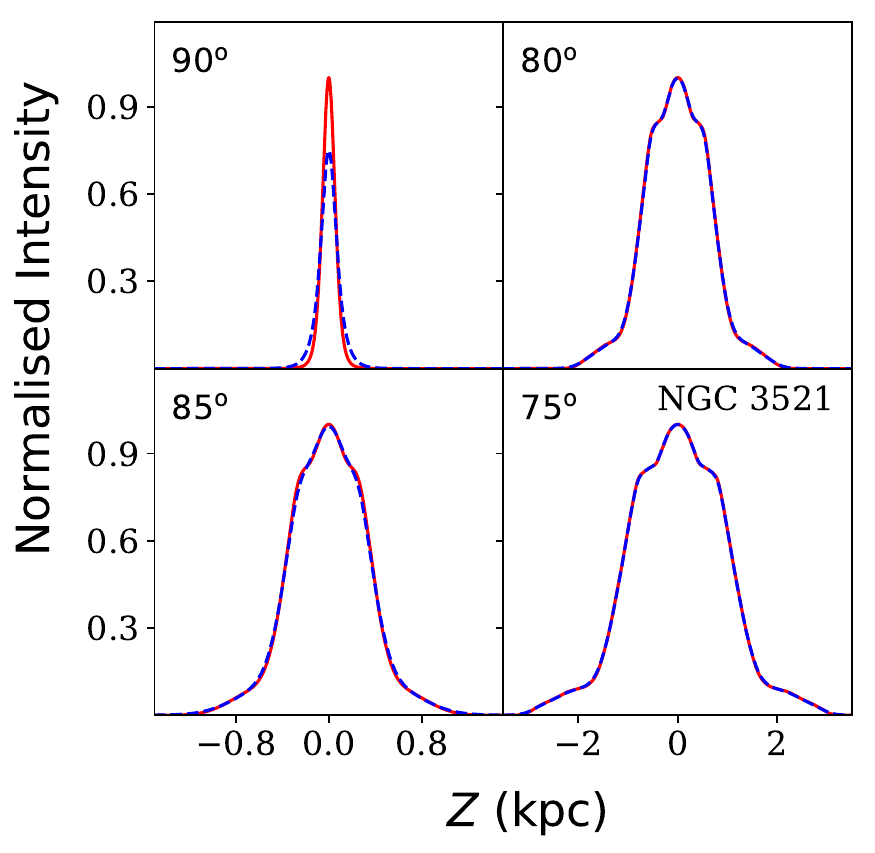}} &
\resizebox{0.23\textwidth}{!}{\includegraphics{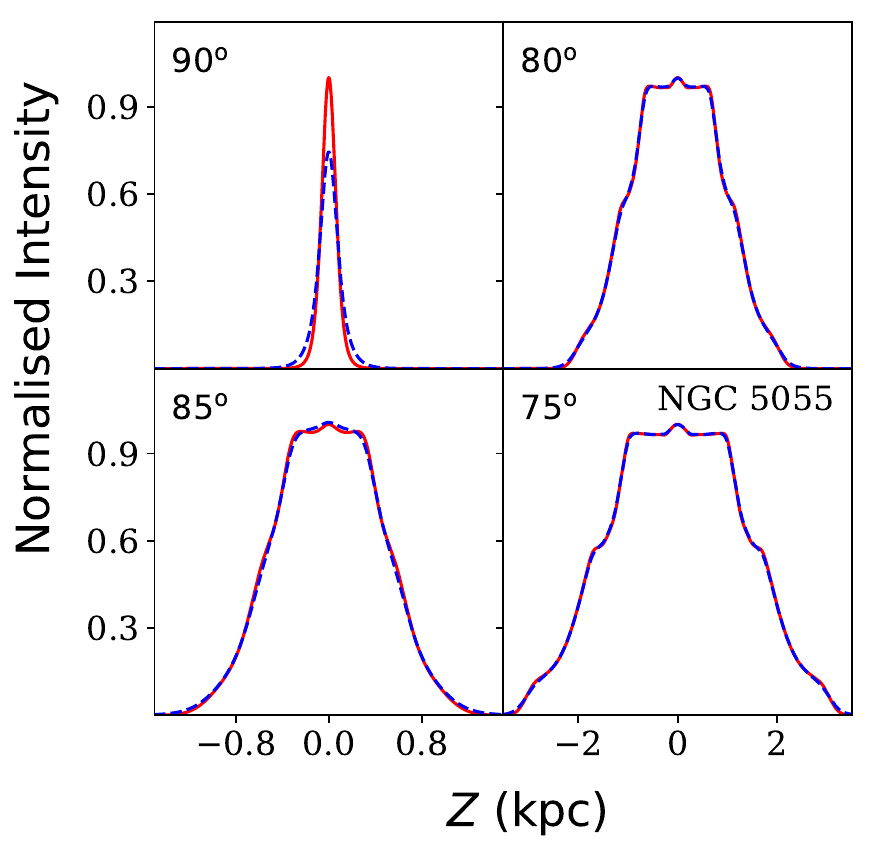}} &
\resizebox{0.23\textwidth}{!}{\includegraphics{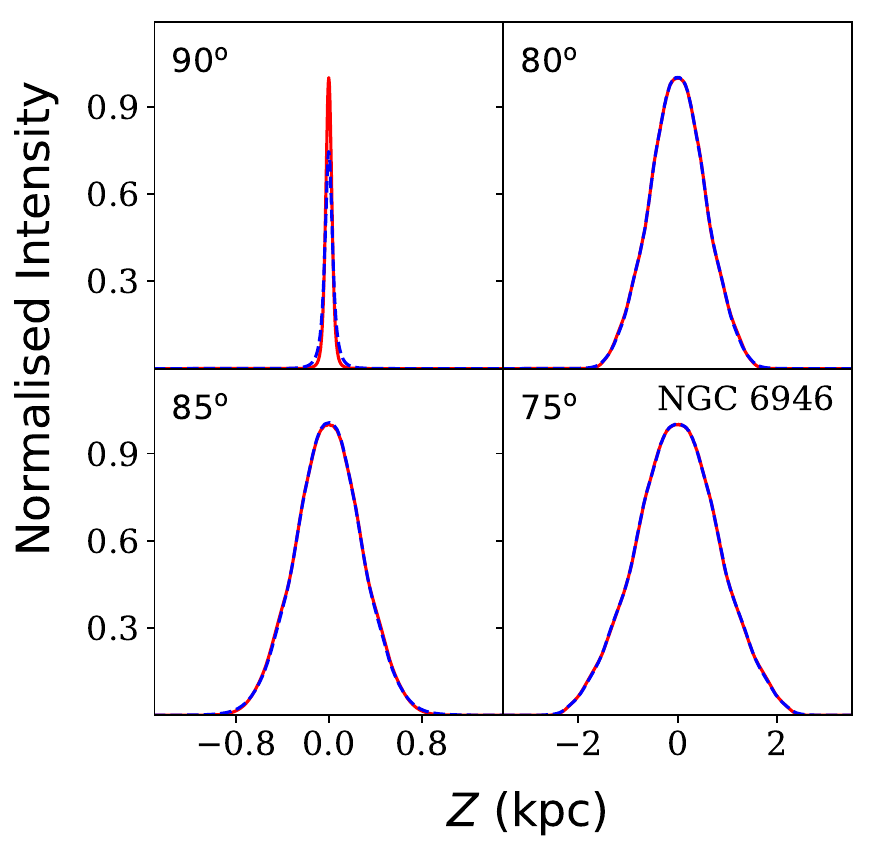}} &
\resizebox{0.23\textwidth}{!}{\includegraphics{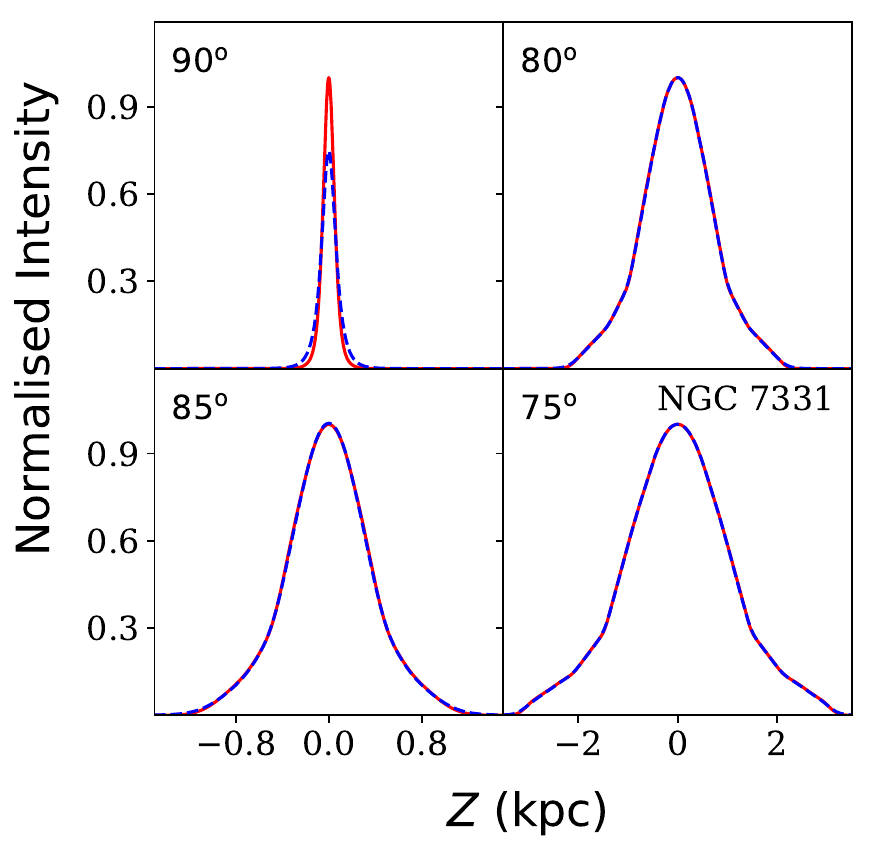}} \\

\end{tabular}
\end{center}
\caption{Comparison of the vertical column density profiles for a single-component and a two-component molecular disc in our sample galaxies (same as Fig.~\ref{vprof_incl90}). The first and third rows show the vertical column density profiles for different spatial resolutions, whereas the second and the fourth column show the same for different inclinations. Different sub-panels represent different spatial resolutions or inclinations, as quoted in the top left of each sub-panel. The vertical profiles were extracted at a radius of 4 kpc for all of the galaxies except NGC 2976 for which it was computed at 1.5 kpc (as its molecular disc does not extend to a radius of 4 kpc). The solid red lines represent the vertical profiles for single-component discs, whereas the blue dashed lines represent the same for two-component molecular discs.}
\label{vprof_incl90_all}
\end{figure*}

\begin{figure*}[h]
\begin{center}
\begin{tabular}{cccc}
\resizebox{0.23\textwidth}{!}{\includegraphics{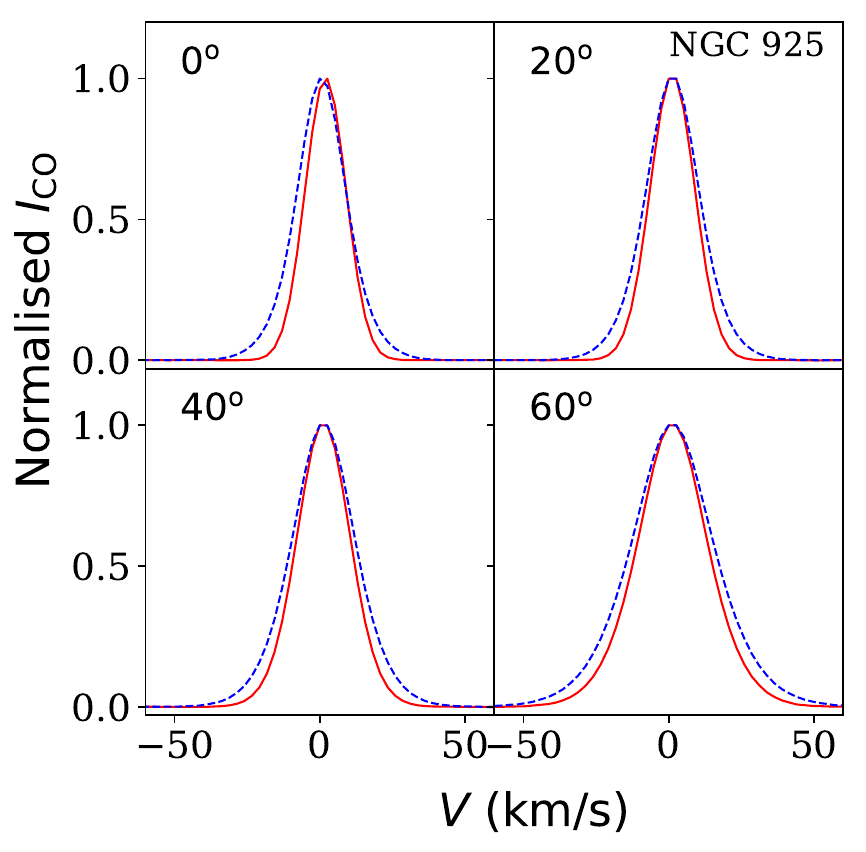}} &
\resizebox{0.23\textwidth}{!}{\includegraphics{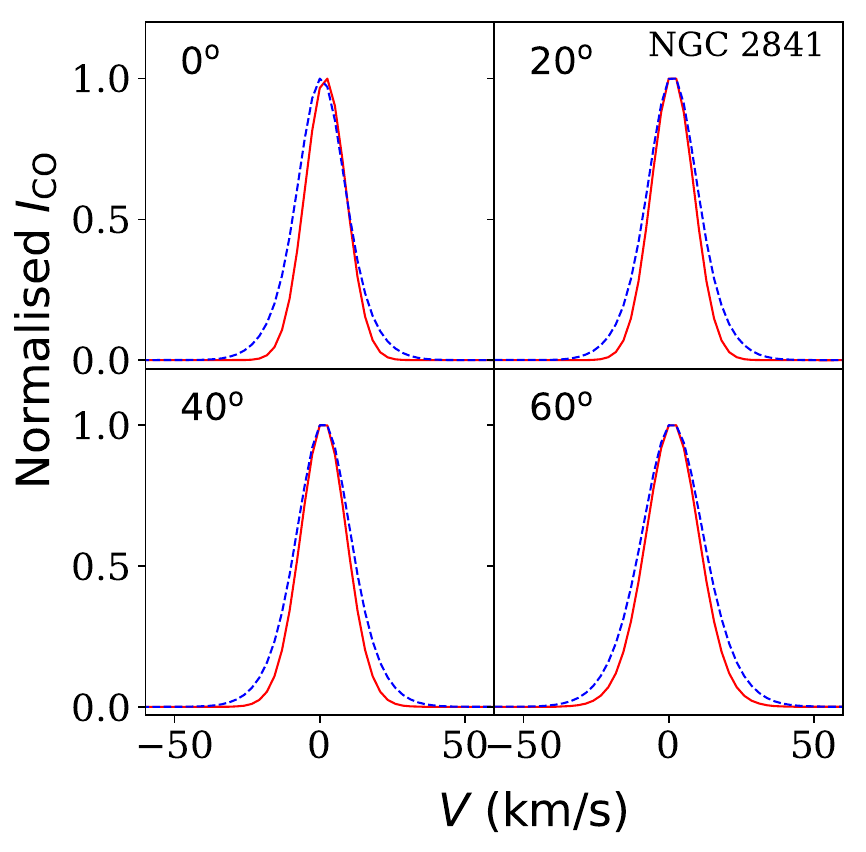}} &
\resizebox{0.23\textwidth}{!}{\includegraphics{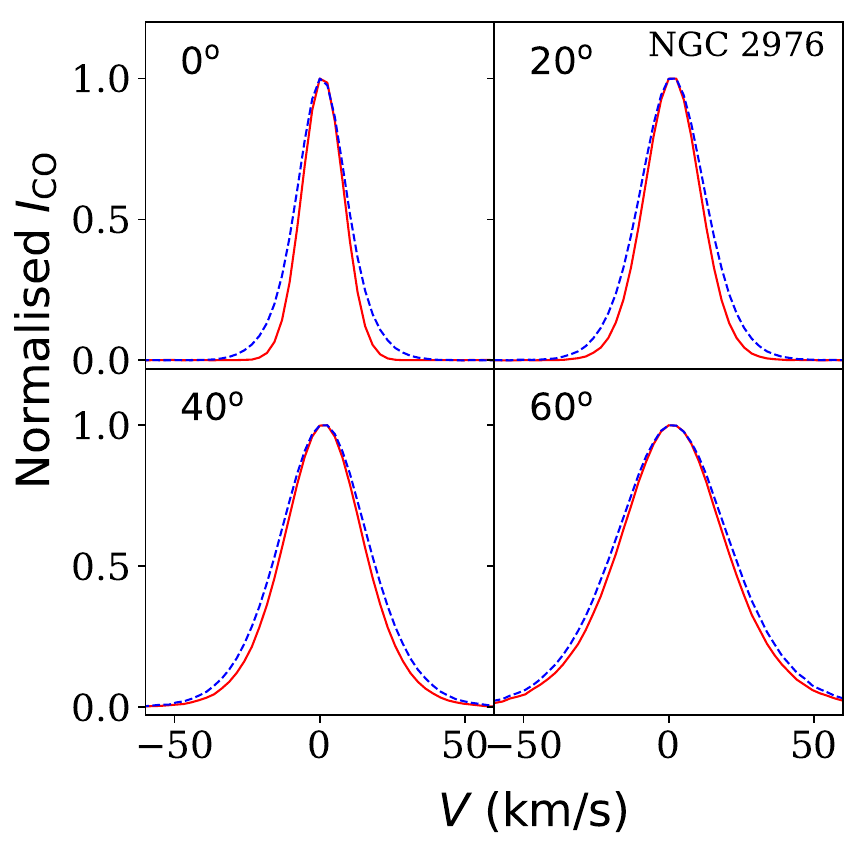}} &
\resizebox{0.23\textwidth}{!}{\includegraphics{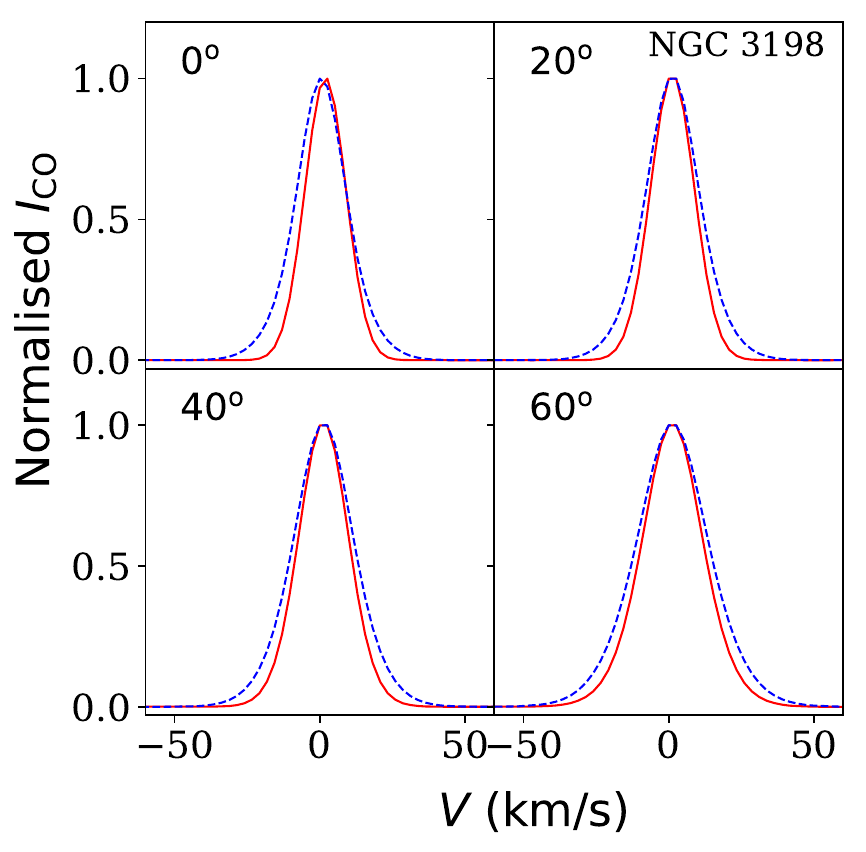}} \\

\resizebox{0.23\textwidth}{!}{\includegraphics{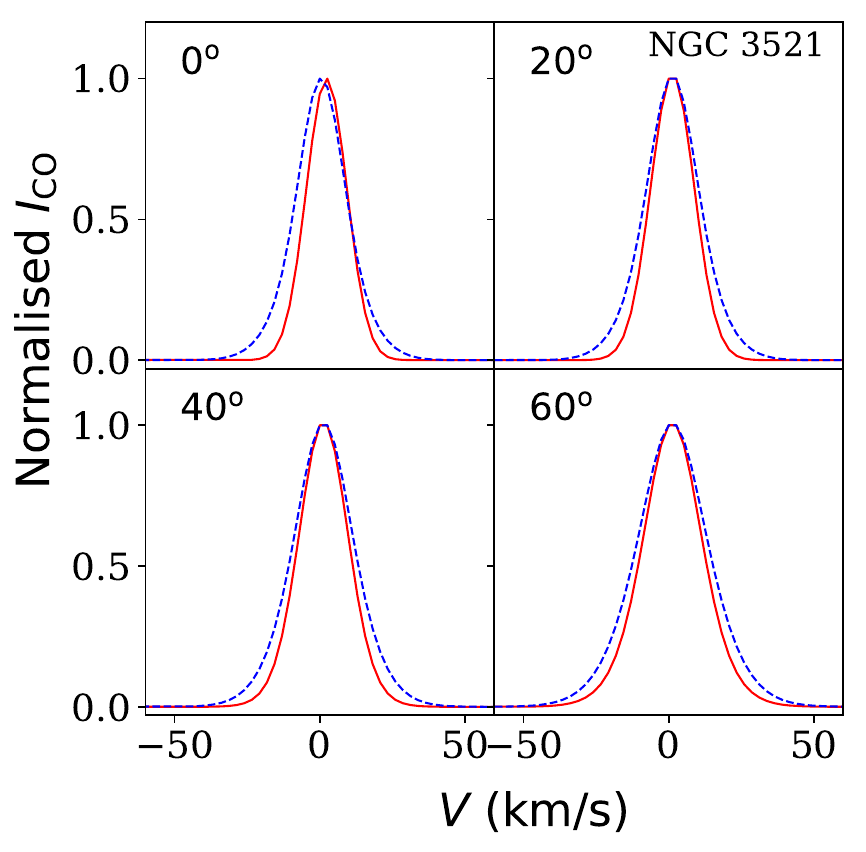}} &
\resizebox{0.23\textwidth}{!}{\includegraphics{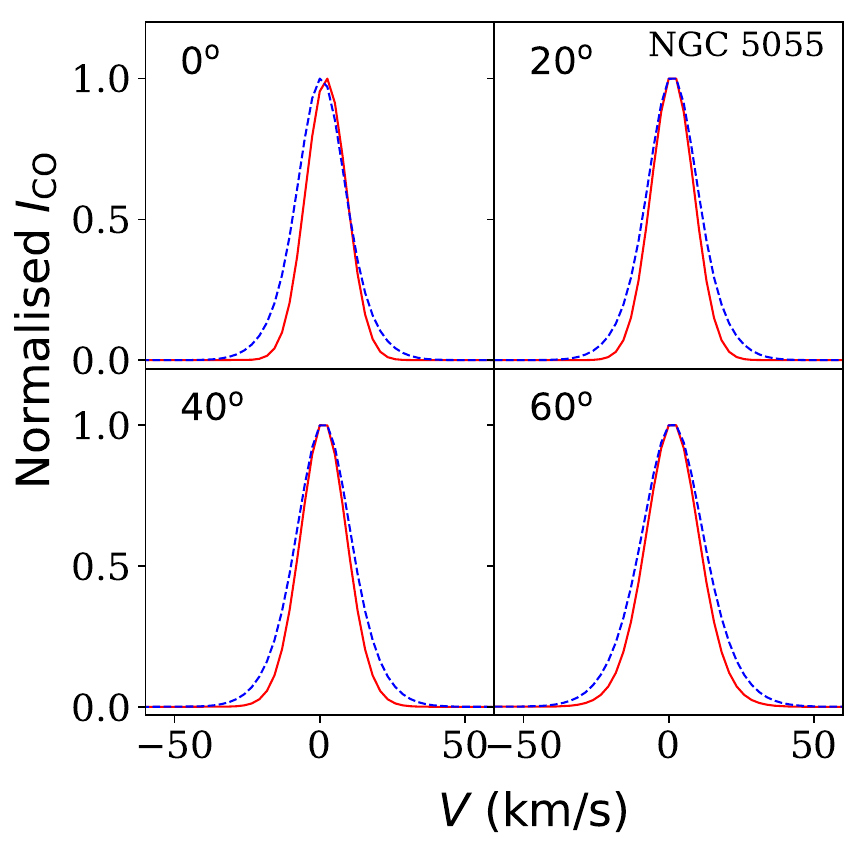}} &
\resizebox{0.23\textwidth}{!}{\includegraphics{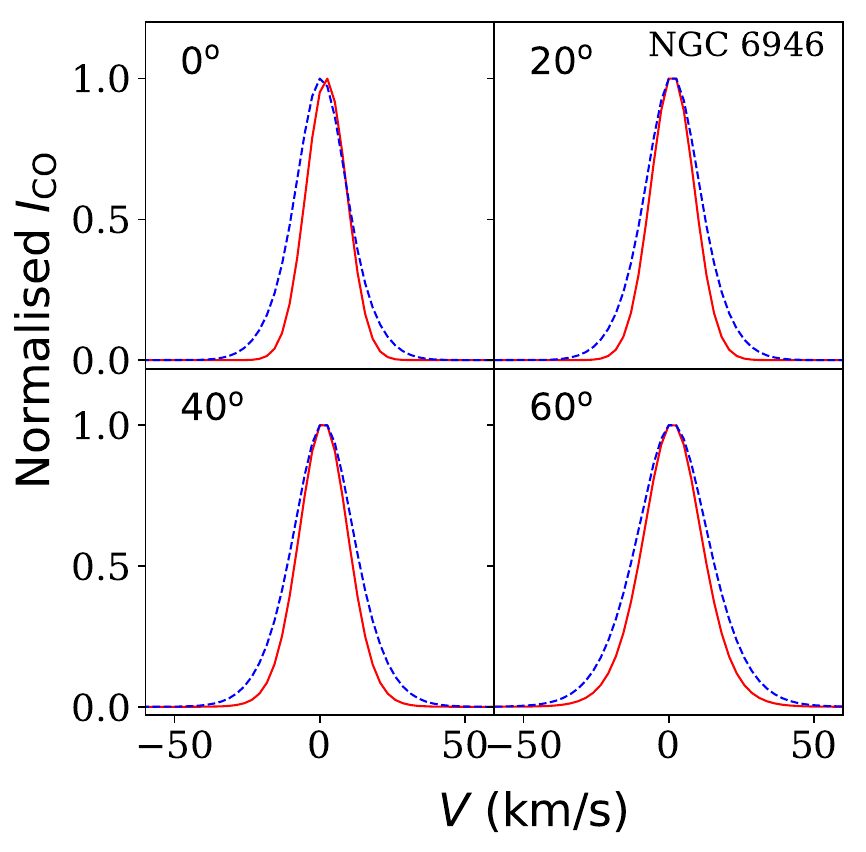}} &
\resizebox{0.23\textwidth}{!}{\includegraphics{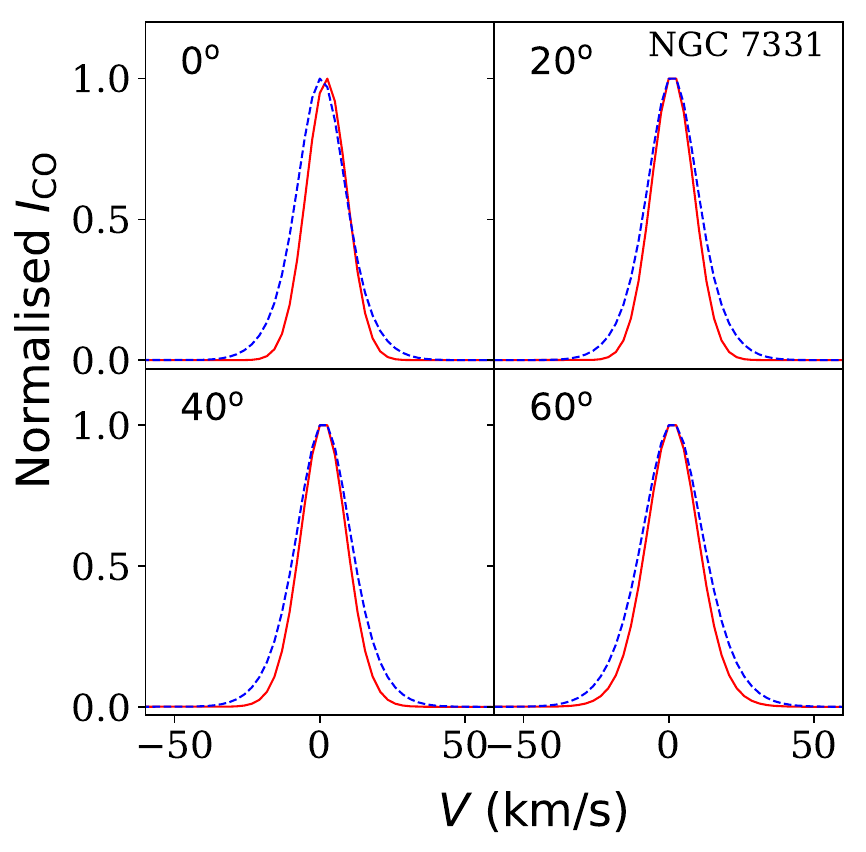}} \\
\end{tabular}
\end{center}
\caption{Comparison of the normalised stacked spectra of our sample galaxies (Same as Fig.~\ref{stack_incl}) for different inclinations. Different panels show the stacked spectra for different galaxies, as quoted in the top right corner of every panel. The solid red lines represent the stacked spectra for single-component molecular discs, whereas the blue dashed lines represent the same for two-component molecular discs. The inclinations are quoted in the top left corners of each sub-panel.}
\label{stack_incl_all}
\end{figure*}

\begin{figure*}[h]
\begin{center}
\begin{tabular}{c}
\resizebox{0.8\textwidth}{!}{\includegraphics{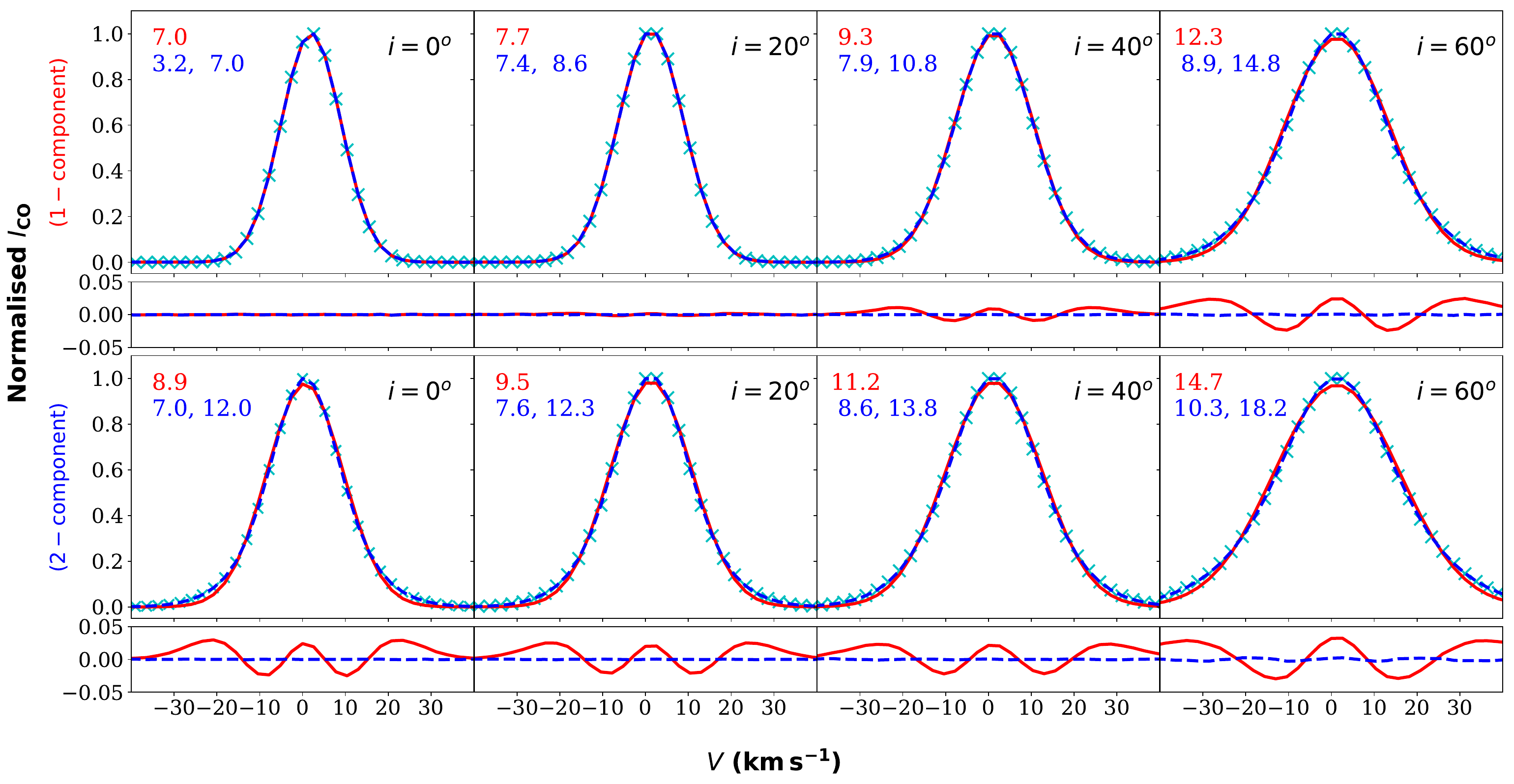}}\\
\resizebox{0.8\textwidth}{!}{\includegraphics{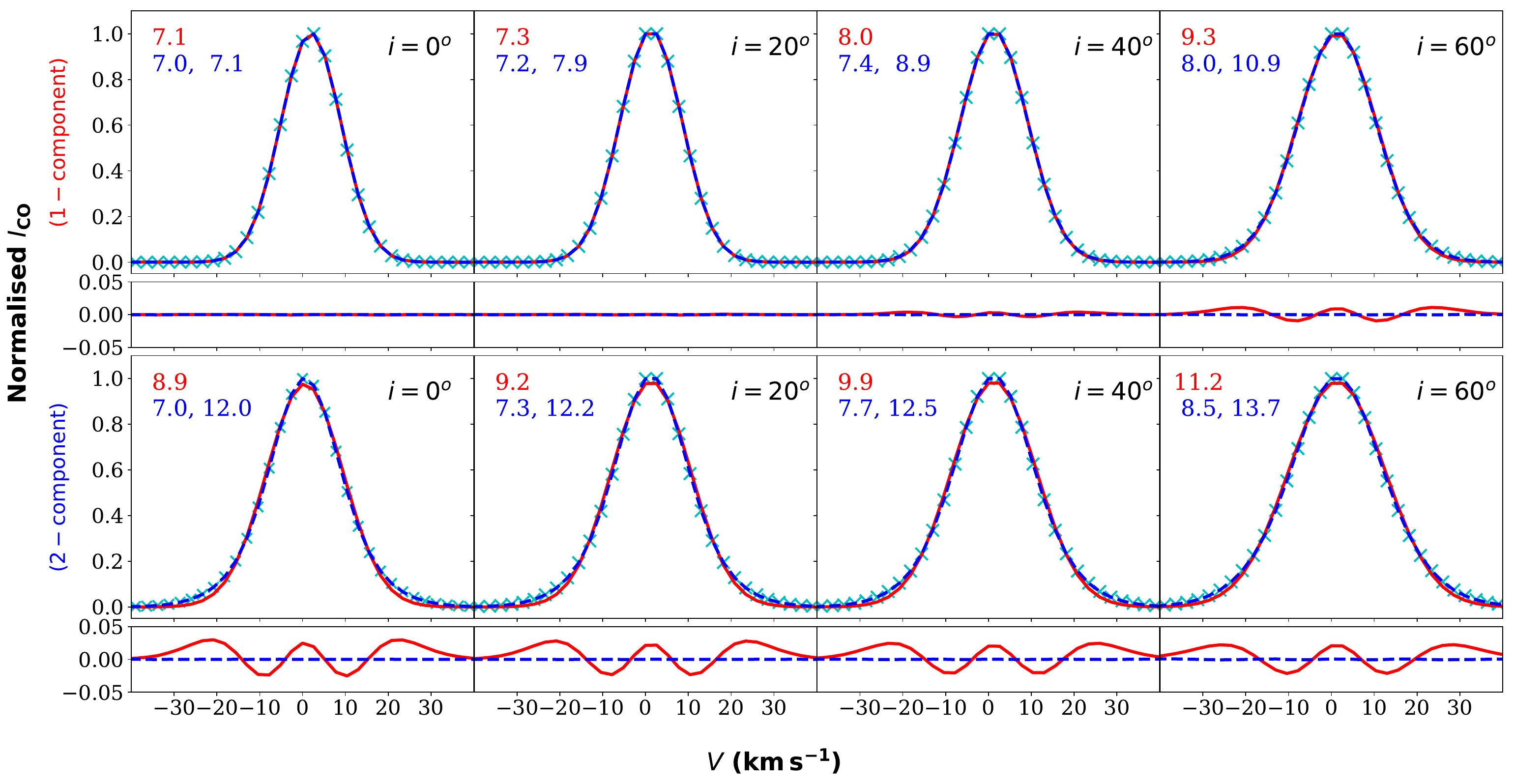}}\\
\resizebox{0.8\textwidth}{!}{\includegraphics{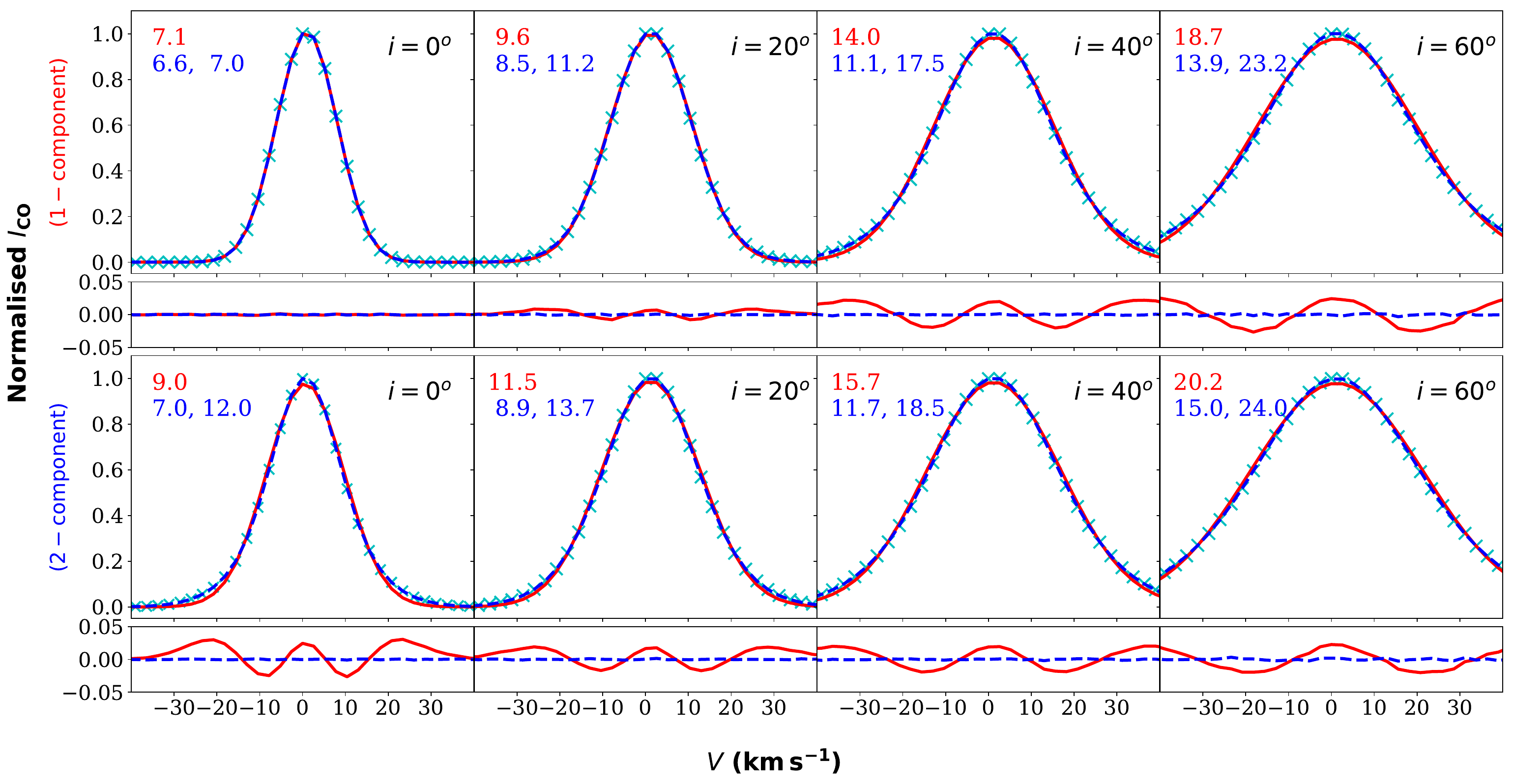}}\\
\end{tabular}
\end{center}
\caption{Same as Fig.~\ref{stack_fitg}, but for galaxies NGC 925 (top panel), NGC 2841 (middle panel), and NGC 2976 (bottom panel).}
\label{stack_fitg_all}
\end{figure*}

\renewcommand\thefigure{A.5. Continued ..}

\begin{figure*}[h]
\begin{center}
\begin{tabular}{c}
\resizebox{0.8\textwidth}{!}{\includegraphics{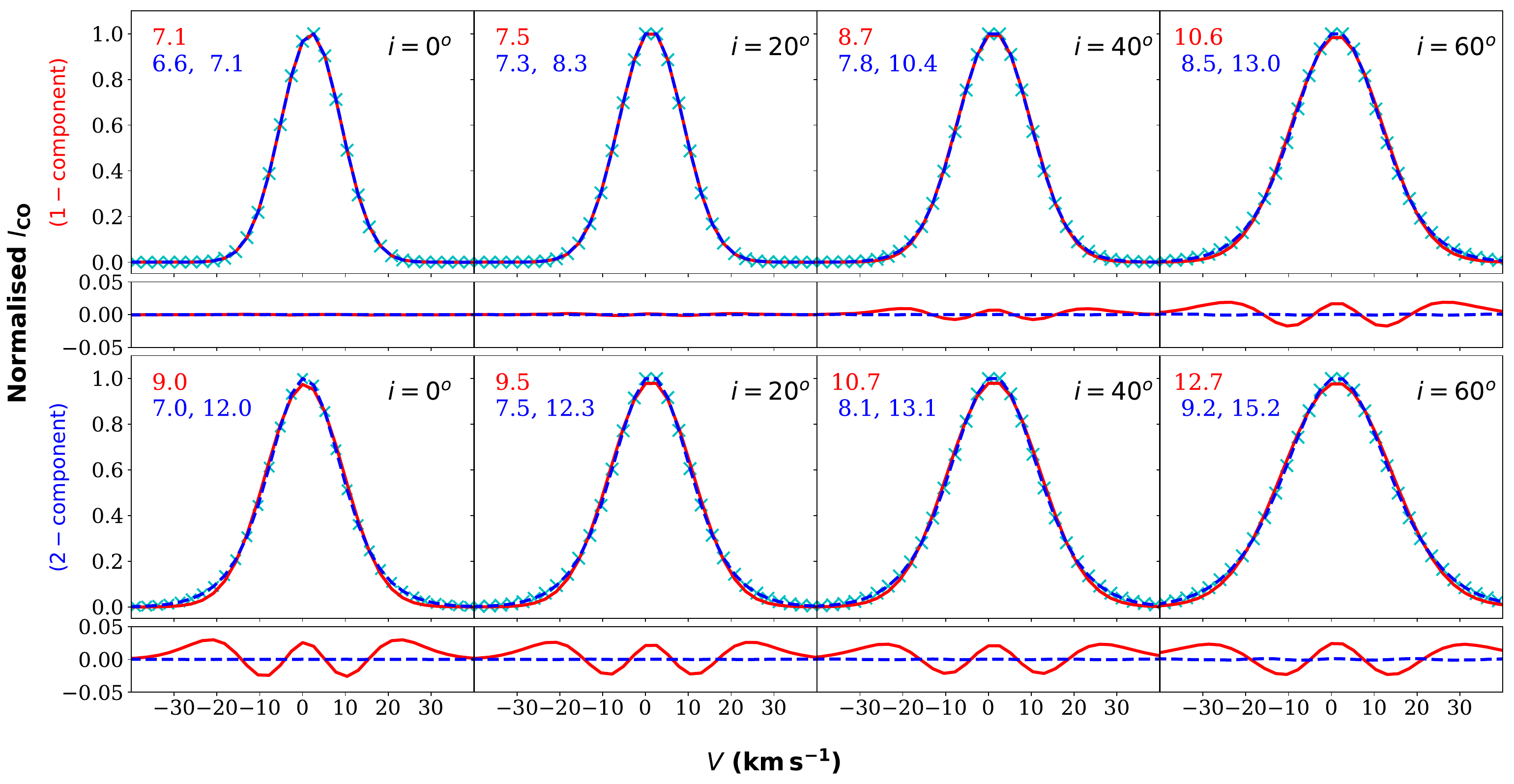}}\\
\resizebox{0.8\textwidth}{!}{\includegraphics{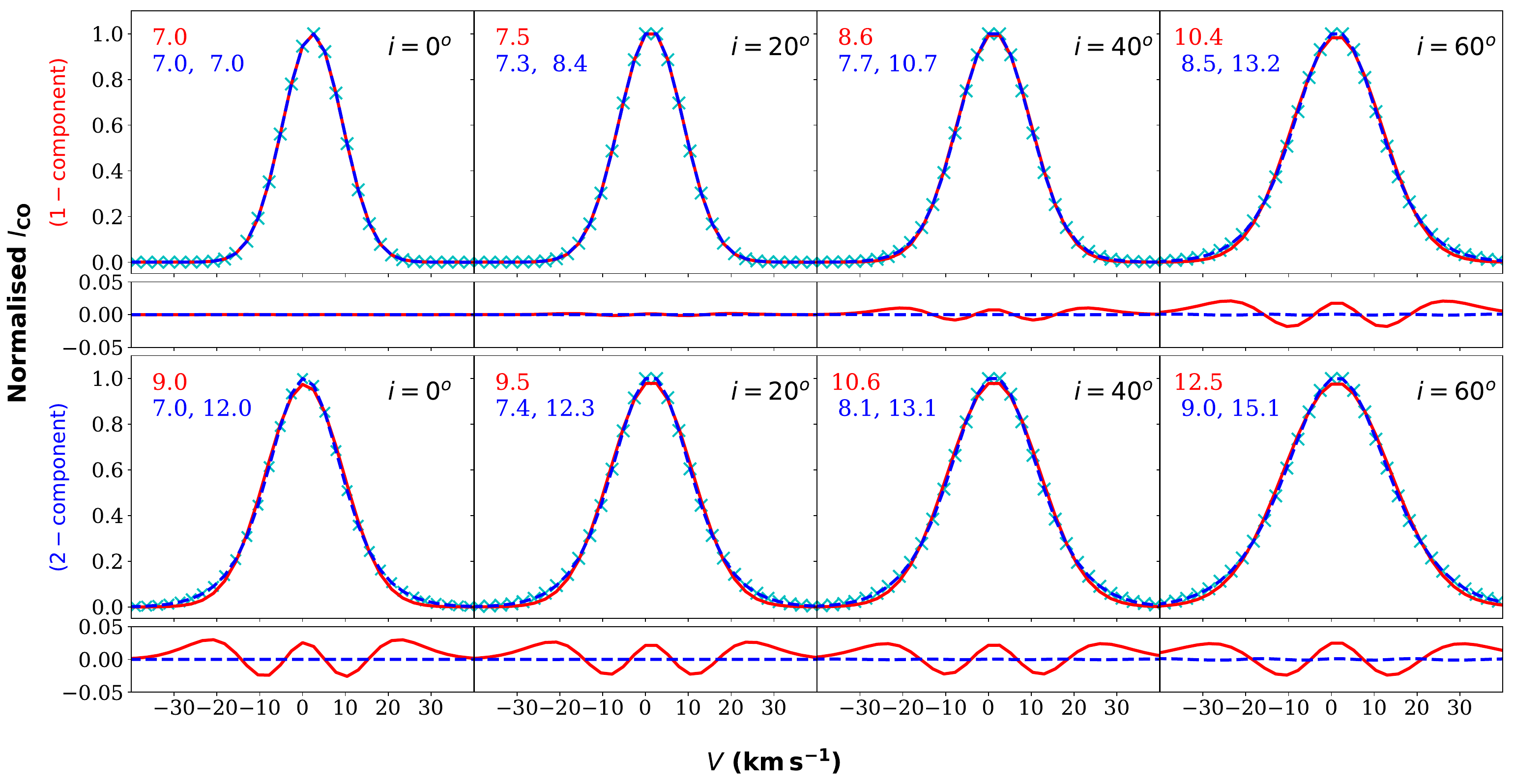}}\\
\resizebox{0.8\textwidth}{!}{\includegraphics{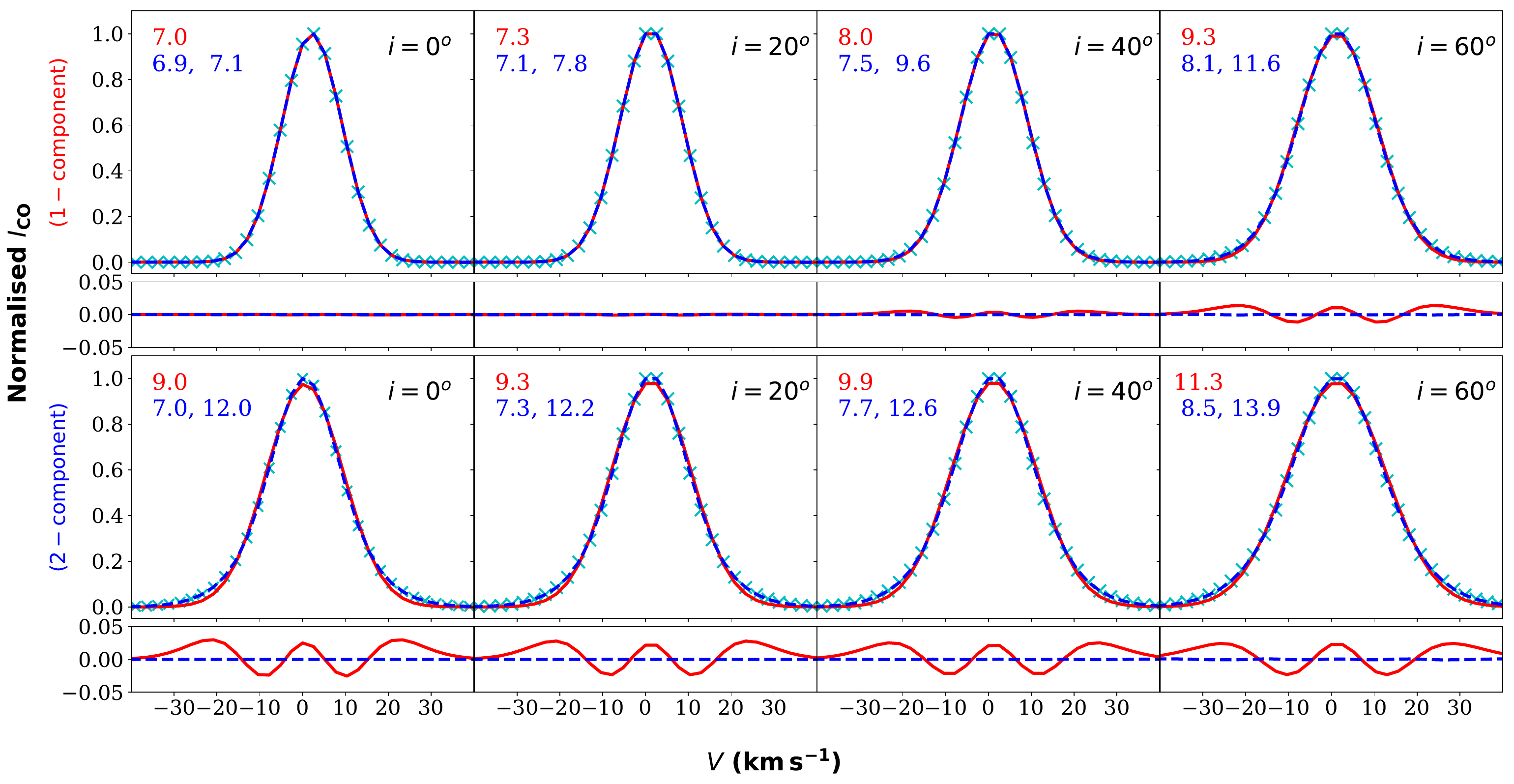}}\\
\end{tabular}
\end{center}
\caption{For galaxies NGC 3198 (top panle), NGC 3521 (middle panel), and NGC 5055 (bottom panel).}
\label{stack_fitg_all2}
\end{figure*}

\begin{figure*}[h]
\begin{center}
\begin{tabular}{c}
\resizebox{0.8\textwidth}{!}{\includegraphics{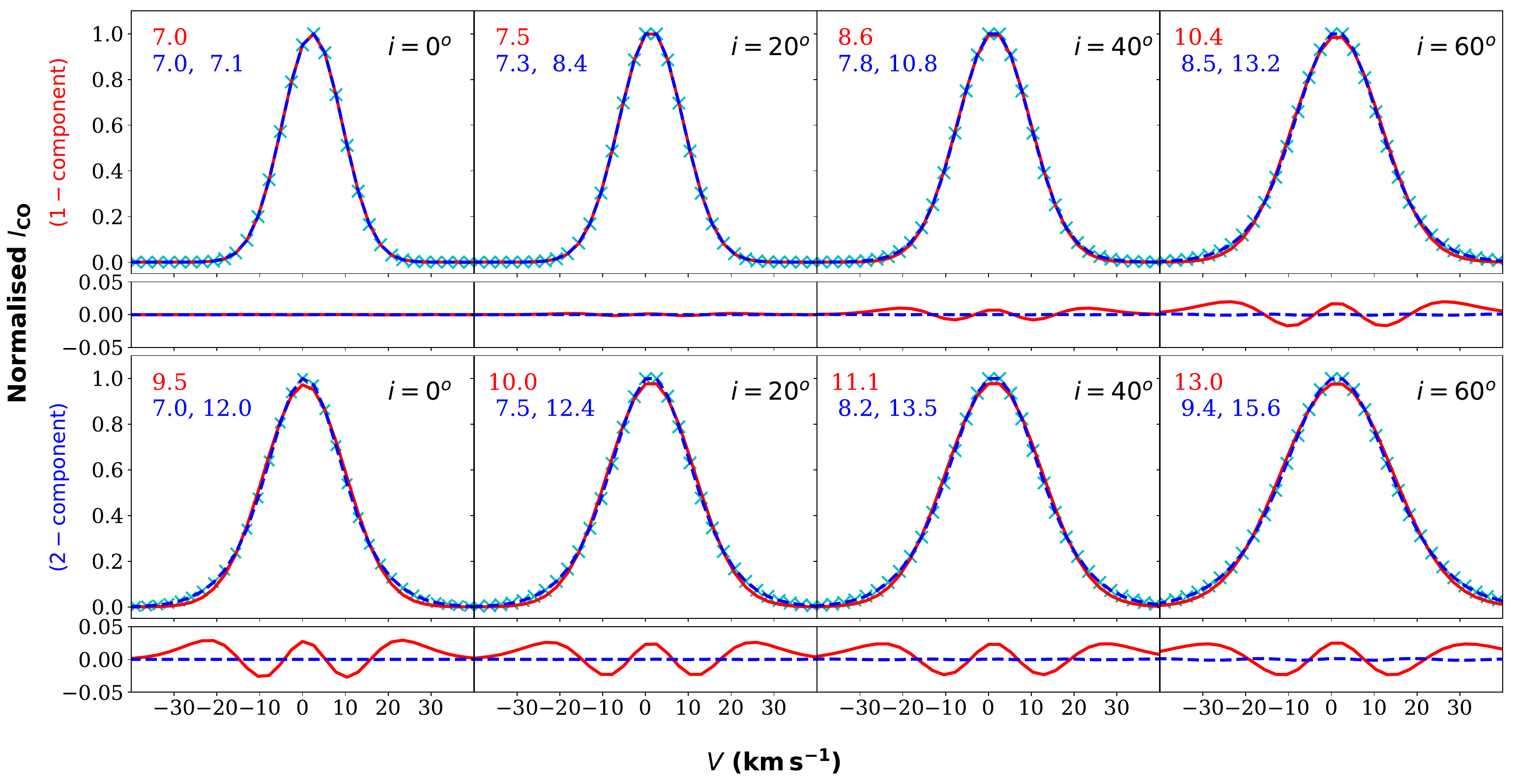}}\\
\resizebox{0.8\textwidth}{!}{\includegraphics{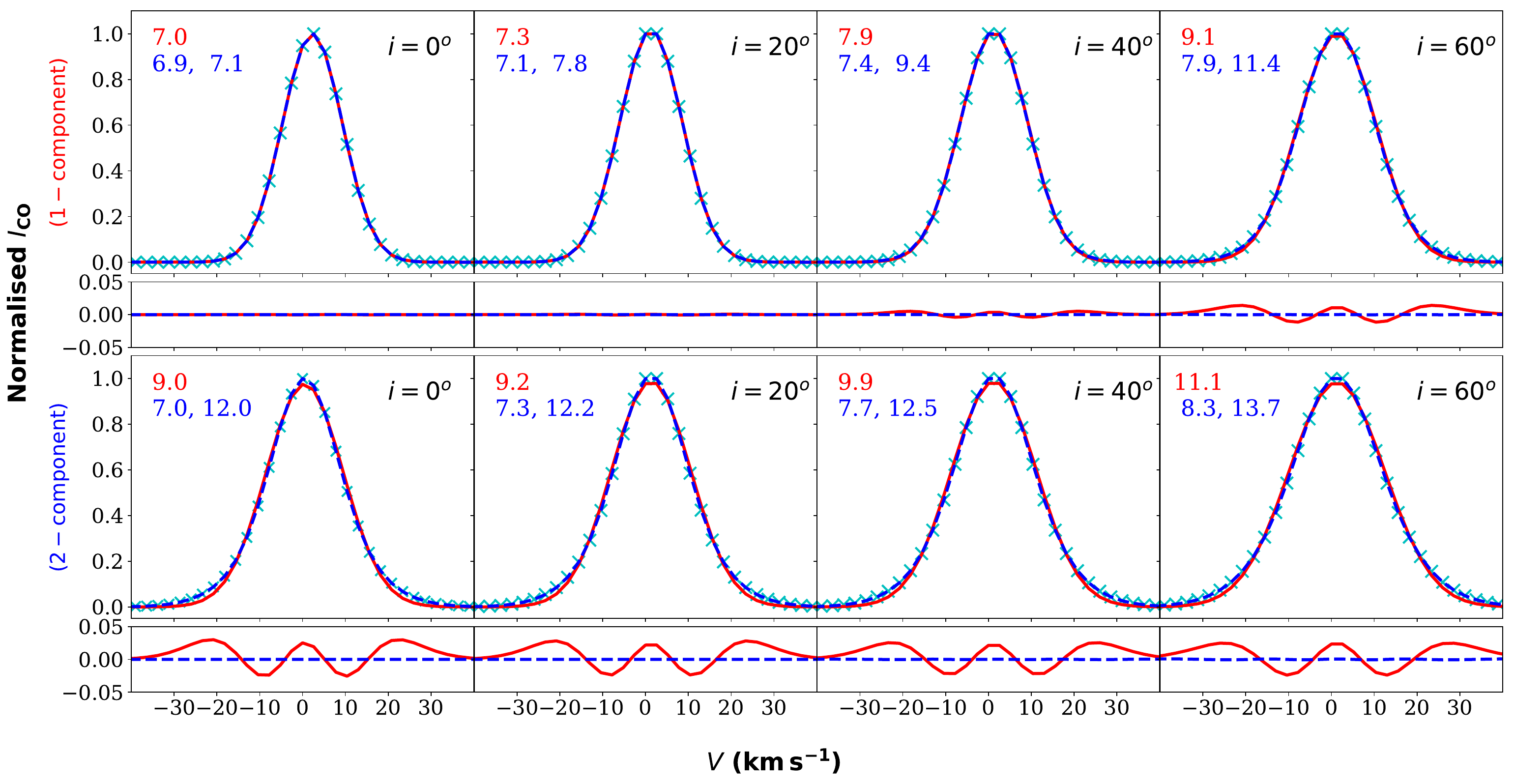}}\\
\end{tabular}
\end{center}
\caption{For galaxies NGC 6946 (top panel) and NGC 7331 (bottom panel).}
\label{stack_fitg_all3}
\end{figure*}

\section{Acknowledgement}
NNP would like to thank the anonymous referee for producing a thoughtful review of the manuscript, which helped to improve the quality of the paper. This paper uses the data products extensively from two large surveys, that is, the THINGS and the HERACLES. NNP would like to acknowledge the respective team members for making the data publicly available. NNP would like to thank Dr. Samir Choudhuri and Dr. Gunjan Verma for their valuable comments and suggestions, which have helped to improve the quality and readability of this manuscript.

\bibliographystyle{aa}
\bibliography{bibliography}

\end{document}